\documentclass[showpacs,aps,pra,twocolumn,superscriptaddress]{revtex4-1}
\usepackage[english]{babel}
\usepackage{graphicx}
\usepackage[colorlinks=true,allcolors=blue]{hyperref}
\usepackage{float}
\usepackage{bbold}
\usepackage{amssymb}
\usepackage{amsmath}
\usepackage{braket}
\usepackage{comment}
\usepackage{xcolor}
\usepackage{microtype}
\usepackage{dsfont}
\usepackage{siunitx}

\begin{document}

\def\equationautorefname~#1\null{Eq.~(#1)\null}
\renewcommand{\figureautorefname}{Fig.}

\title{Classical Approaches to Chiral Polaritonics}

\author{L. Mauro}
\affiliation{Univ. Bordeaux, CNRS, LOMA, UMR 5798, F-33400 Talence, France}
\author{J. Fregoni}
\affiliation{Departamento de Física Teórica de la Materia Condensada and Condensed Matter Physics Center (IFIMAC), Universidad Autónoma de Madrid, 28049 Madrid, Spain}
\author{J. Feist}
\affiliation{Departamento de Física Teórica de la Materia Condensada and Condensed Matter Physics Center (IFIMAC), Universidad Autónoma de Madrid, 28049 Madrid, Spain}
\author{R. Avriller}
\affiliation{Univ. Bordeaux, CNRS, LOMA, UMR 5798, F-33400 Talence, France}

\date{\today}
%
%
\begin{abstract}
We provide a theoretical framework based on classical electromagnetism, to describe optical properties of Fabry-P\'erot cavities, filled with multilayered and linear chiral materials. 
We find a formal link between transfer-matrix, scattering-matrix and Green-function approaches to compute the polarization-dependent optical transmission, and cavity-modified circular dichroism signals.
We show how general symmetries like Lorentz's reciprocity and time-reversal symmetry constrain the modelling of such cavities.
We apply this approach to investigate numerically and analytically the properties of various Fabry-P\'erot cavities, made of either metallic or helicity-preserving dielectric photonic crystal mirrors.
In the latter case, we analyze the onset of \textit{chiral cavity-polaritons} in terms of partial helicity-preservation of electromagnetic waves reflected at the mirrors interfaces. 
Our approach is relevant for designing innovative Fabry-P\'erot cavities for chiral-sensing, and for probing cavity-modified stereochemistry.
\end{abstract}
\maketitle

%
%
\section{Introduction}
The notion of \textit{chirality} is central in many research fields including optics, stereochemistry and biology \cite{Collet2012}.
Usually, an object is said to be chiral if it cannot be superimposed with its mirror image.
The different classes of existing mirror-images of the object with given handedness are called enantiomers.
Chirality is thus defined in terms of a lack of spatial symmetry of the given object, namely with respect to a lack of planar symmetries.
%
%
%
One consequence of chirality at the molecular level arises in optics, where linearly-polarized light propagating in a chiral medium rotates with a different angle, depending on the medium chirality.
This phenomenon, called \textit{optical activity}, is due to the medium refractive index, which is slightly different depending on whether the propagating wave is left circularly polarized (LCP) or right circularly polarized (RCP) \cite{landau2013electrodynamics,Craig1998,Barron2004}.
The phenomenon that is the Kramers-Kronig conjugate of optical activity is called \textit{circular dichroism} (CD).
It stands for the different absorption-rate between LCP and RCP electromagnetic-waves propagating inside the chiral medium \cite{Tang2010,Lozano2018,Purdie1989,snatzke_circular_1968,Hecht1994,Mun2020}.
The measurement and analysis of those chiroptical observables are important in stereochemistry, since they enable to characterize the handedness and degree (purity) of chirality of a given material or solution, namely to get information about the relative concentration of its chiral molecular constituent. 
Many theoretical works have been devoted to understanding and describing the structure-property relationship of molecules constituting a chiral medium. 
The early work of Condon \cite{Condon1937} proposed a simple microscopic model of an isotropic and homogeneous chiral medium, from which he derived macroscopic constitutive relations at the origin of optical activity and circular dichroism.
Slightly different sets of constitutive relations (depending on the material under study) were later proposed \cite{Post,Jaggard1979,Lindell1994}, providing a robust ground for developing a theory of light scattering in multilayered, linear and chiral media \cite{Bassiri1990,Silverman1994,Silverman1986,Silverman1990,Jaggard1992}.
Recently, the study of chirality has been the subject of many developments, in particular due to the extension of the notion of chirality to more general and abstract classes of objects, like the electromagnetic field itself: this extended notion of chirality is however a delicate concept to define and apprehend, due the interaction between light and matter \cite{andrews2018quantum}. 
Regarding experiments, on a practical ground, even for highly concentrated samples, the interaction between chiral molecules and light is unfortunately intrinsically weak, thus inducing chiroptical signals of relative order $10^{-5}-10^{-3}$ \cite{Craig1998,Condon1937,Barron2004}.
There is therefore a strong need for the design of alternative devices that would enhance this interaction strength.
Artificial photonic materials \cite{Choi2013,Park2019,Gao2020,Chen2022,Kondraton2016,Collins2017,Wu2018,Mohammadi2018} are such candidates, using the near-field enhancement of plasmonic nanoparticles to boost their interaction with the chiral molecules \cite{Lieberman2008,Govorov2013,Weiss2016,chiralthiophene,Vestler2019,Kneer2018}.
Planar metasurfaces are other types of structures that were realized to have selective scattering properties with respect to the handedness of the incoming light \cite{Corbaton2019,menzel_advanced_2010,Kuwata2005,solomon_enantiospecific_2019,Li2013,Oh2015,Papakostas2003,Zhang2009,Decker2009,Ni2021}. 
A key role in the design and efficiency of such structures is played by Lorentz's reciprocity \cite{Carminati:98,carminati2000reciprocity,landau2013electrodynamics}, namely the symmetry of electromagnetic fields with respect to swapping the input (source) and output (detector) of a linear optical system \cite{sigwarth_time_2022}.
The general classification of reciprocal structures exhibiting chirality and their related electromagnetic response, was performed extensively by Drezet \textit{et al.} \cite{drezet_reciprocity_2017}.
It enabled to separate and clarify the concepts of time-reversal symmetry and reciprocity. 
Another strategy to enhance light-matter interactions is to confine light and molecules in optical resonators, such as Fabry-P\'{e}rot (FP) cavities \cite{Born1980}. 
In such devices, light undergoes multiple internal reflections, with the direct consequence of increasing the optical path length inside the cavity, opening the possibility to interact resonantly with the embedded material.
The capability of FP optical cavities to manipulate the molecular properties upon achieving light-matter electronic strong coupling has been demonstrated experimentally \cite{ebbesen2016hybrid,schwartz2011reversible}.
In this case, the electric transition dipoles of the embedded organic molecules couple collectively to the confined electromagnetic cavity modes, causing the emergence of hybrid light-matter excitations called \textit{polaritons}. 
However, even for FP cavities in the polaritonic regime, the related enhancement of chiroptical signals is rather limited. 
Indeed, normal (metallic or dielectric) mirrors revert the helicity of light upon reflection \cite{jackson1999classical}, thus resulting in a lack of significant additional imbalance in the polarization content of the electromagnetic modes stored inside the cavity.
For this reason, recent years have witnessed many proposals \cite{voronin2021single,Corbaton2020-2,Sun2022,Corbaton2020,Yoo2015} as well as the first realization of helicity-preserving (HP) mirrors \cite{Maruf2020} and FP cavities \cite{Cao2020,Gautier2021,Yuan2021}.
The latter support chiral cavity-modes, which in turn can couple efficiently to the hosted molecular material.
%
%
\textcolor{black}{Recently, we predicted theoretically that such an HP Fabry-P\'erot cavity can enhance the intrinsic chiroptical signal generated by the embedded chiral molecules, making it potentially useful for chiral-sensing measurements \cite{PhysRevA.107.L021501}.}
%
%

%
In this paper, we complement our previous work \cite{PhysRevA.107.L021501} and provide an in-depth and comprehensive tutorial-style guide of the numerical and analytical methods we developed to investigate chiral light-matter interactions in HP Fabry-P\'erot cavities.
Moreover, we analyze in detail the mechanism responsible for the formation of \textit{chiral cavity-polaritons} reported in \cite{PhysRevA.107.L021501}.
The organisation of this paper is the following. 
In Sec. \ref{genRecipr}, we provide an overview of Lorentz's reciprocity in electromagnetism and its consequences on relevant observable in optics. 
In Sec. \ref{Cmodel}, we present Condon's microscopic model of optical activity in a Pasteur medium, and the associated macroscopic constitutive relations.
We propose in Sec. \ref{chiralTM} a numerical transfer-matrix code for multilayered and linear Pasteur media in presence of losses (absorption), that enables to compute their output optical transmittance and chiroptical signals. 
%
%
%
In Sec. \ref{GenGreen}, we derive an explicit relation between the transmission-matrix of the cavity and its Green's function.
We exemplify the previous approaches in Sec. \ref{FPsilver}, by computing numerically and analytically the chiroptical signals out of a standard FP cavity made of silver mirrors.
In Sec. \ref{HPfabry}, we complement and compare these calculations to the case of HP cavities, for which we analyze the mechanism responsible for the emergence of chiral cavity-polaritons.
Conclusions and perspectives of this work are given in the final Sec. \ref{Conc}.
%

\section{Reciprocity}
\label{genRecipr}
In this section, we generalize the Lorentz reciprocity theorem \cite{landau2013electrodynamics,jackson1999classical} to linear and chiral media.
We show that reciprocity induces some relations among the components of the transmission and reflection electromagnetic matrices.
Consequences on the scattering properties of reciprocal and chiral electromagnetic Fabry-P\'{e}rot cavities are outlined.
%

\subsection{Onsager-Casimir relations for reciprocal systems}
\label{OnsagerCasimir}
We start our derivation by writing the macroscopic Maxwell's equations \cite{landau2013electrodynamics,jackson1999classical}
\begin{equation}    
\begin{aligned}
\vec{\nabla}\wedge \vec{E} &= -\frac{\partial\vec{B}}{\partial t}, &
\vec{\nabla}\cdot  \vec{B} &= 0,\\
\vec{\nabla}\wedge \vec{H} &= \vec{j}+\frac{\partial\vec{D}}{\partial t}, &
\vec{\nabla}\cdot  \vec{D} &= \rho,
\end{aligned}
\label{maxw}
\end{equation}
where $\rho$ and $\vec{j}$ are the distributions of free charges and currents, respectively.
The electromagnetic properties of the medium are described by the constitutive relations
%
%
%
%
\begin{equation}
\begin{aligned}
\vec{D} &= \varepsilon_{0}\vec{E}+\vec{P},\\
\vec{B} &= \mu_{0}\left(\vec{H}+\vec{M}\right),
\label{DisplacementInduction}
\end{aligned}
\end{equation}
linking the electric displacement vector $\vec{D}$ and magnetic induction $\vec{B}$ to the
electric and magnetic polarization vectors $\vec{P}$ and $\vec{M}$.
The vacuum permittivity (permeability) is written as $\varepsilon_{0}$ ($\mu_{0}$).
In a linear and chiral medium, the $\vec{P}$ and $\vec{M}$ vectors are linearly related to the electric $\vec{E}$ and magnetic $\vec{H}$ fields, so that the constitutive relations take the compact form 
\begin{equation}
\vec{\mathcal{D}}=\stackrel{\leftrightarrow}{\alpha}\vec{\mathcal{E}},
\end{equation}
with
\begin{equation}
\vec{\mathcal{D}}=\begin{bmatrix}
\vec{D}\\
\vec{B}\\
\end{bmatrix},\,
\stackrel{\leftrightarrow}{\alpha}=\begin{bmatrix}
\stackrel{\leftrightarrow}{\varepsilon} & \stackrel{\leftrightarrow}{\nu}\\
\stackrel{\leftrightarrow}{\xi} & \stackrel{\leftrightarrow}{\mu}\\
\end{bmatrix},\,
\vec{\mathcal{E}}=\begin{bmatrix}
\vec{E}\\
\vec{H}\\
\end{bmatrix}.
\label{genconst}
\end{equation}
We introduced in eq. \eqref{genconst}, the generalized electromagnetic permittivity tensor $\stackrel{\leftrightarrow}{\alpha}$.
The latter contains the permittivity tensor $\stackrel{\leftrightarrow}{\varepsilon}$ and permeability tensor $\stackrel{\leftrightarrow}{\mu}$, as well as the magneto-electric tensors $\stackrel{\leftrightarrow}{\nu}$ and $\stackrel{\leftrightarrow}{\xi}$ at the origin of optical activity and circular dichroism of the medium.
The components of the $\stackrel{\leftrightarrow}{\alpha}$-tensor are not arbitrary, but are constrained by microscopic reversibility (in the absence of any static external magnetic field), as expressed by the Onsager-Casimir (OC) relations \cite{Onsager1931,Onsager2,RevModPhys.17.343,landau2013electrodynamics}
\begin{equation}
\alpha_{\beta\gamma}=\pm \alpha_{\gamma\beta}
\label{ORL1},
\end{equation}
with a positive (negative) sign if both variables labelled as $\beta,\gamma$ have the same (different) parity with respect to time-reversal symmetry.
From eq. \eqref{ORL1}, it follows that
\begin{eqnarray}
\stackrel{\leftrightarrow}{\varepsilon} &=& \left(\stackrel{\leftrightarrow}{\varepsilon}\right)^{\mathrm{T}}
\label{ORL2},\\
\stackrel{\leftrightarrow}{\mu} &=& \left(\stackrel{\leftrightarrow}{\mu}\right)^{\mathrm{T}}
\label{ORL3},\\
\stackrel{\leftrightarrow}{\nu} &=& - \left(\stackrel{\leftrightarrow}{\xi}\right)^{\mathrm{T}}
\label{ORL4} \, ,
\end{eqnarray}
with $^{\mathrm{T}}$ meaning the transpose of a given tensor.
%
%

\subsection{Lorentz's reciprocity theorem for chiral media}
%
\begin{figure}[ht!]
\begin{center}
\includegraphics[width=0.4\linewidth]{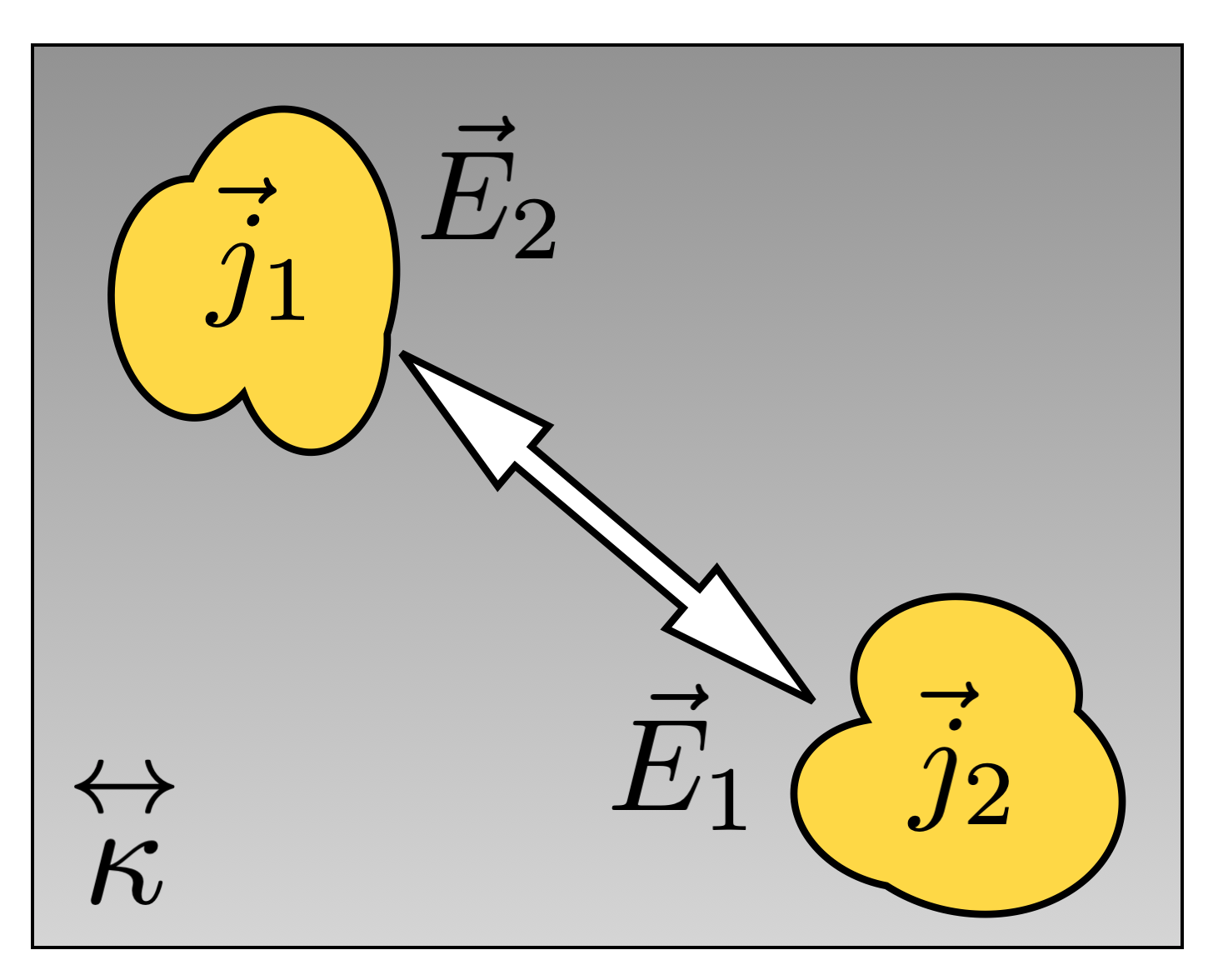}
\caption{Pictorial representation of the reciprocity symmetry obtained by exchanging two current distributions $\vec{j}_1$ and $\vec{j}_2$, at the origin of the two electric fields $\vec{E}_1$ and $\vec{E}_2$, respectively.}
\label{Fig2}
\end{center}
\end{figure}
%
%
%
In this section, we extend the demonstration of Lorentz's reciprocity theorem \cite{landau2013electrodynamics}, to the case of a linear and chiral medium.
To this purpose, we consider two sources of volumic current $\vec{j}_1(\vec{r}_1)$ and $\vec{j}_2(\vec{r}_2)$ embedded in the chiral medium (as depicted in Fig. \ref{Fig2}). 
The volumes associated to each current are $\mathcal{V}_1$ and $\mathcal{V}_2$, respectively.
We denote as $\left( \vec{E}_1, \vec{H}_1\right)$ and $\left( \vec{E}_2, \vec{H}_2\right)$ the electric and magnetic field generated by each source.
Maxwell's equations plus the superposition principle imply that
\begin{equation}
\begin{split}
\vec{\nabla}& \cdot \left\lbrack
\vec{H}_1 \wedge \vec{E}_2 - \vec{H}_2 \wedge \vec{E}_1
\right\rbrack = 
\vec{j}_1 \cdot \vec{E}_2 - \vec{j}_2 \cdot \vec{E}_1
-i\omega \Delta,
\label{LR}
\end{split}
\end{equation}
with
\begin{equation}
\begin{split}
\Delta = \vec{D}_1 \cdot \vec{E}_2 - \vec{D}_2 \cdot \vec{E}_1
+ \vec{H}_1 \cdot \vec{B}_2 - \vec{H}_2 \cdot \vec{B}_1
\label{LR42}.
\end{split}
\end{equation}
Making use of the constitutive relations (see eq. \eqref{genconst}), we obtain
$\Delta = \Delta_{\varepsilon} + \Delta_{\mu} + \Delta_{\nu}$, with
\begin{eqnarray}
\Delta_{\varepsilon} &=& \left\lbrack
\left(\stackrel{\leftrightarrow}{\varepsilon}\vec{E}_1\right) \cdot \vec{E}_2
-
\vec{E}_1 \cdot \left(\stackrel{\leftrightarrow}{\varepsilon} \vec{E}_2\right)
\right\rbrack
\label{LR4} \, , \\
\Delta_{\mu} &=& \left\lbrack
\vec{H}_1 \cdot \left(\stackrel{\leftrightarrow}{\mu}\vec{H}_2\right)
-
\left(\stackrel{\leftrightarrow}{\mu} \vec{H}_1\right) \cdot \vec{H}_2
\right\rbrack
\label{LR5} \, , \\
\Delta_{\nu} &=& \left\lbrack
\left(\stackrel{\leftrightarrow}{\nu} \vec{H}_1\right) \cdot \vec{E}_2
+
\vec{H}_1 \cdot \left(\stackrel{\leftrightarrow}{\xi} \vec{E}_2\right)
\right\rbrack
\nonumber \\
&+& \left\lbrack
-\left(\stackrel{\leftrightarrow}{\xi} \vec{E}_1\right) \cdot \vec{H}_2
-
\vec{E}_1 \cdot \left(\stackrel{\leftrightarrow}{\nu} \vec{H}_2\right)
\right\rbrack
\label{LR6} \, .
\end{eqnarray}
The ``diagonal" terms in eqs. \eqref{LR4}, \eqref{LR5} are associated to the pure electric and pure magnetic electromagnetic response of the medium, respectively. The ``off-diagonal" terms in eq. \eqref{LR6} are absent for an achiral medium, and are associated to optical activity and circular dichroism.
Inserting eqs. \eqref{ORL2}, \eqref{ORL3} and \eqref{ORL4} in eqs. \eqref{LR4}, \eqref{LR5} and \eqref{LR6}, we obtain that $\Delta=0$.
%
%
This implies that eq. \eqref{LR} can be simplified to
\begin{eqnarray}
&&\vec{\nabla} \cdot \left\lbrack
\vec{H}_1 \wedge \vec{E}_2 - \vec{H}_2 \wedge \vec{E}_1
\right\rbrack = 
\vec{j}_1 \cdot \vec{E}_2 - \vec{j}_2 \cdot \vec{E}_1
\label{LR11} \, . 
\end{eqnarray}
We further integrate eq. \eqref{LR11} in a volume $\mathcal{V}$ including $\mathcal{V}_1$ and $\mathcal{V}_2$.
Upon letting $\mathcal{V} \rightarrow\infty$, with proper asymptotic conditions for the fields \cite{Carminati:98}, the surface integrals (generated by the left-hand side of eq. \ref{LR11}) do vanish, and we obtain the final result
\begin{eqnarray}
\int_{\mathcal{V}_1} d^3\vec{r}_1\; \vec{j}_1\left(\vec{r}_1\right) \cdot \vec{E}_2\left(\vec{r}_1\right) 
=
\int_{\mathcal{V}_2} d^3\vec{r}_2\; \vec{j}_2\left(\vec{r}_2\right) \cdot \vec{E}_1\left(\vec{r}_2\right)  
\nonumber \, . \\
\label{LR12} 
\end{eqnarray}
This equation is indeed the extension of Lorentz's reciprocity \cite{landau2013electrodynamics} to the case of linear and chiral systems.
It expresses a relation between the electric fields generated upon exchange between two different sources.
This nontrivial result is a deep consequence of microscopic reversibility expressed by the OC's relations in eq. \eqref{ORL1}.
%

%
\subsection{Beyond dipole approximation}
We now suppose that the volumes $\mathcal{V}_1$ and $\mathcal{V}_2$ are sufficiently far away from each other, so that the electric field $\vec{E}_2\left(\vec{r}_1\right)$ radiated by the source 2 does not vary much spatially on the scale of the other current distribution $1$.
Defining an arbitrary point $\vec{r}^{(0)}_1\equiv\vec{0}\in \mathcal{V}_1$, we expand the electric field $\vec{E}_2\left(\vec{r}_1\right)$ around that point:
\begin{eqnarray}
\vec{E}_{2}\left(\vec{r}_1\right) &\approx &
\vec{E}_{2}(\vec{0})
+ 
\left(\vec{r}_{1} \cdot \vec{\nabla}\right) \vec{E}_{2}(\vec{0})
\label{DA1} \, .
\end{eqnarray}
We evaluate the left-hand side of eq. \eqref{LR12}:
\begin{eqnarray}
\int_{\mathcal{V}_1} d^3\vec{r}_1\; \vec{j}_1\left(\vec{r}_1\right) \cdot \vec{E}_2\left(\vec{r}_1\right)  \approx  
-i\omega \left( \vec{\mu}_1\cdot \vec{E}_{2}(\vec{0})
-
\vec{m}_1\cdot \vec{B}_{2}(\vec{0}) \right)
\, , \nonumber \\
\label{DA2} 
\end{eqnarray}
where $\vec{\mu}_1$ and $\vec{m}_1$ are the electric and magnetic dipole moments of the distribution $1$.
A similar equation is obtained by exchanging $1$ and $2$.
The first term on the right-hand side of eq. \eqref{DA2} results from the standard electric-dipole approximation, while the second one is the magnetic contribution coming from next order corrections in the spatial variation of the electric field \cite{Condon1937,Craig1998}.
Note that the contribution of the electric quadrupole moment has been neglected in eq. \eqref{DA2}, since it yields a vanishing contribution upon rotational averaging of the molecules orientation \cite{Craig1998} (see Appendix \ref{BeCustoState} for a complete expression including the electric quadrupole). 
We then obtain 
\begin{eqnarray}
\vec{\mu}_1\cdot \vec{E}_{2}\left( \vec{r}_{1} \right) 
-
\vec{m}_1\cdot \vec{B}_{2}\left( \vec{r}_{1} \right)
&=& 
\vec{\mu}_2\cdot \vec{E}_{1}\left( \vec{r}_{2} \right) 
-
\vec{m}_2\cdot \vec{B}_{1}\left( \vec{r}_{2} \right)
\nonumber \, . \\
\label{DA3}  
\end{eqnarray}
This statement expresses the Lorentz reciprocity theorem in optically active media, for which the contribution of the magnetic dipole has to be taken into account.
%
%
%
We note that the minus sign in front of the magnetic term is fundamental, and is due to the fact that the magnetic field is odd under time reversal.
%

\subsection{Green's function}
\label{Greenscatt}
%
%
We now consider a linear, chiral and reciprocal (LCR) scattering medium, in between two free-space regions with two sources of electromagnetic field infinitely far away.
We denote the free-space region hosting one source as 1, and the free-space region hosting the other source as 2. 
The system is sketched in Fig. \ref{Fig3}.
Either region (1 or 2) can play the role of source or detector of electromagnetic fields.
The statement of reciprocity given by eq. \eqref{DA3} expresses a symmetry upon exchanging source and detector in the electromagnetic scattering problem.
The latter can be written in a compact form
\begin{eqnarray}
\left(\vec{\mathcal{P}}_{1}\right)^{\mathrm{T}} \hat{\sigma}_z \vec{\mathcal{E}}_2(1) =\left(\vec{\mathcal{P}}_2\right)^{\mathrm{T}} \hat{\sigma}_z \vec{\mathcal{E}}_1(2)
\label{EGM2} \, , 
\end{eqnarray} 
with $\left(\vec{\mathcal{P}}\right)^{\mathrm{T}}\equiv\left(\vec{\mu},\mu_{0}\vec{m}\right)$ a vector made of the electric and magnetic dipole moments acting as a source, $\vec{\mathcal{E}}$ the vector of generated electric and magnetic fields defined in eq. \eqref{genconst}, and $\hat{\sigma}_z=\mathrm{diag}\left(\mathds{Id},-\mathds{Id}\right)$ a tensor made of the $2\times 2$ identity tensor $\mathds{Id}$.
We now introduce the electromagnetic Green's tensor $\stackrel{\leftrightarrow}{\mathcal{G}}\left(2,1\right)$, which connects the source $\vec{\mathcal{P}}_1$ (at its position in region 1) to the field $\vec{\mathcal{E}}_1(2)$ radiated by the \emph{same} source evaluated in region 2:
\begin{eqnarray}
\vec{\mathcal{E}}_1(2) &=& \stackrel{\leftrightarrow}{\mathcal{G}}\left(2,1\right) \vec{\mathcal{P}}_1
\label{EGM3} \, .
\end{eqnarray}
In other words, the Green's tensor accounts for all the scattering events (multiple transmissions and reflections) that modify the field $\vec{\mathcal{E}}_1$ when traveling from region 1 to region 2 through the LCR medium.
Inserting eq. \eqref{EGM3} in eq. \eqref{EGM2} leads to
\begin{eqnarray}
\left(\vec{\mathcal{P}}_1\right)^{\mathrm{T}} \hat{\sigma}_z \stackrel{\leftrightarrow}{\mathcal{G}}\left(1,2\right)  \vec{\mathcal{P}}_2 &=& \left(\vec{\mathcal{P}}_1\right)^{\mathrm{T}} \left\lbrack \stackrel{\leftrightarrow}{\mathcal{G}}\left(2,1\right)\right\rbrack^{\mathrm{T}} \hat{\sigma}_z \vec{\mathcal{P}}_2
\, , \nonumber \\
\label{EGM4} 
\end{eqnarray}
which is valid for any sources $\vec{\mathcal{P}}_1,\vec{\mathcal{P}}_2$.
Hence, the electromagnetic Green's tensor fulfills the following relation
\begin{eqnarray}
\left\lbrack\stackrel{\leftrightarrow}{\mathcal{G}}\left(2,1\right)\right\rbrack^{\mathrm{T}} &=& \hat{\sigma}_z \stackrel{\leftrightarrow}{\mathcal{G}}\left(1,2\right)\hat{\sigma}_z
\label{EGM5} \, ,
\end{eqnarray}
which expresses electromagnetic reciprocity at the level of the Green's functions. 
%

\subsection{Scattering matrix}
\label{SC}
%
\begin{figure}[ht!]
\begin{center}
\includegraphics[width=0.7\linewidth]{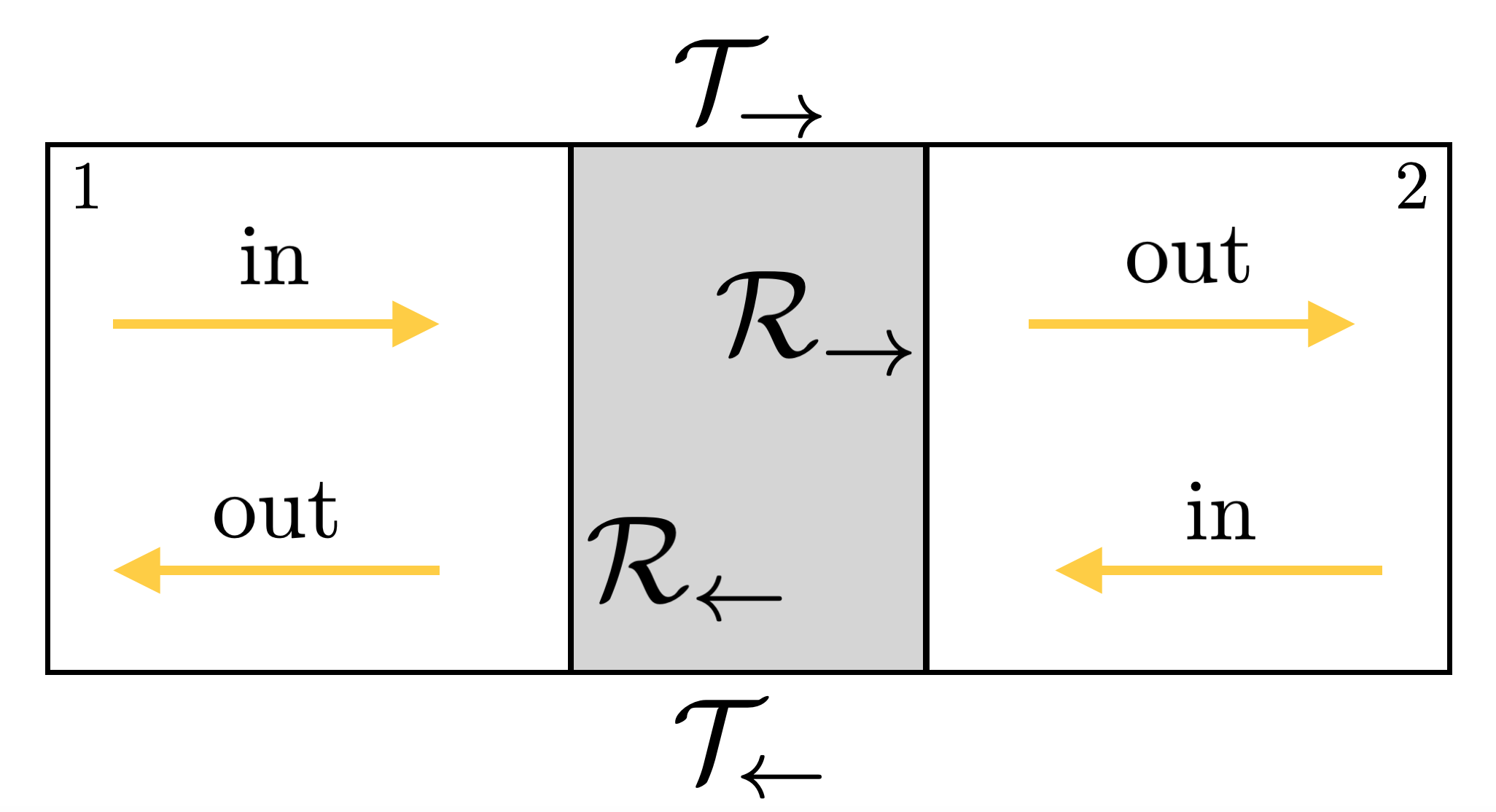}
\caption{Sketch of the scattering problem: input waves from the regions $1$ and $2$ are scattered by the central medium, leading to output waves.}
\label{Fig3}
\end{center}
\end{figure}
%
%
%
Here, we reformulate the previous scattering problem in terms of an input-output formalism \cite{drezet_reciprocity_2017,newton2013scattering}.
Input electromagnetic fields (in the far field) from the left $\vec{\mathcal{E}}_{1,\mathrm{in}}$ and from the right $\vec{\mathcal{E}}_{2,\mathrm{in}}$, are scattered by a LCR medium into output electromagnetic fields to the left $\vec{\mathcal{E}}_{1,\rm{out}}$ and to the right $\vec{\mathcal{E}}_{2,\rm{out}}$ (see Figure \ref{Fig3}).
We can relate the input and output fields by introducing the scattering matrix $\stackrel{\leftrightarrow}{\mathcal{S}}$:
\begin{equation}
\vec{\mathrm{E}}_{\rm{out}}=\stackrel{\leftrightarrow}{\mathcal{S}} \vec{\mathrm{E}}_{\rm{in}},
\label{SC3}
\end{equation}
with
\begin{equation}
\vec{\mathrm{E}}_{\rm{out}}=\begin{bmatrix}
\vec{\mathcal{E}}_{1,\rm{out}} \\
\vec{\mathcal{E}}_{2,\rm{out}}
\end{bmatrix},\,
\stackrel{\leftrightarrow}{\mathcal{S}}=\begin{bmatrix}
\mathcal{R}_{\leftarrow} & \mathcal{T}_{\leftarrow}  \\
\mathcal{T}_{\rightarrow} & \mathcal{R}_{\rightarrow}
\end{bmatrix},\,
\vec{\mathrm{E}}_{\rm{in}}=\begin{bmatrix}
\vec{\mathcal{E}}_{1,\rm{in}} \\
\vec{\mathcal{E}}_{2,\rm{in}}
\end{bmatrix},
\label{scattMat}
\end{equation}
where $\mathcal{T}_{\rightarrow}$, $\mathcal{T}_{\leftarrow}$ are the transmission matrices for electromagnetic waves traveling from left to right and right to left, respectively. 
$\mathcal{R}_{\rightarrow}$, $\mathcal{R}_{\leftarrow}$ are reflection matrices for electromagnetic waves reflected from right to right and left to left, respectively.
We suppose that the waves arriving on the scatterer are accurately described within the paraxial approximation, for which the electromagnetic field is transverse to the wave propagation and has two degrees of freedom given by the polarization. 
As in Sec. \ref{Greenscatt}, we consider the case $\vec{\mathcal{E}}_{2,\rm{in}}=\vec{0}$, for which the source is on the left of the scatterer (region 1), while the detector is on the other side (region 2).
The transmission matrix for a wave traveling from left to right then connects the output field in region 2 to the input field in region 1, according to $\vec{\mathcal{E}}_{2,\rm{out}}= \mathcal{T}_{\rightarrow}\vec{\mathcal{E}}_{1,\rm{in}}$.
This output wave can also be expressed in terms of the electromagnetic Green's function $\vec{\mathcal{E}}_{2,\rm{out}} = \stackrel{\leftrightarrow}{\mathcal{G}}(2,1)\vec{\mathcal{P}}_1$.
Since the input (source) field $\vec{\mathcal{E}}_{1,\rm{in}}$ is a linear function of the dipolar source $\vec{\mathcal{P}}_1$, we can define a matrix $\mathcal{N}$ such that $\vec{\mathcal{E}}_{1,\rm{in}} = \mathcal{N}_1\vec{\mathcal{P}}_1$.
For far fields in the paraxial approximation, the matrices $\mathcal{N}_1$ and $\mathcal{N}_2$ are diagonal and filled by direction cosines, that is cosines of the angles formed by the direction of the wavevector relative to the cartesian axis, that go to 1 \cite{Carminati:98}, thus are proportional to the identity matrix.
%
%
Using this relation, we are finally able to formally relate the Green's tensor with the scattering properties of the LCR medium as
\begin{eqnarray}
\mathcal{T}_{\rightarrow} \mathcal{N}_1 &=& \stackrel{\leftrightarrow}{\mathcal{G}}(2,1)
\label{RSC1}.
\end{eqnarray}
Exchanging the source and detector (region 1 becomes region 2 and the arrow is inverted), we find
\begin{eqnarray}
\mathcal{T}_{\leftarrow} \mathcal{N}_2 &=& \stackrel{\leftrightarrow}{\mathcal{G}}(1,2)
\label{RSC2}.
\end{eqnarray}
Eqs. \eqref{RSC1}, \eqref{RSC2} connect the forward and backward transmission matrices to the electromagnetic Green's functions.
Thus, the transmission matrices inherit the symmetry of the Green's functions under reciprocity written in eq. \eqref{EGM5}.
This implies that $\left( \mathcal{T}_{\leftarrow} \mathcal{N}_2 \right)^{\mathrm{T}} = \hat{\sigma}_z \left( \mathcal{T}_{\rightarrow} \mathcal{N}_1 \right) \hat{\sigma}_z$, which simplifies further in the paraxial approximation to
\begin{equation}
\left( \mathcal{T}_{\leftarrow} \right)^{\mathrm{T}} =
\hat{\sigma}_z 
\left( \mathcal{T}_{\rightarrow} \right)
\hat{\sigma}_z 
\label{RSC3} \, .
\end{equation}
A similar argument holds for the reflection matrices, leading to 
\begin{eqnarray}
\left( \mathcal{R}_{\leftarrow} \right)^{\mathrm{T}} &=&
\hat{\sigma}_z 
\left(\mathcal{R}_{\leftarrow} \right)
\hat{\sigma}_z 
\label{RSC4} \, , \\
\left( \mathcal{R}_{\rightarrow} \right)^{\mathrm{T}} &=&
\hat{\sigma}_z 
\left( \mathcal{R}_{\rightarrow} \right)
\hat{\sigma}_z 
\label{RSC5} \, .
\end{eqnarray}
Those relations are another important result of this section.
They can be subsumed in a reciprocity symmetry-property of the scattering matrix
\begin{equation}
\mbox{LCR}
\Leftrightarrow\left(\stackrel{\leftrightarrow}{\mathcal{S}}\right)^{\mathrm{T}} =\stackrel{\leftrightarrow}{\sigma}_z\, \stackrel{\leftrightarrow}{\mathcal{S}} \,\stackrel{\leftrightarrow}{\sigma}_z
\label{RSC9} \, ,
\end{equation}
with
\begin{equation}
\stackrel{\leftrightarrow}{\sigma}_z =\begin{bmatrix}
\hat{\sigma}_z & \hat{0} \\
\hat{0} & \hat{\sigma}_z
\end{bmatrix}
\label{RSC10}.
\end{equation}
%

\section{Condon's microscopic model}
\label{Cmodel}
%
\begin{figure}[ht!]
\begin{center}
\includegraphics[width=0.6\linewidth]{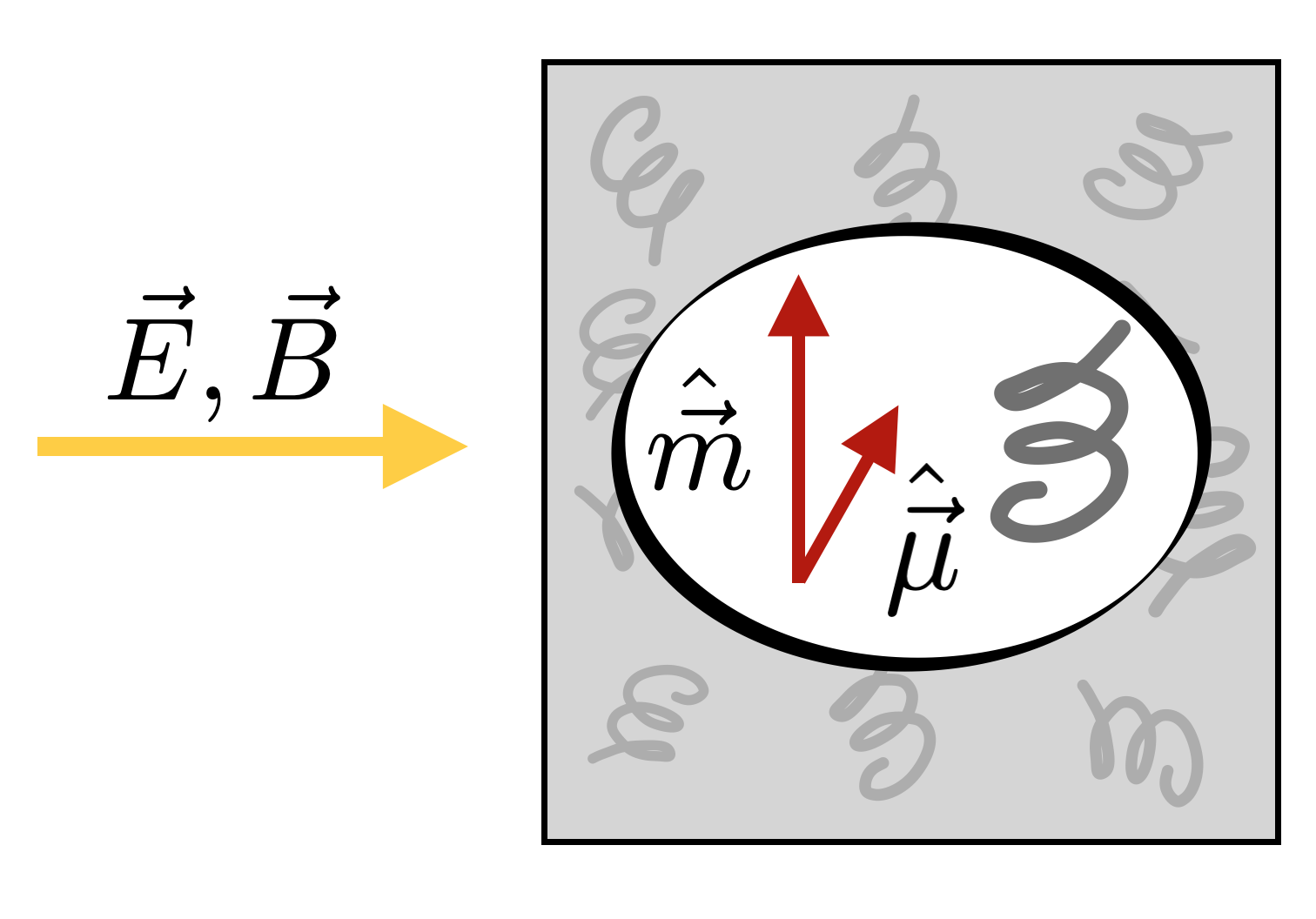}
\caption{Sketch of a Pasteur medium probed by an external and classical electromagnetic field $(\vec{E},\vec{B})$. Light-matter interaction generates an induced electric dipole $\hat{\vec{\mu}}$ and a magnetic-dipole $\hat{\vec{m}}$ at the microscopic level.}
\label{Fig1}
\end{center}
\end{figure}
%
%
In this section, we describe a microscopic model of chirality and optical activity, originally due to
Condon \cite{Condon1937}, in which molecular electric and magnetic dipoles interact with classical light.
We derive the macroscopic constitutive relations arising from this microscopic model, and specialize them to the case of linear, homogeneous, isotropic and chiral (LHIC) reciprocal media, called \emph{Pasteur media} \cite{Lindell1994,Jaggard1979}.
We did not consider the case of bi-isotropic \emph{Tellegen media} \cite{Lindell1994}, which contain an additional parameter in the constitutive relations breaking electromagnetic reciprocity, and are thus beyond the scope of the present paper.
%
%

\subsection{Microscopic polarization}
\label{Micropo}
We start with an expression of the Hamiltonian $\mathcal{\hat{H}}(t)$ written in the minimal-coupling scheme, describing the interaction of an atom or a molecule with a classical electromagnetic field \cite{Craig1998,Condon1937}.
We write $\mathcal{\hat{H}}(t) = \mathcal{\hat{H}}_0 + \mathcal{\hat{V}}(t)$, with $\mathcal{\hat{H}}_0 \equiv \frac{\|\hat{\vec{p}}\|^{2}}{2m} + V(\hat{\vec{x}})$ the bare Hamiltonian of a single electron of mass $m$, charge $q=-e$, momentum operator $\hat{\vec{p}}$, and position operator $\hat{\vec{x}}$.
This electron also interacts with the mean-field potential energy $V(\hat{\vec{x}})$ induced by other nuclei and electrons. 
We suppose that the bare Hamiltonian $\mathcal{\hat{H}}_0$ has been fully diagonalized, leading to atomic or molecular eigenstates $\ket{a}$ and eigenenergies $\epsilon_a$.
The interaction Hamiltonian with the time-dependent electromagnetic field is written as
\begin{equation}
\begin{split}
\mathcal{\hat{V}}(t) &= -\frac{q \vec{A}(\hat{\vec{x}},t)\cdot\hat{\vec{p}}}{m},
\end{split}
\label{qedham}
\end{equation}
where $\vec{A}(\hat{\vec{x}},t)$ is the classical time-dependent potential-vector field (depending on the position operator of the electron),  written in Coulomb gauge ($\nabla\cdot\vec{A}=0$).
The quadratic term proportional to $||\vec{A}||^{2}$ is usually neglected on the right-hand side of eq. \eqref{qedham}, since it is usually smaller than the linear one.
This will limit \textcolor{black}{the range of validity} of our investigations when dealing with Fabry-P\'{e}rot cavities, to the regime of \textcolor{black}{weak to moderately strong} (but not ultrastrong) light-matter coupling, for which the electromagnetic coupling strength is weaker than the cavity optical frequency (although it might be larger than the overall losses) \cite{Schafer2020}.
%
%
For an optically active (chiral) medium, the dipolar approximation is not sufficient, and one has to correct it with the spatial derivatives of $\vec{A}(\hat{\vec{x}},t)$ around the origin of the coordinate system (in analogy with eq. \eqref{DA1}) fixed at a position $\vec{0}$ inside the molecule
\begin{equation}
\vec{A}(\hat{\vec{x}},t)\approx \vec{A}(\vec{0},t)+\left(\hat{\vec{x}}\cdot \vec{\nabla}\right)\vec{A}(\vec{0},t).
\label{varfirstorder}
\end{equation}
Injecting eq. \eqref{varfirstorder} in eq. \eqref{qedham}, we obtain the interaction Hamiltonian as
\begin{equation}
\mathcal{\hat{V}}(t)=-\frac{q}{m}\left[\vec{A}(\vec{0},t)\cdot\hat{\vec{p}}+\frac{1}{2}\left(\vec{\nabla}\wedge\vec{A}(\vec{0},t)\right)\cdot\left(\hat{\vec{x}}\wedge\hat{\vec{p}}\right)+\cdots\right].
\label{inter}
\end{equation}
The first term stands for the contribution of the electric-dipole moment, while the second one encodes the magnetic-dipole moment contribution. 
Similarly to eq. \eqref{DA2}, we removed terms corresponding to the quadrupole moments, that vanish upon averaging over the molecular orientation \cite{Craig1998}.
We consider the case of monochromatic fields $\vec{A}(\hat{\vec{x}},t)= \mbox{Re} \left( \vec{A}(\hat{\vec{x}},\omega)e^{-i\omega t} \right)$, $\vec{E}(\vec{0},t)= \mbox{Re} \left( \vec{E}(\omega)e^{-i\omega t} \right)$ and $\vec{H}(\vec{0},t)= \mbox{Re} \left( \vec{H}(\omega)e^{-i\omega t} \right)$ applied to the molecule.
Using time-dependent perturbation theory at first-order in $\mathcal{\hat{V}}(t)$ (see Appendix \ref{appInducMicr} for a detailed derivation) enables to find an expression for the perturbed wave function $\ket{\psi_a(t)}$ of the electron that was initially in the eigenstate $\ket{a}$ of $\hat{\mathcal{H}}_0$.
The induced electric dipole $\langle\vec{\mu}_{a}\rangle(\omega)$ and magnetic-dipole $\langle\vec{m}_{a}\rangle(\omega)$ in the state $\ket{\psi_a(t)}$, averaged over the molecular dipole orientations are obtained as
%
%
%
%
\begin{equation}
\begin{split}
\langle\vec{\mu}_{a}\rangle(\omega) &\approx \alpha_{a}(\omega) \vec{E}(\omega)+i\omega\frac{g_{a}(\omega)}{c}\mu_{0}\vec{H}(\omega),\\
\langle\vec{m}_{a}\rangle (\omega) &\approx\beta_{a}(\omega)\mu_{0}\vec{H}(\omega)-i\omega\frac{g_{a}(\omega)}{c}\vec{E}(\omega),
\label{constAver}
\end{split}
\end{equation}
with
\begin{equation}
\begin{split}
\alpha_{a}(\omega) &=\frac{2}{3\hbar}\sum_{b}\frac{\Omega_{ba}}{\Omega_{ba}^{2}-\omega^{2}}\lVert\bra{a}\hat{\vec{\mu}}\ket{b}\rVert^{2},\\
\beta_{a}(\omega)&=\frac{2}{3\hbar}\sum_{b}\frac{\Omega_{ba}}{\Omega_{ba}^{2}-\omega^{2}}\lVert\bra{a}\hat{\vec{m}}\ket{b}\rVert^{2},\\
g_{a}(\omega)&=\frac{2 c}{3\hbar}\sum_{b}\frac{\mathrm{Im}\{\bra{a}\hat{\vec{\mu}}\ket{b}\cdot\bra{b}\hat{\vec{m}}\ket{a}\}}{\Omega_{ba}^{2}-\omega^{2}}.
\end{split}
\label{polariz}
\end{equation}
The Bohr frequency of the $a-b$ transition is written as $\Omega_{ba}=\left( \epsilon_b - \epsilon_a\right)/\hbar$.
The polarizabilities $\alpha_{a}(\omega)$ and $\beta_{a}(\omega)$ describe the response of the electric and magnetic dipoles when the molecule is exposed to an electromagnetic perturbation. 
By effect of this perturbation, the optical activity emerges as the dynamic magneto-electric coupling $g_{a}(\omega)$.
%
%

%
\subsection{Condon constitutive relations}
\label{CondConst}
The macroscopic constitutive relations (Condon relations) of a LHIC medium follow directly from homogeneity and isotropy of the medium in eq. \eqref{constAver}.
To this effect, we consider $N$ molecules per unit volume, where each molecule has an associated probability $p_a$ (for instance a thermal Boltzmann distribution) to be in the state $\ket{a}$.
We obtain the macroscopic polarization $\vec{P}=N\sum_{a}p_{a}\langle\vec{\mu}_{a}\rangle$ and magnetization $\vec{M}=N\sum_{a}p_{a}\langle\vec{m}_{a}\rangle$, such that
%
%
\begin{equation}
\begin{split}
\vec{P}&=\varepsilon_0 \chi_e \vec{E}+i\omega\frac{g}{c}\vec{H},\\
\vec{M}&=\chi_m\vec{H}-i\omega\frac{g}{\mu_{0}c}\vec{E},
\end{split}
\end{equation}
%
%
%
where $\chi_e=N\sum_{a}p_{a}\alpha_{a}/\varepsilon_0$ is the electric susceptibility, $\chi_m=\mu_{0} N\sum_{a}p_{a}\beta_{a}$ is the magnetic susceptibility, and $g=\mu_{0}N\sum_{a}p_{a}g_{a}$.
These relations can be rewritten in terms of the electric displacement $\vec{D}$ and magnetic induction $\vec{B}$, using eq. \eqref{DisplacementInduction} as
\begin{equation}
\begin{split}
\vec{D}=\varepsilon\vec{E}+i\frac{\kappa}{c}\vec{H},\\
\vec{B}=\mu\vec{H}-i\frac{\kappa}{c}\vec{E},
\end{split}
\label{ccr}
\end{equation}
where $\varepsilon=\varepsilon_{0}\left(1+\chi_{e}\right)$ is the permittivity, and $\mu=\mu_{0}\left(1+\chi_{m}\right)$ is the permeability. 
The dimensionless magneto-electric coupling $\kappa(\omega) = \omega g(\omega)$ is the \textit{Pasteur coefficient} of the LHIC medium, and takes typical values in the range $10^{-5}-10^{-3}$ \cite{Craig1998,Condon1937,Barron2004}.
The macroscopic Condon constitutive relations, reported in eq. \eqref{ccr}, are valid for any LHIC medium, or \emph{Pasteur medium} \cite{Lindell1994,Jaggard1979} (see Figure \ref{Fig1}), and can be recovered from the more general eq. \eqref{genconst}, in the specific case of a scalar tensor $\stackrel{\leftrightarrow}{\nu} \equiv i\frac{\kappa}{c} \mathds{Id}$.
%
%
%
The Condon constitutive relations obviously fulfill the more general reciprocity relations discussed in Sec. \ref{genRecipr}.
%

%
\section{Transfer-matrix approach for chiral media}
\label{chiralTM}
In this section, we show how to compute the optical properties of any scattering chiral medium.
We derive analytically the transmission and reflection matrices for an interface between two chiral media following the transfer-matrix approach of ref. \cite{Jaggard1992}.
This approach considers the propagation of left-handed (LCP, $+$) and right-handed (RCP, $-$) circularly polarized light-rays through different layers, at arbitrary incident angles. 
We extend these results to the case of transfer-matrices for multilayer Pasteur media. 
In contrast to the original work by Jaggard \cite{Jaggard1992}, here we use Condon's constitutive relations (see eq. \eqref{ccr}) for describing the electromagnetic response of the medium, rather than the ones derived by Post \cite{Post}.
This choice is dictated both by simplicity of the available microscopic model developed in Sec. \ref{Cmodel}, and by the fact that the obtained results do not depend crucially on it.
The results derived in this section are fully compatible with classical Maxwell equations, and are independent of the approximations made in Sec.\ref{Cmodel}: the light-matter interaction is fully encoded into the parameterized dielectric permittivity and Pasteur coefficients written explicitly in Sec.\ref{FPS}.
The question of how to derive such macroscopic dielectric constant \cite{PhysRev.112.1555} and Pasteur coefficient from a realistic microscopic model of the material strongly interacting with the cavity electromagnetic field is beyond the scope of the present paper. 
%
%
%
We implemented a numerical code based on this transfer-matrix approach, which is able to simulate the optical properties of an arbitrary number of layers.
Several examples are shown in Sec. \ref{FPsilver} and Sec. \ref{HPfabry}.
%
%
%
%
\subsection{Light propagation in a bulk Pasteur medium}
%
\begin{figure}[ht!]
\begin{center}
\includegraphics[width=0.6\linewidth]{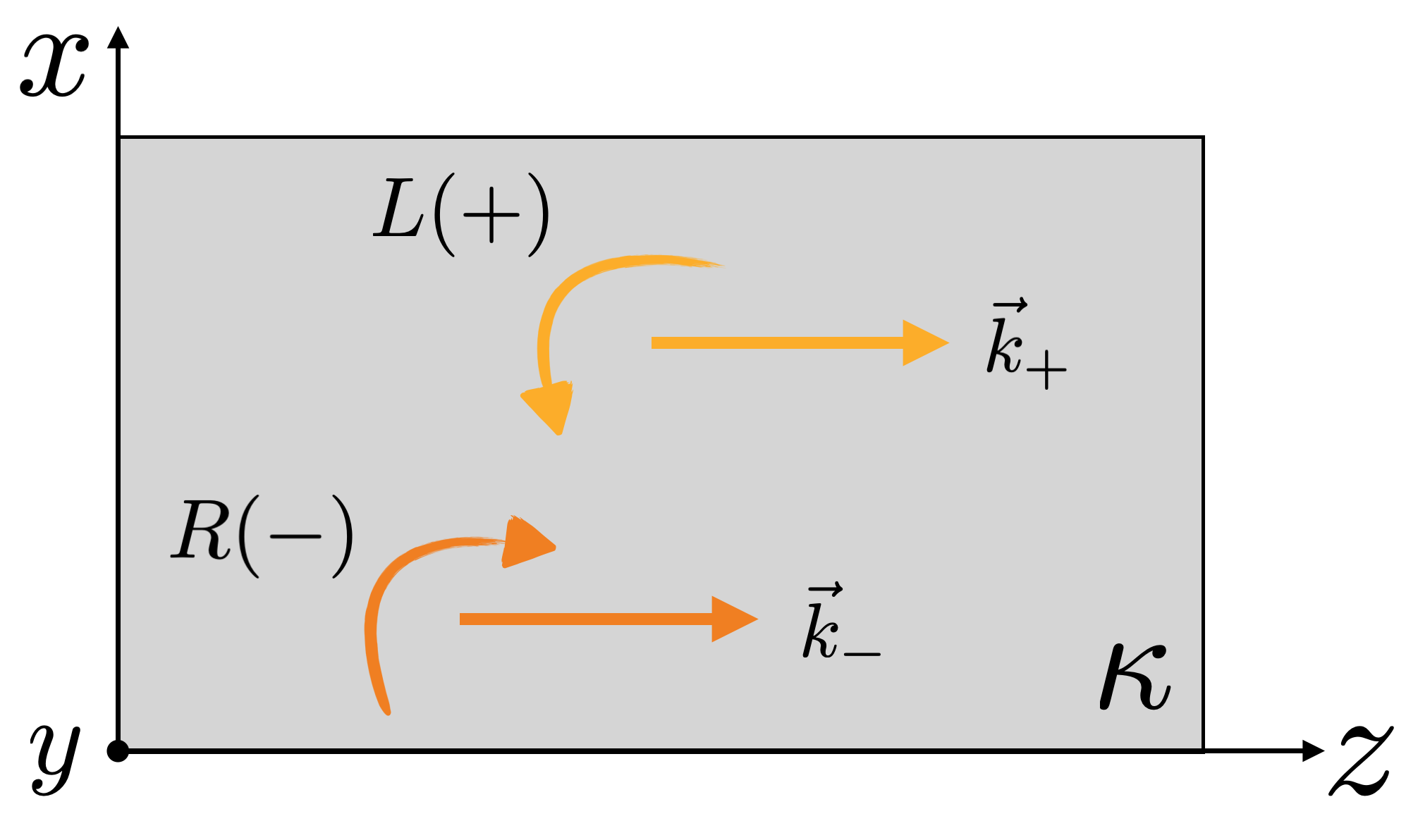}
\caption{Sketch of LCP ($+$) and RCP ($-$) waves with wave-vectors $\vec{k}_+$ and $\vec{k}_-$ that propagate in a LHIC medium with Pasteur coefficient $\kappa$.}
\label{Fig4}
\end{center}
\end{figure}
%
We study the propagation of electromagnetic waves in a bulk Pasteur medium.
We derive the related wave-equation from Maxwell's eqs. \eqref{maxw} (assuming a harmonic time-dependence for the fields, $\rho=0$ and $\vec{j}=0$), and Condon constitutive relations (see Sec. \ref{Cmodel})  
\begin{equation}
\vec{\nabla}\wedge\vec{\nabla}\wedge \vec{E}-2\omega\frac{\kappa}{c}\vec{\nabla}\wedge \vec{E}-\omega^{2}\left(\mu\varepsilon-\frac{\kappa^{2}}{c^{2}}\right)\vec{E} = \vec{0}.
\label{Helm}
\end{equation}
The usual Helmholtz equation for achiral media is retrieved in the limit $\kappa\rightarrow 0$.
We obtain eigenmodes of eq. \eqref{Helm} as forward and backward propagating electric fields, given respectively by $\vec{E}^{\alpha}_{\rightarrow}(\vec{r})=\vec{E}^{\alpha}_{\rightarrow}e^{i\vec{k}_{\alpha}\cdot\vec{r}}$ and 
$\vec{E}^{\alpha}_{\leftarrow}(\vec{r})=\vec{E}^{\alpha}_{\leftarrow}e^{-i\vec{k}_{\alpha}\cdot\vec{r}}$, with $\vec{k}_\alpha$ the wave-vectors that differ for the $\alpha=+$ (LCP) and 
$\alpha=-$ (RCP) waves.
%
%
%
%
%
%
The amplitudes of the waves are given by $\vec{E}^{\alpha}_{\rightarrow(\leftarrow)}=E_{\rightarrow(\leftarrow)}\left(\vec{e}_x + i \alpha \vec{e}_y\right)/\sqrt{2}$,
with $E_{\rightarrow(\leftarrow)}$ the arbitrary amplitude of the forward (backward) propagating field, and $\vec{e}_x$ and $\vec{e}_y$ the unit vectors perpendicular to $\vec{k}_\alpha$ (see Figure \ref{Fig4}).
%
%
%
%
We obtain from eq. \eqref{Helm} the dispersion relations $||\vec{k}_\alpha||\equiv k_{\alpha}=\frac{\omega}{c}n_{\alpha}$ with the polarization-dependent refraction index
\begin{equation}
\begin{split}
n_{\alpha}= n + \alpha \kappa,
\end{split}
\label{refindex}
\end{equation}
where $n=c\sqrt{\mu\varepsilon}$ is the average refractive index of the material when $\kappa=0$.
The complex and frequency-dependent refractive indices $n_{+}$ and $n_{-}$ encode the polarization-dependent dispersion and absorption of LCP ($+$) and RCP ($-$) waves propagating inside the Pasteur medium.
%
%
\subsection{Transfer-matrix at a single interface}
\label{SingleInt}
%
\begin{figure}[ht!]
\begin{center}
\includegraphics[width=0.6\linewidth]{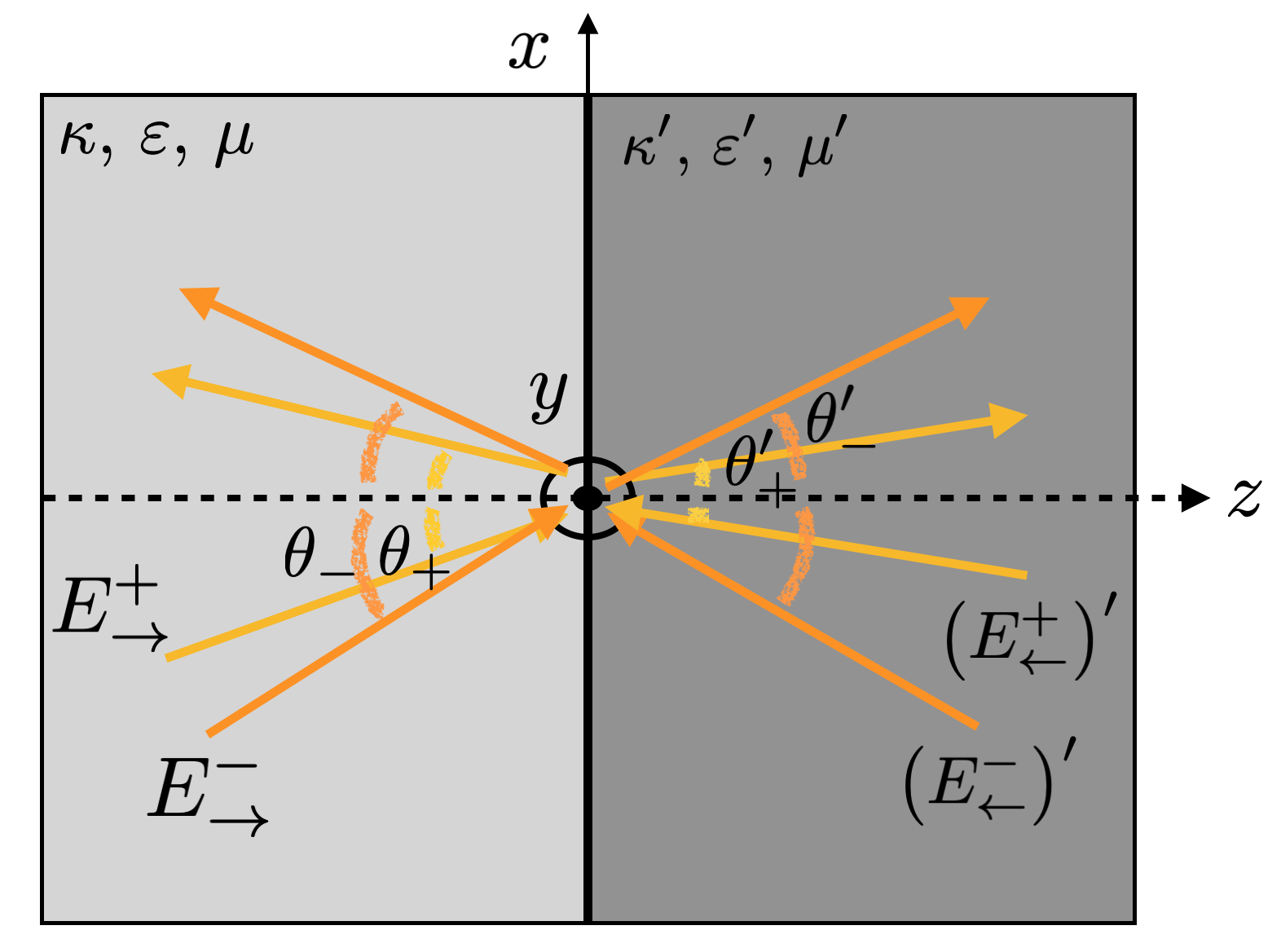}
\caption{Pictorial representation of the electric fields at the interface between two LHIC media of different Pasteur coefficients $\kappa$ and $\kappa'$.}
\label{Fig5}
\end{center}
\end{figure}
%
%
We now consider an interface between two different LHIC media (see Figure \ref{Fig5}), characterized 
by $\kappa$, $\varepsilon$ and $\mu$ for the first medium and $\kappa^{\prime}$, $\varepsilon^{\prime}$ and $\mu^{\prime}$ for the second one.
We also introduce the relative electromagnetic impedances for each medium, $\eta\equiv\sqrt{\frac{\mu}{\varepsilon}}$ and $\eta^{\prime}\equiv\sqrt{\frac{\mu^{\prime}}{\varepsilon^{\prime}}}$.
After expressing the continuity relations of the $\vec{D}$, $\vec{B}$, $\vec{E}$ and $\vec{H}$ fields  at the interface between the two media, we obtain the transfer-matrix $\stackrel{\leftrightarrow}{\mathrm{T}}$ relating the input (in) and output (out) electric fields in each material (see Appendix \ref{TmatrixBoundary} for a detailed derivation), 
\begin{equation}
\begin{split}
\left[\begin{array}{c}
\left(\begin{array}{c}
E^{+}_{\rightarrow}\\
E^{-}_{\rightarrow}\\
\end{array}\right)_{\mathrm{in}}\\
\left(\begin{array}{c}
E^{+}_{\leftarrow}\\
E^{-}_{\leftarrow}\\
\end{array}\right)_{\mathrm{out}}
\end{array}\right]&=\stackrel{\leftrightarrow}{\mathrm{T}}\left[\begin{array}{c}
\left(\begin{array}{c}
E^{+}_{\rightarrow}\\
E^{-}_{\rightarrow}\\
\end{array}\right)_{\mathrm{out}}\\
\left(\begin{array}{c}
E^{+}_{\leftarrow}\\
E^{-}_{\leftarrow}\\
\end{array}\right)_{\mathrm{in}}
\end{array}\right]^{\prime},
\end{split}
\label{tma}
\end{equation}
with
\begin{equation}
\begin{split}
\stackrel{\leftrightarrow}{\mathrm{T}}&=\left(
\begin{array}{cc}
\mathcal{M}_{\mathcal{T}} & \mathcal{M}_{\mathcal{R}}\\
\mathcal{M}_{\mathcal{R}} & \mathcal{M}_{\mathcal{T}}\\
\end{array}\right).
\end{split}
\label{tma42}
\end{equation}
We introduced in eq. \eqref{tma42}, the submatrices
\begin{equation}
\scriptscriptstyle{
\mathcal{M}_{\mathcal{T}(\mathcal{R})}=\left[
\begin{array}{cc}
\frac{1}{4}\left(1+\frac{\eta}{\eta^{\prime}}\right)\left(1\pm\frac{\cos\theta^{\prime}_{+}}{\cos\theta_{+}}\right) & \frac{1}{4}\left(\frac{\eta}{\eta^{\prime}}-1\right)\left(1\mp\frac{\cos\theta^{\prime}_{-}}{\cos\theta_{+}}\right)\\
\frac{1}{4}\left(\frac{\eta}{\eta^{\prime}}-1\right)\left(1\mp\frac{\cos\theta^{\prime}_{+}}{\cos\theta_{-}}\right) & \frac{1}{4}\left(1+\frac{\eta}{\eta^{\prime}}\right)\left(1\pm\frac{\cos\theta^{\prime}_{-}}{\cos\theta_{-}}\right)\\
\end{array}\right]},\label{eq:mt} 
\end{equation}
where $\cos\theta^{\prime}_{\pm}=\sqrt{1-\left(\frac{k_{\pm}}{k^{\prime}_{\pm}}\right)^{2}\sin^{2}\theta_{\pm}}$, and $\theta_{\pm}$, $\theta^{\prime}_{\pm}$ are the incidence angles of light in each medium.
We extract from eq. \eqref{tma}, the transmission and reflection matrices already defined in Sec. \ref{SC}.
For instance, by imposing $\left(E^{+}_{\leftarrow},E^{-}_{\leftarrow}\right)=\left(0,0\right)$, we obtain the $2\times 2$ transmission matrix $\mathcal{T}_{\rightarrow}$ and reflection matrix $\mathcal{R}_{\leftarrow}$ (see eq. \eqref{scattMat}) defined in the circular-polarization basis
\begin{equation}
\begin{split}
\left(\begin{array}{c}
E^{+}_{\rightarrow}\\
E^{-}_{\rightarrow}\\
\end{array}\right)^{\prime}_{\mathrm{out}}=&\,\mathcal{T}_{\rightarrow}\left(\begin{array}{c}
E^{+}_{\rightarrow}\\
E^{-}_{\rightarrow}\\
\end{array}\right)_{\mathrm{in}},\mbox{with}\hspace{0,2cm}\mathcal{T}_{\rightarrow}=\mathcal{M}^{-1}_{\mathcal{T}},\\
\left(\begin{array}{c}
E^{+}_{\leftarrow}\\
E^{-}_{\leftarrow}\\
\end{array}\right)_{\mathrm{out}}=&\,\mathcal{R}_{\leftarrow}\left(\begin{array}{c}
E^{+}_{\rightarrow}\\
E^{-}_{\rightarrow}\\
\end{array}\right)_{\mathrm{in}},\mbox{with}\hspace{0,2cm}\mathcal{R}_{\leftarrow}=\mathcal{M}_{\mathcal{R}}\mathcal{M}^{-1}_{\mathcal{T}}.
\end{split}
\label{trforw}
\end{equation}
The remaining transmission matrix $\mathcal{T}_{\leftarrow}$ and reflection matrix $\mathcal{R}_{\rightarrow}$ are obtained with an analogous argument as
\begin{equation}
\begin{split}
\mathcal{T}_{\leftarrow}&=\mathcal{M}_{T}-\mathcal{M}_{\mathcal{R}}\mathcal{M}_{T}^{-1}\mathcal{M}_{\mathcal{R}},\\
\mathcal{R}_{\rightarrow}&=-\mathcal{M}^{-1}_{\mathcal{T}}\mathcal{M}_{\mathcal{R}}.
\end{split}
\label{trback}
\end{equation}
%

\subsection{Transfer-matrix for multilayers}
\label{Transnum}
%
\begin{figure}[ht!]
\begin{center}
\includegraphics[width=0.45\linewidth]{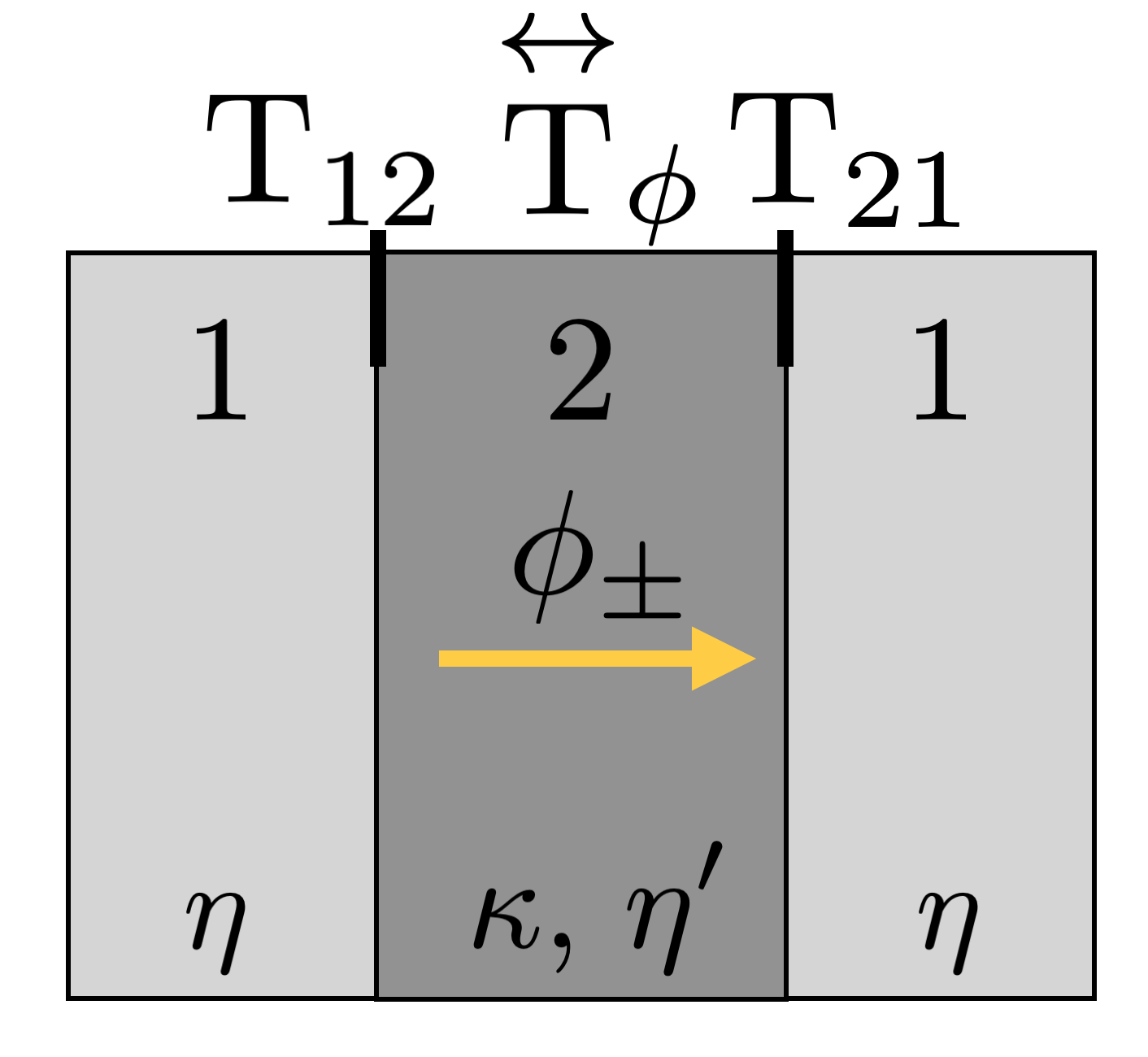}
\caption{Sketch of the transfer matrix approach for a LHIC medium $(2)$ (in dark grey) embedded in an achiral medium $(1)$ (in light grey).}
\label{FigLhic}
\end{center}
\end{figure}
%
The main advantage of resorting to the previous $\stackrel{\leftrightarrow}{\mathrm{T}}$-matrix technique relies on its simple implementation for numerical calculations. 
Indeed, to extend it to stratified media, it is sufficient to multiply from left to right the transfer-matrices of the interfaces (see eq. \eqref{tma}) and bulk layers encountered along the light propagation axis (as is conventional in optics \cite{Jaggard1992}). 
In this section, we illustrate in Figure \ref{FigLhic} the transfer-matrix method, by considering a LHIC medium (labelled as $(2)$), characterized by impedance $\eta^{\prime}$ and thickness $L$, embedded in an achiral material (labelled as $(1)$) of impedance $\eta$.
The medium is illuminated at normal incidence ($\theta^{\prime}_{\pm}=\theta_{\pm}=0$ in eq. \eqref{eq:mt}). 
The full transfer-matrix $\stackrel{\leftrightarrow}{\mathrm{T}}_{LHIC}$ of the system is then the product of one transfer-matrix $\mathrm{T}_{12}$ for the $(1)$-$(2)$ interface, one transfer-matrix $\stackrel{\leftrightarrow}{\mathrm{T}}_{\phi}$ for the bulk-layer $(2)$, and 
one transfer-matrix $\mathrm{T}_{12}$ for the $(2)$-$(1)$ interface
\begin{equation}
\stackrel{\leftrightarrow}{\mathrm{T}}_{LHIC}=\mathrm{T}_{12}\stackrel{\leftrightarrow}{\mathrm{T}}_{\phi}\mathrm{T}_{21}, 
\label{tlhic}
\end{equation}
with the $\stackrel{\leftrightarrow}{\mathrm{T}}_{\phi}$-matrix given by
%
%
\begin{equation}\label{eq:pmat}
\begin{split}
\stackrel{\leftrightarrow}{\mathrm{T}}_{\phi}&=
\begin{bmatrix}
e^{-i\phi_{+}} & 0 & 0 & 0\\
0 & e^{-i\phi_{-}} & 0 & 0\\
0 & 0 & e^{i\phi_{+}} & 0 \\
0 & 0 & 0 & e^{i\phi_{-}} \\
\end{bmatrix},
\end{split}
\end{equation}
%
%
and $\phi_{\pm}=k_{\pm}L$. 
We highlight that the $\stackrel{\leftrightarrow}{\mathrm{T}}_{LHIC}$-matrix is not in general symmetric, which is instead the case for the transfer-matrix of a single interface (see eq. \eqref{tma}). 
From eq. \eqref{tlhic}, as in Sec. \ref{SingleInt}, we derive all the transmission and reflection matrices.
The latter fulfill the relations $\mathcal{T}_\rightarrow =\mathcal{T}_\leftarrow \equiv \mathcal{T}$ 
and $\mathcal{R}_\rightarrow =\mathcal{R}_\leftarrow \equiv \mathcal{R}$, with
\begin{equation}
\mathcal{T} = 4 \frac{\bar{\eta}}{\Xi_1} \begin{bmatrix}
e^{i\phi_{+}} & 0\\ 
0 & e^{i\phi_{-}} \\
\end{bmatrix},
\, 
\mathcal{R}=\frac{\Xi_2}{\Xi_1} \begin{bmatrix}
0 & 1 \\
1 & 0
\end{bmatrix}
\label{trchiachi} \, ,
\end{equation}
and
\begin{eqnarray}
&&\Xi_1 = \left(1+\bar{\eta}\right)^{2}-\left(\bar{\eta}-1\right)^{2}e^{i\left(\phi_{+}+\phi_{-}\right)}
\label{trchiachi2} \, , \\
&&\Xi_2 = \left(1-\bar{\eta}^{2}\right) \left( e^{i\left(\phi_{+}+\phi_{-}\right)} - 1 \right)
\label{trchiachi3} \, , \\
&&\phi_{+}+\phi_{-} = 2\omega L\sqrt{\mu\varepsilon} \, ; \mbox{   } \bar{\eta}\equiv\eta/\eta^{\prime}
\label{trchiachi4} \, .
\end{eqnarray}
Under illumination at normal incidence, the polarization of the incoming wave is preserved in transmission ($\mathcal{T}_{-+}=\mathcal{T}_{+-}=0$) while it is perfectly reversed in reflection ($\mathcal{R}_{++}=\mathcal{R}_{--}=0$).
We also notice that the reflection matrix is polarization-independent, while the dependence of the transmission matrix with polarization comes from the complex numbers $e^{i\phi_\pm}$.
The latter encode the difference in phase velocity (real part of the refraction index $n_\pm$) and absorption coefficient (imaginary part of $n_\pm$) for LCP and RCP propagating waves inside the Pasteur medium $(2)$.
%
%
%
%
%

\section{Green's function approach for multilayered systems}
\label{GenGreen}
\begin{figure}[ht!]
\begin{center}
\includegraphics[width=0.5\linewidth]{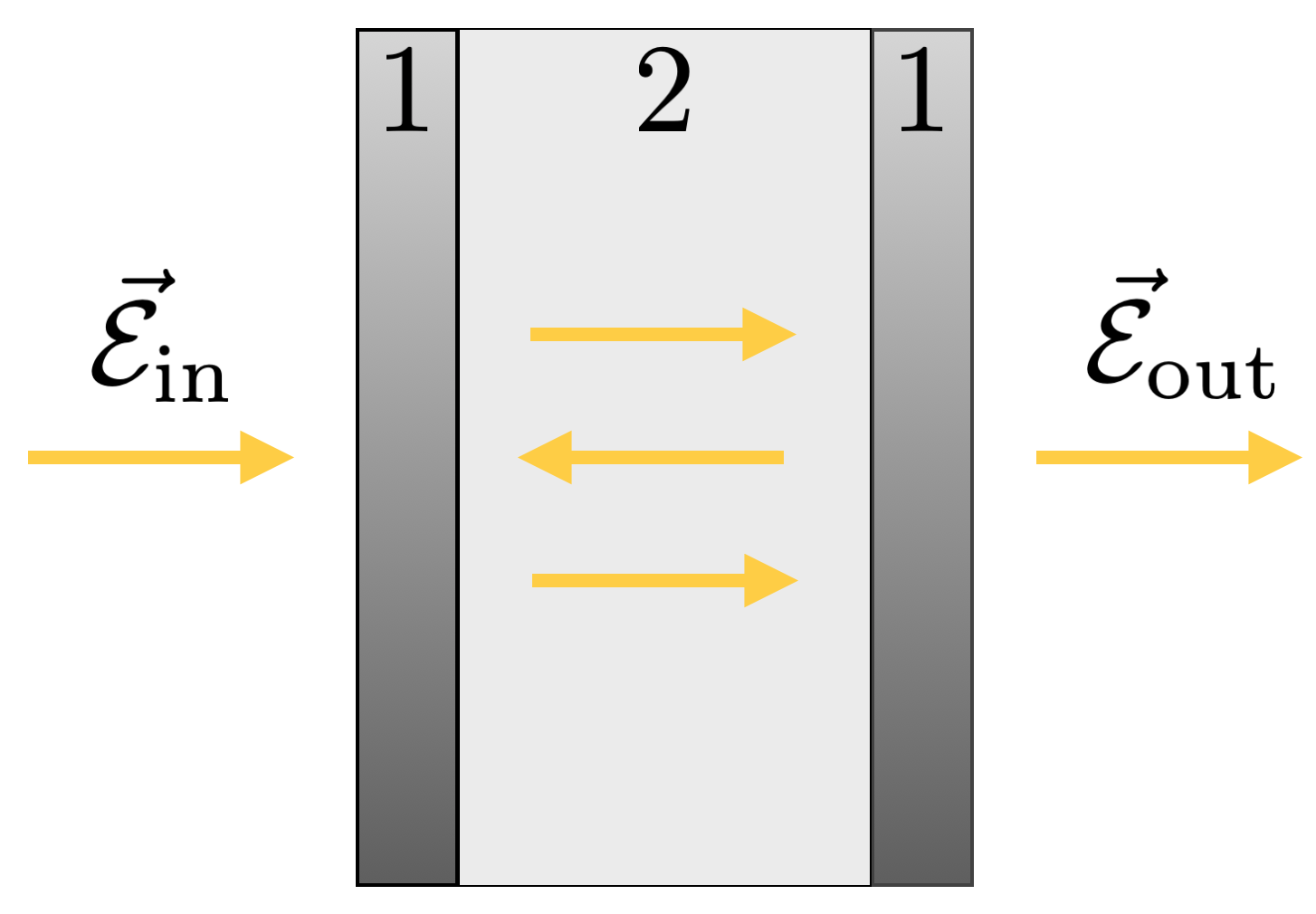}
\caption{Sketch of a FP cavity made by arbitrary mirrors $(1)$, with incoming (outgoing) electric field $\vec{\mathcal{E}}_{\rm{in}}$ ($\vec{\mathcal{E}}_{\rm{out}}$). 
Multiple reflection processes occur in the embedded medium $(2)$.
}
\label{Fig6}
\end{center}
\end{figure}
%
%
In this section, we reformulate and link the scattering-matrix approach of Sec. \ref{SC} to the Green function one introduced in Sec. \ref{Greenscatt}.
%
%
We derive an explicit expression of the optical forward transmission for an incoming LCP or RCP electric field $\vec{\mathcal{E}}_{\mathrm{in}}$ (the source) propagating across the Fabry-P\'erot (FP) cavity shown in Figure \ref{Fig6}.
The mirrors are labelled as $(1)$, while the cavity-embedded Pasteur medium is labelled as $(2)$.
We suppose that no input-field (no source) is applied coming from the right.
The transmitted output field $\vec{\mathcal{E}}_{\mathrm{out}}$ out of the cavity is provided by $\vec{\mathcal{E}}_{\mathrm{out}} = \mathcal{T}_\rightarrow \vec{\mathcal{E}}_{\mathrm{in}}$ (see eq. \eqref{scattMat}), with 
\begin{equation}
\mathcal{T}_\rightarrow \equiv \begin{bmatrix}
t_{++} & t_{-+}\\
t_{+-} & t_{--}\\
\end{bmatrix}
\label{Jon} .
\end{equation}
We define $T_{+}$ ($T_{-}$), the total transmission or normalized intensity of the output-wave $\vec{\mathcal{E}}_{+,\mathrm{out}}$ ($\vec{\mathcal{E}}_{-,\mathrm{out}}$) generated upon illuminating the cavity by an LCP (RCP) incoming wave $\vec{\mathcal{E}}_{+,\mathrm{in}}\equiv(1,0)$ ($\vec{\mathcal{E}}_{-,\mathrm{in}}\equiv(0,1)$).
We find in the circularly polarized basis that
\begin{equation}
T_{\pm}=\lvert\lvert \vec{\mathcal{E}}_{\pm,\mathrm{out}}\rvert\rvert^{2}=\lvert  t_{\pm\pm}\rvert^{2}+\lvert t_{\pm\mp}\rvert^{2}.
\label{intensity}
\end{equation}
The different dispersion (optical activity) and absorption (circular dichroism) of the propagating wave across the cavity induces an asymmetry in transmission $\Delta T = T_+ - T_-$.
\textcolor{black}{We define naturally the differential circular transmission $\mathcal{DCT}$, as the relevant chiroptical signal that quantifies this asymmetry with respect to the average transmission $\overline{T} = \left( T_+ + T_- \right)/2$ as}
%
%
\begin{equation}
\mathcal{DCT}=\frac{\Delta T}{\overline{T}}.
\label{dci}
\end{equation}
We collect and sum at all orders the series of transmitted and reflected wave amplitudes at each round-trip (multiple-scattering) inside the FP cavity (see detailed derivation in Appendix \ref{GFappro}).
From now, we adopt the usual convention in condensed matter (suited for analytical calculations), that matrices and Green functions are multiplied from the right to the left, upon propagation of the wave along the optical axis. 
This is in contrast to the transfer-matrix approach in Sec. \ref{chiralTM}, where the opposite convention of optics was adopted. 
We obtain an exact expression for the forward transmission matrix
\begin{equation}
\mathcal{T}_\rightarrow = \mathcal{T}^{\rightarrow}_{12}\stackrel{\leftrightarrow}{\mathcal{G}}_{22}\mathcal{T}^{\rightarrow}_{21},
\label{calcJ1}
\end{equation}
where $\mathcal{T}^{\rightarrow}_{12}$ ($\mathcal{T}^{\rightarrow}_{21}$) is the forward transmission matrix across the right (left) mirror $(1)$, and $\stackrel{\leftrightarrow}{\mathcal{G}}_{22}$ is the Green function of the Pasteur medium $(2)$ ``dressed" by multiple reflections at the left and right mirrors $(1)$.
The former is given by
\begin{equation}
\stackrel{\leftrightarrow}{\mathcal{G}}_{22}=\frac{1}{\left(\stackrel{\leftrightarrow}{M}_{\phi}\right)^{-1}-\stackrel{\leftrightarrow}{\Sigma}_{22}}
\label{calcJ2} \, ,
\end{equation}
where $\stackrel{\leftrightarrow}{M}_{\phi}=\mathrm{diag}\left(e^{i\phi_{+}},e^{i\phi_{-}}\right)$ is the bare Green function of medium $(2)$ alone (without mirror reflections), and
\begin{equation}
\stackrel{\leftrightarrow}{\Sigma}_{22}=\mathcal{R}^{\rightarrow}_{22}\stackrel{\leftrightarrow}{M}_{\phi}\mathcal{R}^{\leftarrow}_{22},
\label{calcJ22}
\end{equation}
is the self-energy associated to forward (backward) reflection matrices of the wave $\mathcal{R}^{\rightarrow}_{22}$ ($\mathcal{R}^{\leftarrow}_{22}$) at the interfaces between media $(1)$ and $(2)$.
We note that the Green's function $\stackrel{\leftrightarrow}{\mathcal{G}}_{22}$ is symmetric for reciprocal media, and thus fulfills the general constraints presented in Sec. \ref{Greenscatt}.
Finally, using eq. \eqref{calcJ1}, we obtain the following expressions of the average transmission $\overline{T}$ and $\Delta T$ transmission imbalance across this optical junction
\begin{eqnarray}
\overline{T} &=& \mathrm{Tr}\left[ \mathcal{T}^\dagger_{\rightarrow} \mathcal{T}_{\rightarrow} \right]
\label{dci1} \, , \\ 
\Delta{T} &=& \mathrm{Tr}\left[ \hat{\sigma}_{z} \mathcal{T}^\dagger_{\rightarrow} \mathcal{T}_{\rightarrow} \right]
\label{dci2} \, .
\end{eqnarray}
These last equations are the main results of this paper. 
They provide a general and compact expression of $\mathcal{DCT}$ that is valid for any reciprocal and layered medium (with dispersion and losses), any kind of mirrors and every incident angle of the incoming wave.
%
%
Moreover, eq. \eqref{dci1} and eq. \eqref{calcJ1} share a nice analogy with the expression of the Landauer electric conductance across a quantum coherent mesoscopic conductor \cite{Landauer1957,Buttiker1986,fisher1981relation}.
%

%
\section{Fabry-P\'{e}rot cavity with silver mirrors}
\label{FPS}
%
\label{FPsilver}
\begin{figure}[ht!]
\begin{center}
\includegraphics[width=0.45\linewidth]{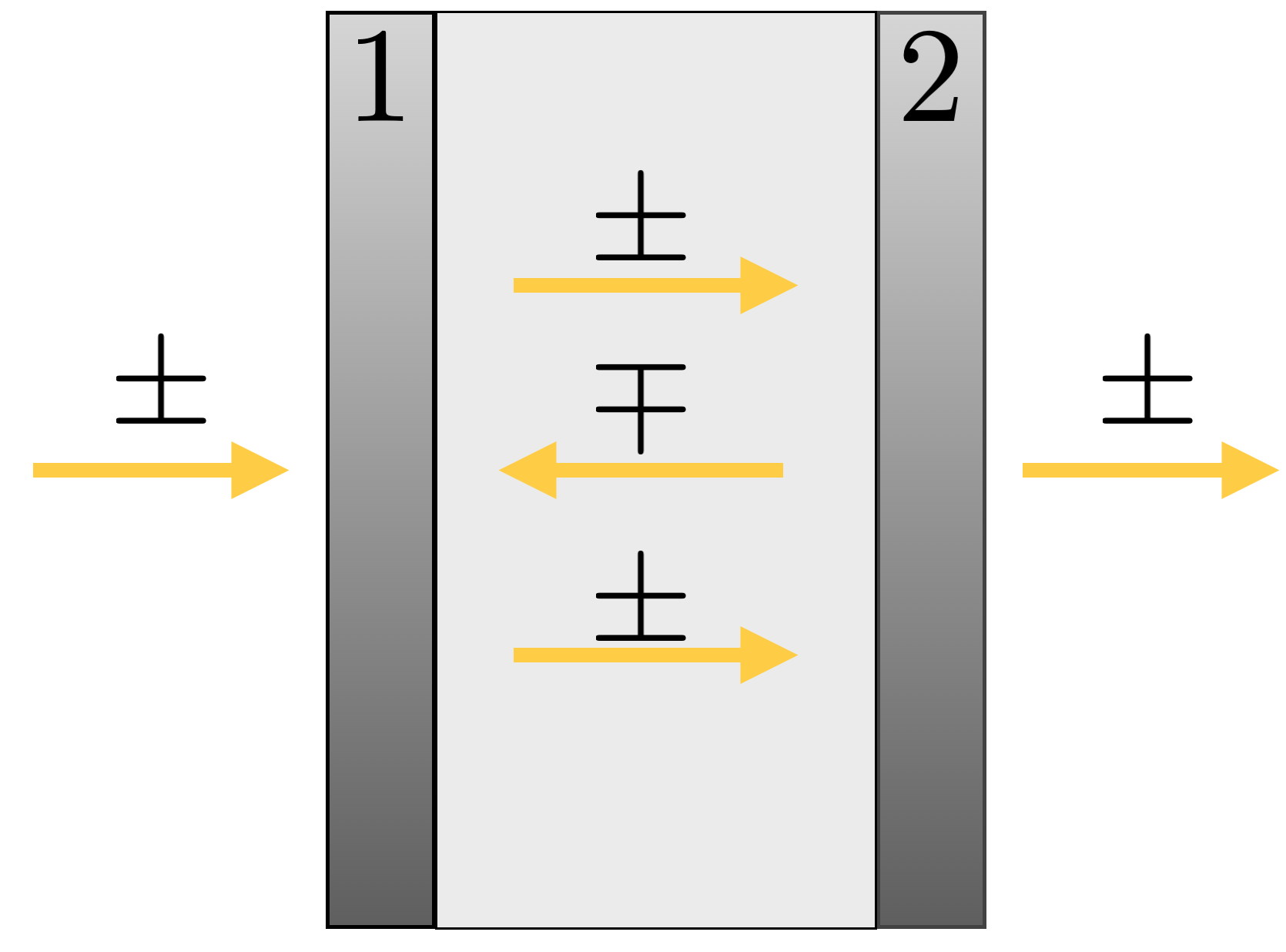}
\caption{Sketch of a FP cavity made by silver mirrors, that is illuminated at normal incidence by an  LCP ($+$) or RCP ($-$) wave.
The helicity of the wave inside the cavity is reversed upon each successive reflection at the mirror interfaces.
}
\label{Fig7}
\end{center}
\end{figure}
%
%
In the rest of the paper, we exemplify the outcome of the transfer-matrix numerical code developed in Sec. \ref{chiralTM}, and complement it with analytical calculations made with the Green function approach of Sec. \ref{GenGreen}.
For this purpose, in this section, we study chiroptical properties of a Fabry-P\'{e}rot cavity made of dispersive and lossy silver mirrors (labelled as $(1)$ and $(2)$ in Figure \ref{Fig7}) and filled with a non-magnetic LHIC Pasteur medium ($\mu=\mu_0$) of thickness $L$.
The metallic mirrors are modelled by a standard dielectric permittivity \cite{jackson1999classical}, with parameters chosen to correspond to silver.
The relative permittivity $\varepsilon_r\left(\omega\right)=\varepsilon\left(\omega\right)/\varepsilon_0$ of the LHIC medium is described by a broad Lorentzian absorption lineshape \cite{jackson1999classical}
\begin{equation}
\varepsilon_r\left(\omega\right) = \varepsilon_{\infty}+\frac{\omega_p^2f}{\omega^{2}_{0}-\omega^{2}-i\gamma\omega},
\label{ConstRel3}
\end{equation}
where $\varepsilon_{\infty}$ is the background relative permittivity, $\omega_p$ is the medium plasma frequency (proportional to the square root of the molecular concentration), $\omega_{0}$ is the 
material resonance (absorption) frequency, and $f\in\lbrack 0,1\rbrack$ is the oscillator strength for the corresponding electronic transition.
The material losses are modelled by the $\gamma$ damping factor and its (complex) Pasteur coefficient $\kappa(\omega)$ is chosen to be also frequency-dependent as
\begin{equation}
\kappa\left(\omega\right)=\kappa\frac{\omega^{2}_{p}\,\omega f}{\omega_{0}\left[\left(\omega^{2}_{0}-\omega^{2}\right)-i\gamma\omega\right]}.
\label{kappapdisp}
\end{equation}
We note that eq. \eqref{ConstRel3} and eq. \eqref{kappapdisp} are consistent with the quantum mechanical calculation performed in Sec. \ref{Micropo} (see also Appendix \ref{ModKappa}).
The effective refractive indices $n_{\pm}(\omega)$ of the LHIC medium are given by eq. \eqref{refindex}.
%

\subsection{Normal incidence}
%
\subsubsection{Weak-coupling regime (no polaritons)}
%
\begin{figure}[ht!]
\begin{center}
\includegraphics[width=\linewidth]{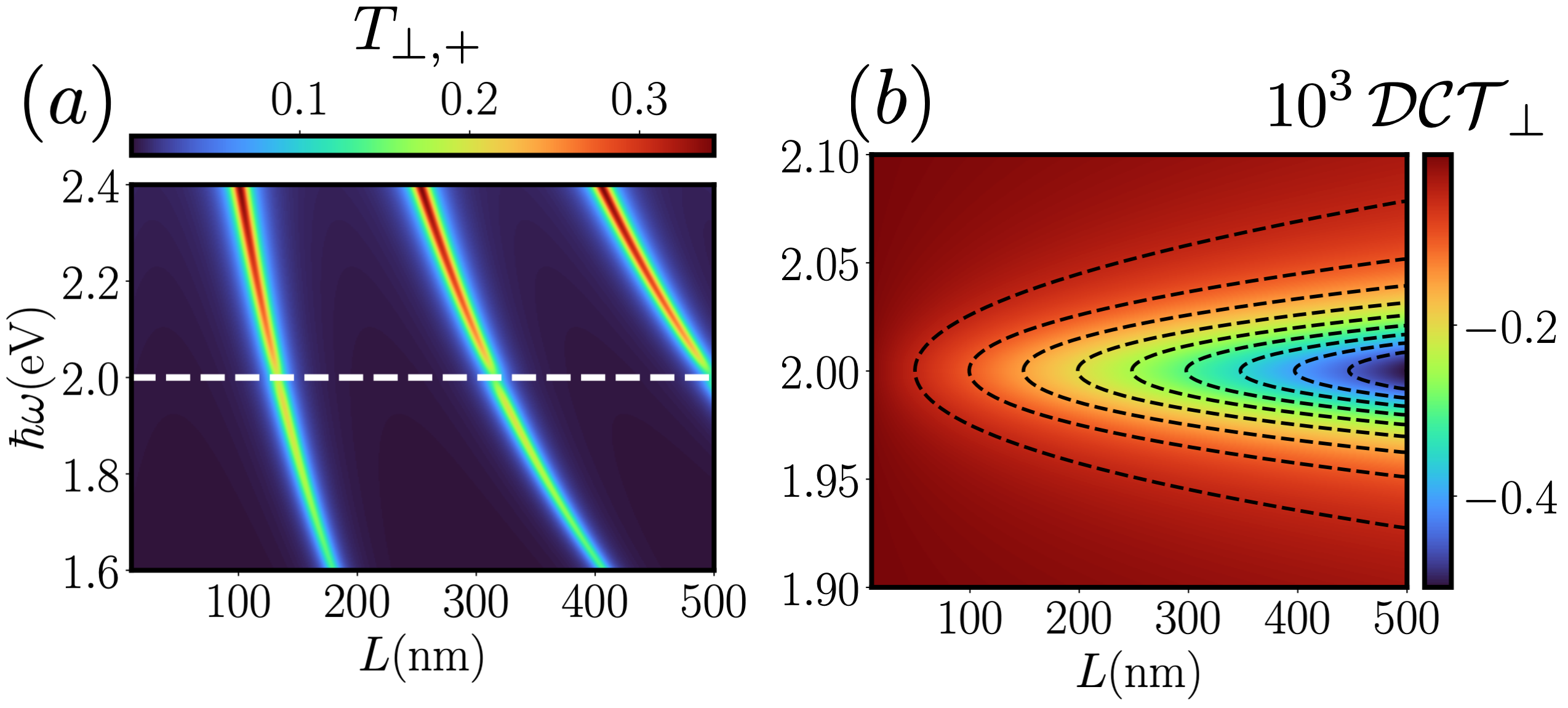}
\caption{($a$) Transmittance $T_{\perp,+}$ at normal incidence ($\theta=0$) of a FP cavity made by [Ag mirror-LHIC-Ag mirror] as a function of $\omega$ in eV and the thickness $L$ of a LHIC medium in nm. The dashed white line indicates the LHIC resonance frequency $\omega_{0}$ (see eq. \eqref{ConstRel3}). ($b$) Differential circular transmission at normal incidence $\mathcal{DCT}_{\perp}$  of the same system. The black dashed lines indicate the intensity regions. The parameters for the silver mirror are : $\varepsilon_{\infty,Ag}=4.8$, $\hbar\omega_{p,Ag}\sqrt{f}=\SI{9.5}{eV}$, $\hbar\omega_{0,Ag}=\SI{0}{eV}$, $\hbar\gamma_{Ag}=\SI{0.17}{eV}$,  $d_{Ag}=\SI{30}{nm}$. The parameters for the LHIC medium are : $\varepsilon_{\infty}=2.89$, $\hbar\omega_{p}\sqrt{f}=\SI{0.05}{eV}$, $\hbar\omega_{0}=\SI{2.0}{eV}$, $\hbar\gamma=\SI{0.05}{eV}$ and $\kappa=10^{-3}$.}
\label{Plot1}
\end{center}
\end{figure}
%
We consider in this subsection the situation of normal incidence under illumination of the FP cavity.
In this case, the helicity of incident circularly polarized waves reverse perfectly at each reflection on the metallic mirrors (see Figure \ref{Fig7}).
The transmission-matrix $\mathcal{T}$ and reflection-matrix $\mathcal{R}$ for this cavity was computed exactly in eq. \eqref{trchiachi}.
Using eq. \eqref{calcJ1}, we compute analytically the diagonal matrix elements of the transmission matrix $\mathcal{T}^\dagger_{\rightarrow} \mathcal{T}_{\rightarrow}$ leading to the transmission intensities (or transmittances) $T_{\perp,\pm}(\omega)$ (see also eq. \eqref{intensity})
\begin{equation}
T_{\perp,\pm}(\omega)=T_{\perp}(\omega) e^{-\alpha_{\pm}(\omega)L},
\label{ChiralBeer}
\end{equation}
where $\alpha_{\pm}(\omega)=2 \frac{\omega}{c}\mathrm{Im}(n_{\pm}(\omega))$ is the (polarization-dependent) inverse attenuation length in the Pasteur medium, and
\begin{equation}
T_{\perp}(\omega)=\frac{T_{1}(\omega)T_{2}(\omega)}{1+R^{2}(\omega)e^{-2\alpha(\omega) L}-2 e^{-\alpha(\omega) L}\mathrm{Re}\left(r^{2}(\omega)e^{i\beta(\omega)L}\right)},
\label{backgroundterm}
\end{equation}
is the standard transmittance of a FP cavity with $\kappa=0$ \cite{IntroOptics,Born1980}.
We wrote $T_{1(2)}(\omega)$ the transmission factor across the mirror $1(2)$, $R(\omega)\equiv\lvert r(\omega)\rvert^{2}$ the reflection factor inside the cavity at the mirror interfaces, $\alpha(\omega)=2\frac{\omega}{c}\mathrm{Im}(n(\omega))$ the (polarization-independent) attenuation factor, and $\beta(\omega)=2\frac{\omega}{c}\mathrm{Re}(n(\omega))$ the dispersion factor of the medium.
Equation \eqref{ChiralBeer} is a generalization of the well-known Beer-Lambert Law of absorption (recovered when $\kappa=0$) to the case of Pasteur media.
Since the absorption factor is different for each polarization of the incoming light $(\alpha_+\ne\alpha_-)$, circular dichroism is induced by the Pasteur medium.
Using eqs. \eqref{ChiralBeer}, and \eqref{dci}, we find the following analytical formula
for the differential circular transmission
\begin{equation}
\mathcal{DCT}_{\perp}(\omega)=2\tanh\left[\frac{L}{2}\left(\alpha_{-}(\omega)-\alpha_{+}(\omega)\right)\right].
\label{dctnorm}
\end{equation}
This expression is the main result of this subsection.
It has a deep implication, namely it shows that the properties of the metallic mirrors do not influence the $\mathcal{DCT}_{\perp}$ of the whole system.
Remarkably, if we compute the $\mathcal{DCT}_{\perp}$ for the bare LHIC medium (without mirrors), we find exactly the same expression. 
This means that embedding a LHIC Pasteur medium inside a FP cavity made of standard mirrors does not yield a significant increase of its chiroptical response \cite{Baranov2020,Munkhbat2021}. 
This phenomenon arises from the reversal of the helicity of the wave inside the cavity at each successive reflection on the metallic mirrors.
The only contribution to circular dichroism thus arises from direct transmission across the Pasteur medium, as it is actually described by eq. \eqref{ChiralBeer}.
We report in Figure \ref{Plot1}-($a$), the 2D-map of the transmittance $T_{\perp,+}$ as a function of the medium thickness $L$ and input light frequency $\omega$, computed numerically by the transfer-matrix approach of Sec. \ref{Transnum}. 
The first three FP modes located at \textcolor{black}{$\omega=p \pi c/nL$} with $p=1,2,3$ are visible.
Due to the losses in the mirrors and material, the transmittance decreases exponentially with the cavity length, in full accordance with eq. \eqref{ChiralBeer}. 
We plot in Figure \ref{Plot1}-($b$) the related 2D-map for $\mathcal{DCT}_{\perp}$, in a range of frequency close to the material  absorption at $\omega_{0}$ (dashed white line in Figure \ref{Plot1}-($a$)).
We find that $\mathcal{DCT}_{\perp}$ takes the largest values for $\omega$ in a band centered around $\omega_0$ and of width $\gamma$, due to the chosen Lorentzian frequency-dependence of $\kappa(\omega)$ in our model (see eq. \eqref{kappapdisp}).
At fixed frequency, $\mathcal{DCT}_{\perp}$ increases together with the cavity thickness.
This is consistent with eq. \eqref{dctnorm}, meaning that indeed the only factor influencing $\mathcal{DCT}_{\perp}$ is the asymmetric absorption of LCP and RCP waves along a direct transmission path across the Pasteur material.
Finally, the $\mathcal{DCT}_{\perp}$-signal is negative, due to the fact that for our chosen positive value of $\kappa=10^{-3}$, the chiral Pasteur material absorbs LCP waves more than RCP ones.
%

%
\subsubsection{Polaritonic regime}
\label{PolForm}
%
\begin{figure}[ht!]
\begin{center}
\includegraphics[width=\linewidth]{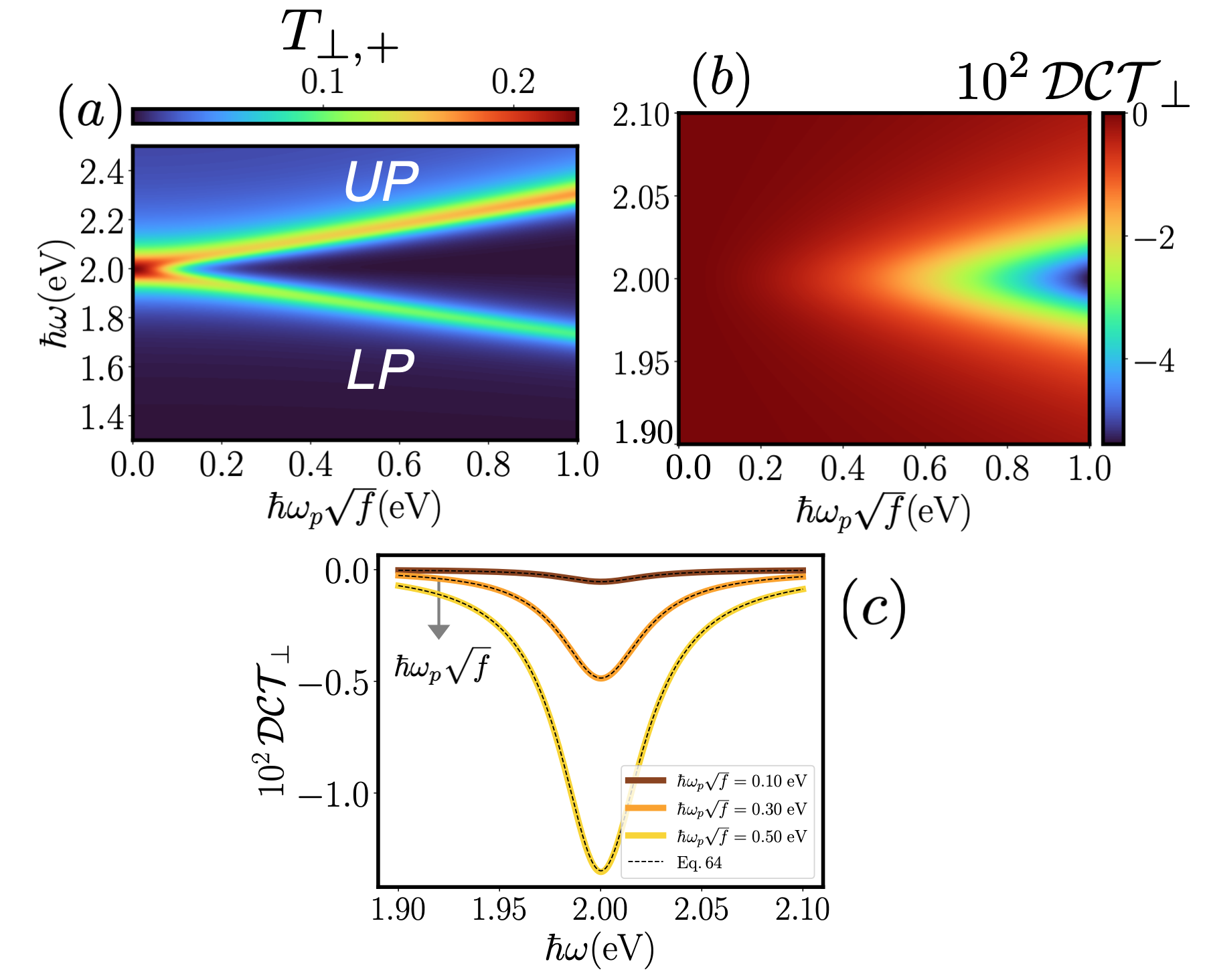}
\caption{($a$) Transmittance $T_{\perp,+}$ at normal incidence ($\theta=0$) of a FP cavity made by [Ag mirror-LHIC-Ag mirror] as a function of $\omega$ and plasma frequency $\omega_{p}\sqrt{f}$ of the LHIC medium. 
\textcolor{black}{The cavity mode is at resonance with the material absorption frequency $\omega_{0}$}. 
($b$) Differential circular transmission at normal incidence $\mathcal{DCT}_{\perp}$  of the same system. The parameters are those of Figure \ref{Plot1} except that $L=\SI{133}{nm}$. \textcolor{black}{($c$) Differential circular transmission at normal incidence $\mathcal{DCT}_{\perp}$ as a function of $\omega$ for three values of $\hbar\omega_{p}\sqrt{f}$ (\SI{0.1}{eV} in brown, \SI{0.3}{eV} in orange, \SI{0.5}{eV} in yellow) and comparison to the analytical formula in eq. \eqref{dctnorm} (dashed black lines).} 
%
}
\label{Plot2}
\end{center}
\end{figure}
%
In this subsection, we study the effect on chiroptical properties of increasing the light-matter coupling strength given by the plasma frequency $\hbar\omega_{p}\sqrt{f}$ in eq. \eqref{ConstRel3}.
When $\hbar\omega_{p}\sqrt{f}$ becomes significantly larger than the total cavity losses, the FP cavity modes do mix with the material properties to generate new hybrid light-matter eigenmodes: one at lower energies, the \textit{lower polariton} (LP), and the other at higher energies, the \textit{upper polariton} (UP) \cite{Zhu1990,PhysRevLett.69.3314}.
The appearance of the LP and UP has a direct signature in the analytical properties of the background transmittance $T_{\perp}(\omega)$ (see eq. \eqref{ChiralBeer}) seen as a function of the complex-valued frequency $\omega\equiv z_\omega$. 
The LP and UP are associated to the emergence of new poles in $T_{\perp}(z_\omega)$ (or zeros of the transmittance denominator) in the complex plane of $z_\omega$.
This regime is the so-called strong-coupling or \textit{polaritonic} regime. 
The splitting between the polaritonic states is centered at cavity frequencies that are resonant with the material absorption frequency $\omega_{0}$.
The onset of polaritons is well-resolved in Figure \ref{Plot2}-($a$), by computing numerically the transmittance $T_{\perp,+}$ as a function of $\hbar\omega_{p}\sqrt{f}$.
The latter is varied by changing the concentration of molecules in the Pasteur medium.
\textcolor{black}{Within the model of dielectric constant provided in Eq.~\ref{ConstRel3}, we find that the observed splitting between polaritonic states scales linearly with $\hbar\omega_{p}\sqrt{f}/n_\infty$, with $n_\infty=\sqrt{\varepsilon_\infty}$.} 
%
%
We observe that the LP is brighter than the UP, due to asymmetry in the frequency-dependent response of the cavity.
We show in Figure \ref{Plot2}-($b$) the related evolution of $\mathcal{DCT}_{\perp}$ with $\hbar\omega_{p}\sqrt{f}$, at frequencies close to $\omega_0$.
Similarly to Figure \ref{Plot1}-($b$), we observe that the $\mathcal{DCT}_{\perp}$-signal is maximum around the material absorption frequency $\omega_0$ and increases with $\hbar\omega_{p}\sqrt{f}$, due to the corresponding increase of $\kappa(\omega)$ in eq. \eqref{kappapdisp}.
%
%
\textit{However, there is no signature of the LP and UP branches in the $\mathcal{DCT}_{\perp}$-signal.}
As described by eq. \eqref{dctnorm}, this is due to the only dependence of $\mathcal{DCT}_{\perp}$ with the difference of LCP and RCP absorption coefficients in the Pasteur medium (and not with the background transmission).
\textcolor{black}{As shown in Figure~\ref{Plot2}-($c$), the $\mathcal{DCT}_{\perp}$-signals computed numerically (plain curves) for given values of $\hbar\omega_{p}\sqrt{f}$ (\SI{0.1}{eV} in brown, \SI{0.3}{eV} in orange, \SI{0.5}{eV} in yellow), perfectly match with the analytical formula written in eq. \eqref{dctnorm} (dashed black curves).}
%
\subsection{Finite incidence angle}
%
\begin{figure}[ht!]
\begin{center}
\includegraphics[width=\linewidth]{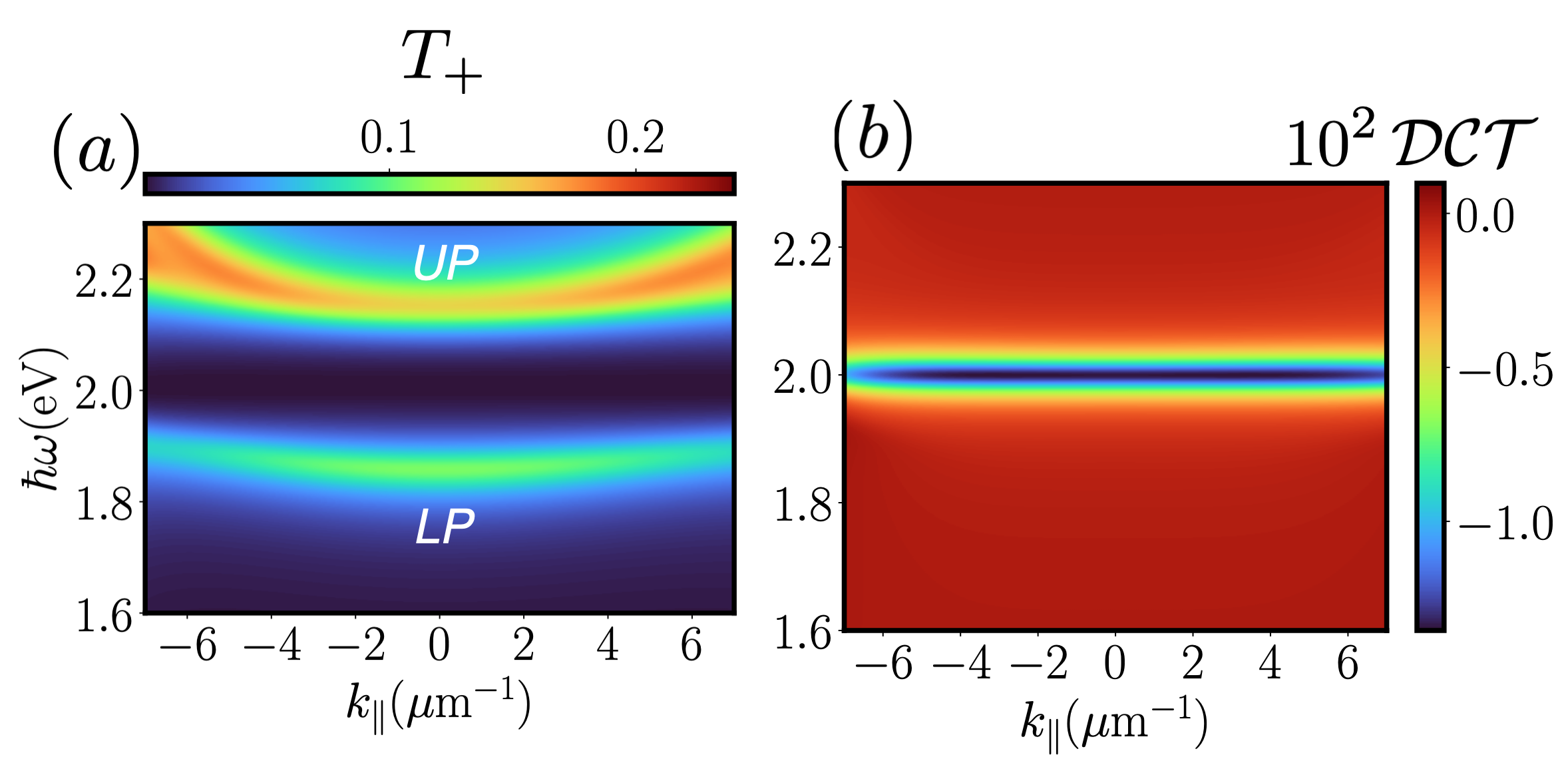}
\caption{($a$) Transmittance $T_{+}$ of a FP cavity made by [Ag mirror-LHIC-Ag mirror] as a function of $\omega$ in eV and $k_{\parallel}=n\omega \sin\theta/c$ in $\mu m^{-1}$.
%
%
($b$) Differential circular transmission $\mathcal{DCT}$ of the same system.
The parameters are those of Figure \ref{Plot1} except that $L=\SI{133}{nm}$ and $\hbar\omega_{p}\sqrt{f}=\SI{0.5}{eV}$.
}
\label{Plot3}
\end{center}
\end{figure}
%
%
We complement the results of the previous section to the case of finite incidence angle ($\theta \neq 0$), and investigate their dependence with the parallel component of the incident wave-vector $k_{\parallel}=n\frac{\omega}{c}\sin\theta$.
In comparison to the case of normal incidence, the transmission and reflection matrices have now both diagonal and off-diagonal matrix elements in the circularly polarized basis (as in Sec. \ref{SingleInt}). 
In contrast to eqs. \eqref{ChiralBeer} and \eqref{dctnorm}, the resulting transmittances and $\mathcal{DCT}$ at finite incident angles, are expected now to depend not only on the Pasteur medium absorption and dispersion, but also on the polarization-dependent multiple reflection processes arising at the mirror interfaces.
We show that this is the case in Figure \ref{Plot3}-$(a)$ for the computed 2D-map of the transmittance $T_{+}$ as a function of $k_{\parallel}$ and $\omega$, where we resolve the dispersion of the LP and UP polariton.
%
%
%
This is in strong contrast with the corresponding $\mathcal{DCT}$ map at finite incident angles, shown in Figure \ref{Plot3}-($b$).
We observe an increased (and negative) $\mathcal{DCT}$-signal for frequencies close to $\omega_0$, due to the frequency-dependence of the Pasteur coefficient $\kappa(\omega)$, but without the expected dispersion behavior (encoded in the $k_{\parallel}$-dependence of the transmission and reflection matrices).
We still detect no signature of the UP and LP polariton in the $\mathcal{DCT}$-signal, suggesting that upon varying the incidence angle $\theta$, the asymmetry in the polarization-content of the wave stored inside the cavity is still too weak to have a significant effect beyond the trivial eq. \eqref{dctnorm}. 
%
%
%
%
This justifies the search for different materials with which to build the cavity and that would enhance its intrinsic chiroptical properties.
 For this purpose, we propose and study in the last Sec. \ref{HPfabry}, a FP cavity made of dielectric photonic crystal (PC) mirrors.
%

%
%
\section{Fabry-P\'{e}rot cavity with dielectric photonic crystal mirrors}
\label{HPfabry}
%
%
\begin{figure}[ht!]
\begin{center}
\includegraphics[width=1.0\linewidth]{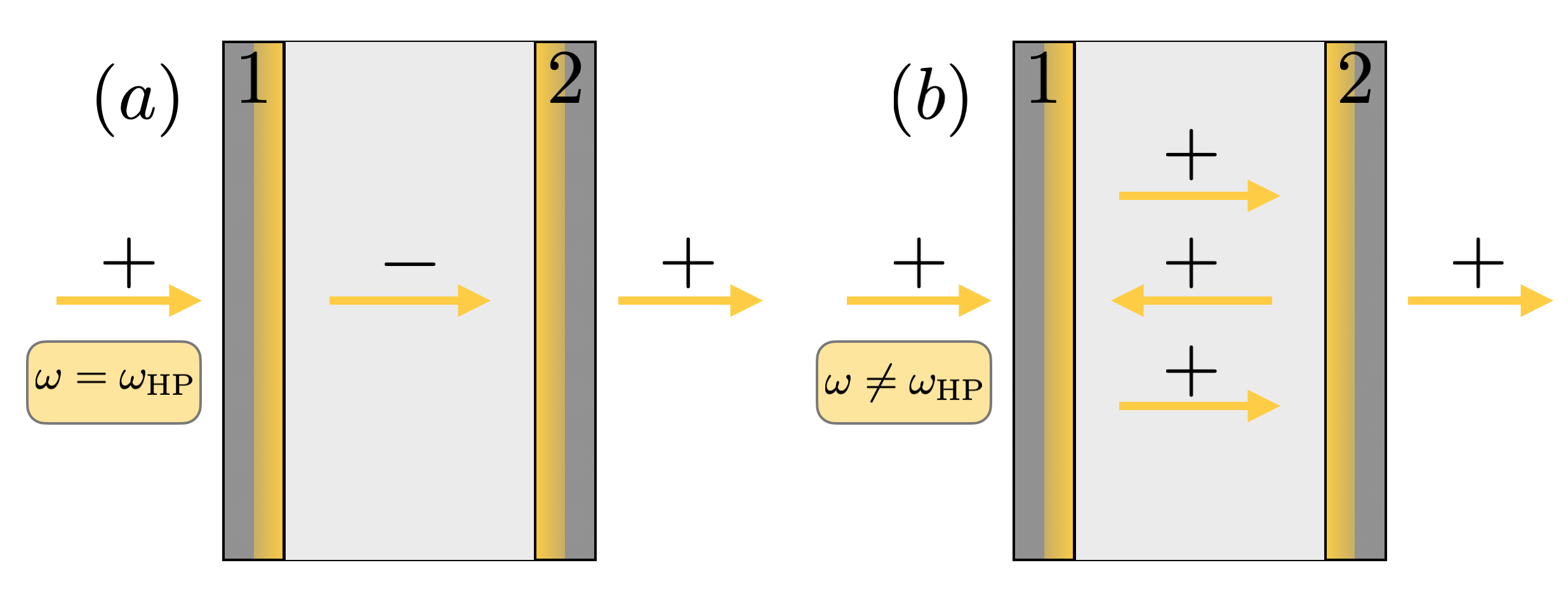}
\caption{Sketch of a FP cavity made by two dielectric photonic crystal mirrors $(1)$ and $(2)$, illuminated at normal incidence. 
The mirror $(2)$ is obtained by flipping the mirror $(1)$, as shown by the flipped position of the yellow band on the scheme.
\textcolor{black}{This cavity transmits and converts perfectly one polarization of the waves inside the cavity for a frequency equal to the helicity-preserving band frequency $\omega_{\rm{HP}}$ (see panel $(a)$), while preserving partially the helicity of the reflected waves at the internal medium-mirror interfaces when slightly detuned in frequency from $\omega_{\rm{HP}}$ (see panel $(b)$).}
%
%
%
%
%
}
\label{Fig8}
\end{center}
\end{figure}
%
In this section, we consider an alternative Fabry-P\'{e}rot cavity, made of two helicity-preserving (HP) dielectric photonic crystal mirrors (see Fig. \ref{Fig8}), instead of the previous standard metallic mirrors. 
The optical response of such mirrors was reported experimentally in ref. \cite{Maruf2020}, and their theoretical modelling was proposed in refs. \cite{voronin2021single,PhysRevA.107.L021501}.
In the following, we provide further details about the specificity of the modelling and chiroptical response of such an HP cavity under illumination at normal incidence, thus complementing our recent ref. \cite{PhysRevA.107.L021501}.
%

%
\subsection{Time-reversal symmetric dielectric photonic crystal mirrors}
\label{TRSMirror}
We consider a single lossless PC mirror, labelled $(1)$ in Fig. \ref{Fig8}.
%
%
%
\textcolor{black}{Upon normal illumination at nominal frequency $\omega = \omega_{\mathrm{HP}}$ (HP band frequency) of the photonic structure, this mirror is designed to perfectly transmit and cross-convert one polarization of the incoming wave (LCP (+)$\rightarrow$ RCP (-) for the mirror $(1)$ of Fig.~\ref{Fig8}-$(a)$), while reflecting and preserving entirely the other polarization (RCP (-)$\rightarrow$ RCP (-)).} 
%
%
A standard metallic mirror would have reversed the handedness of the reflected wave \cite{jackson1999classical}.
%
%
Outside of the nominal HP band ($|\omega-\omega_{\mathrm{HP}}| \gg \gamma_{\mathrm{HP}}$, with $\gamma_{\mathrm{HP}}$ the bandwidth), the mirror behaves as a standard and quasi lossless dielectric mirror both in reflection and transmission.
%
%
\textcolor{black}{The target helicity-preserving properties of this mirror for optics are achieved in the neighborhood of the HP band, at which frequencies ($|\omega-\omega_{\mathrm{HP}}| \approx \gamma_{\mathrm{HP}}$) the helicity of light is partially conserved upon internal reflection (see LCP (+)$\rightarrow$ LCP (+) for internal reflections on mirror $(1)$ in Fig.~\ref{Fig8}-$(b)$).}
%
%
Following our previous work \cite{PhysRevA.107.L021501}, we model a single HP mirror \textcolor{black}{(the left mirror $(1)$ in Fig.~\ref{Fig8})} with the following transmission ($\mathcal{T}_{\rightarrow}$) and reflection matrices ($\mathcal{R}_{\leftarrow}$) 
\begin{equation}
\begin{split}
\mathcal{T}_{\rightarrow}&=
\begin{bmatrix}
t & 0\\
t_{+-} & t\\
\end{bmatrix}=
\begin{bmatrix}
|t|\cdot e^{i\phi_{t}} & 0\\
|t_{+-}|\cdot e^{i\phi_{\mathrm{HP}}} & |t|\cdot e^{i\phi_{t}}\\
\end{bmatrix}\\
\mathcal{R}_{\leftarrow}&=
\begin{bmatrix}
0 & r\\
r & r_{--}\\
\end{bmatrix}=
\begin{bmatrix}
0 & |r|\cdot e^{i\phi_{r}}\\
|r|\cdot e^{i\phi_{r}} & |r_{--}|\cdot e^{i\phi_{\mathrm{HP}}}\\
\end{bmatrix},
\end{split}
\label{firstAssumption}
\end{equation}
that are parametrized by the transmission and reflection amplitudes $(|t|,|t_{+-}|,|r|,|r_{--}|)$, and by the phases $(\phi_t,\phi_r,\phi_{\mathrm{HP}})$.
Those parameters are constrained to reproduce various known limits.
First, eq. \eqref{firstAssumption} should reproduce the behavior of a standard dielectric mirror for frequencies out of the HP band ($|\omega-\omega_{\mathrm{HP}}| \gg \gamma_{\mathrm{HP}}$), namely it should coincide with the transmission and reflection matrices (with $\kappa=0$) derived in eq. \eqref{trchiachi}. 
This condition is fulfilled by choosing equal diagonal transmission coefficients and equal off-diagonal reflection coefficients, namely that $t_{++}=t_{--}=|t|\cdot e^{i\phi_{t}}$ and $r_{-+}=r_{+-}=|r|\cdot e^{i\phi_{r}}$.
Then, the parametrization of eq. \eqref{firstAssumption}, should also reproduce the absolute values of the experimental transmission and reflection amplitudes reported in ref. \cite{Maruf2020}, for frequencies in the HP band ($|\omega-\omega_{\mathrm{HP}}| \leq \gamma_{\mathrm{HP}}$).
%
%
In particular, the matrix elements $t_{-+}$ and $r_{++}$ are chosen to be zero, to reproduce the strong imbalance of the mirror in cross-transmission and in reflection \cite{Maruf2020}.
In order to reproduce the experimental frequency-dependence of $|t_{+-}|$ in the HP band, we adopt a Lorentzian-like model \cite{PhysRevA.107.L021501} with resonance frequency $\omega_{\mathrm{HP}}$ and decay rate $\gamma_{\mathrm{HP}}$
\begin{equation}
|t_{+-}|\cdot e^{i\phi_{\mathrm{HP}}}=\frac{\gamma_{\mathrm{HP}}}{i(\omega-\omega_{\mathrm{HP}})+\gamma_{\mathrm{HP}}}
\label{CMT} \, .
\end{equation}
This amounts to describe approximately the optical response of the HP mirrors by a \emph{coupled-mode theory} \cite{Gorkunov2020,Kondraton2016,Fan2003} with one single-mode of the photonic mirror.
%
%
Since the mirrors have negligible losses, the conjunction of energy-conservation and reciprocity of the dielectric mirror, imposes time-reversal symmetry \cite{Fink1993,Rosny2010} (TR) on its scattering-matrix $\stackrel{\leftrightarrow}{\mathcal{S}}$, and thus the following additional constraints (see Appendix \ref{Treversability})
\begin{equation}
\mbox{TR}
\Leftrightarrow\,\stackrel{\leftrightarrow}{\mathcal{S}}=\stackrel{\leftrightarrow}{\sigma}_z\,\left(\stackrel{\leftrightarrow}{\mathcal{S}^{\star}}\right)^{-1} \,\stackrel{\leftrightarrow}{\sigma}_z
\label{TRscattering} \, ,
\end{equation}
where $\stackrel{\leftrightarrow}{\sigma}_z$ is defined in eq. \eqref{RSC10}.
Using eq. \eqref{TRscattering} and the parametrization in eq. \eqref{firstAssumption} and eq. \eqref{CMT}, we can reconstruct the full scattering-matrix of the mirror, containing the following transmission and reflection matrices
\begin{equation}
\begin{split}
\mathcal{T}_{\rightarrow}&=
\begin{bmatrix}
|t|\cdot e^{i\phi_{t}} & 0\\
|t_{+-}|\cdot e^{i\phi_{\mathrm{HP}}} & |t|\cdot e^{i\phi_{t}}\\
\end{bmatrix}=\left(\mathcal{T}_{\leftarrow}\right)^{\mathrm{T}},\\
\mathcal{R}_{\leftarrow}&=
\begin{bmatrix}
0 & |t|\cdot e^{i(2\phi_{\mathrm{HP}}-\phi_{t})}\\
|t|\cdot e^{i(2\phi_{\mathrm{HP}}-\phi_{t})} & -|t_{+-}|\cdot e^{i\phi_{\mathrm{HP}}}\\
\end{bmatrix}=\left(\mathcal{R}_{\leftarrow}\right)^{\mathrm{T}},\\
\mathcal{R}_{\rightarrow}&=
\begin{bmatrix}
|t_{+-}|\cdot e^{i(4\phi_{t}-3\phi_{\mathrm{HP}})} & -|t|\cdot e^{i(3\phi_{t}-2\phi_{\mathrm{HP}})}\\
-|t|\cdot e^{i(3\phi_{t}-2\phi_{\mathrm{HP}})} & 0\\
\end{bmatrix}=\left(\mathcal{R}_{\rightarrow}\right)^{\mathrm{T}} ,
\end{split}
\label{Smetafinal}
\end{equation}
with $|t_{+-}|^{2}+2|t|^{2}=1$ standing for energy-conservation.
%

\subsection{Empty HP Fabry-P\'erot cavity}
%
%
\begin{figure}[ht!]
\begin{center}
\includegraphics[width=\linewidth]{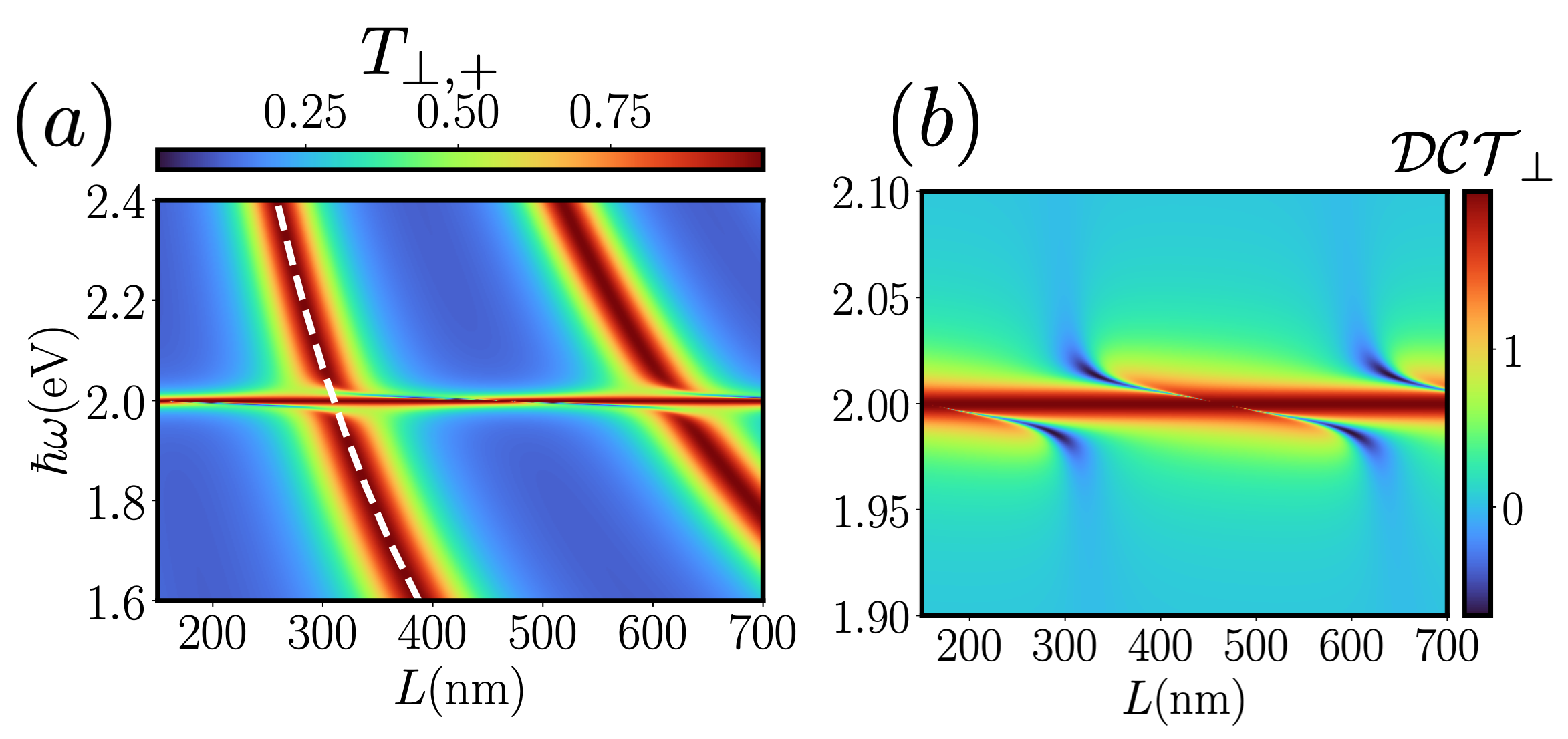}
\caption{($a$) Transmittance at normal incidence ($\theta=0$) $T_{\perp,+}$ of a FP cavity made by [HP mirror-Air-HP mirror] as a function of $\omega$ in eV and the thickness $L$ of a layer of air in nm. The dashed white line indicates the fundamental FP mode, obtained by the relation $L[\mathrm{nm}]=1239.8/(2\hbar\omega[\mathrm{eV}])$. ($b$) Differential circular transmission at normal incidence $\mathcal{DCT}_{\perp}$ of the same system. The parameters for the dielectric photonic crystal mirrors are as follows: $\hbar\omega_{\mathrm{HP}}=\SI{2.0}{eV}$, $\hbar\gamma_{\mathrm{HP}}=\SI{0.01}{eV}$ and $\phi_{t}=\pi/2$.}
\label{Plot4}
\end{center}
\end{figure}
%
%
In this subsection, we consider the optical response of an empty HP Fabry-P\'erot cavity.
The first mirror $(1)$ is presented and studied in Sec. \ref{TRSMirror}, while the second mirror $(2)$ is obtained by flipping the first one (see Fig. \ref{Fig8}). 
The scattering matrix of the flipped mirror yields the same coefficients as the ones in eq. \eqref{Smetafinal}, but with exchanged polarization indices.
We present in Figure \ref{Plot4}-($a$) the transmittance $T_{\perp,+}$ of the HP cavity as a function of frequency and cavity length. 
The main features of this 2D map are the presence of standard FP modes as in Fig. \ref{Plot1}-($a$) (see dashed white line in Fig. \ref{Plot4}-($a$)), and the emergence of an helicity-preserving and non-dispersive band located at $\hbar\omega=\hbar\omega_{\rm{HP}}=\SI{2.0}{eV}$.
We compute numerically the chiroptical response of this cavity close to the HP band, and show on Figure \ref{Plot4}-($b$) the related 2D map of $\mathcal{DCT}_{\perp}$.
The HP region appears as a red non-dispersive band with $\mathcal{DCT}_{\perp}=2$, while blue regions  with negative $\mathcal{DCT}_{\perp}$ are generated when detuning the cavity away from the HP band.
In order to interpret those plots, we obtain analytically (following the Green function approach developed in Sec. \ref{GenGreen}) the average transmission $\overline{T}_\perp$ (see eq. \eqref{dci1}) and transmission imbalance $\Delta T_\perp$ (see eq. \eqref{dci2}) along the fundamental FP mode (hence for $L=\lambda/2$), leading to
%
%
%
\begin{eqnarray}
\overline{T}_\perp &=& \frac{32(1 + \delta^2) + 8 \delta^4 + \delta^8}{64(1 + \delta^2) + 16\delta^4 + \delta^8},
\label{TAvchiralMir} \\
\mathcal{DCT}_{\perp} &=& \frac{16(4-\delta^{4})}{32\left(1+\delta^{2}\right)+8\delta^{4}+\delta^{8}},
\label{DCTchiralMir}
\end{eqnarray} 
where $\delta \equiv \left(\omega-\omega_{\mathrm{HP}}\right)/\gamma_{\mathrm{HP}}$ is the relative detuning with respect to the HP region. 
As we have shown in our recent ref. \cite{PhysRevA.107.L021501}, eq. \eqref{DCTchiralMir} perfectly describes the shape of the $\mathcal{DCT}_{\perp}$-signal.
At the center of the HP region ($\delta=0$), $\mathcal{DCT}_{\perp}$ reaches its maximum value of 2 for an averaged transmittance $\overline{T}_\perp = 0.5$.
This is due to the specificity in transmission and reflection of the cavity dielectric photonic crystal mirrors.
Indeed, an incoming LCP wave at $\omega=\omega_{\mathrm{HP}}$ is perfectly transmitted and converted by mirror $(1)$ to a RCP wave inside the cavity. 
The latter is further perfectly transmitted by mirror $(2)$ as an LCP wave out of the cavity \textcolor{black}{(see Fig.~\ref{Fig8}-$(a)$)}, thus resulting in a total transmittance $T_{\perp,+}=1$.  
In contrast to that behavior, an incoming RCP wave is perfectly reflected by mirror $(1)$ and thus does not enter into the cavity.
This implies a vanishing transmittance $T_{\perp,-}=0$ and leads to $\overline{T}_\perp = 0.5$ and $\Delta T_\perp=1$. 
At the center of the HP band, the cavity achieves the largest imbalance in LCP and RCP transmission, reaching a dual point of ``maximal electromagnetic chirality" \cite{Corbaton2016}.
Far from the HP band $(|\delta|\rightarrow\infty)$, the $DCT_\perp$ signal vanishes with the inverse of the fourth power of the detuning $DCT_\perp \approx - 16/\delta^4$, recovering the behavior of a FP cavity with standard mirrors shown in Fig. \ref{Plot1}.
The related average transmission in this limit is $\overline{T}_\perp \rightarrow 1$, for which $T_{\perp,+}=T_{\perp,-}\approx 1$.
Interestingly, close to the crossover region between standard and helicity-preserving regions $(|\delta| \approx \sqrt{2})$, the $DCT_\perp$-signal acquires a strong negative value, revealed as blue regions in Fig. \ref{Plot4}-($b$). 
There, the electromagnetic field stored inside the cavity acquires a significant imbalance in polarization content (more LCP than RCP wave content in \textcolor{black}{Fig.~\ref{Fig8}-$(b)$}) by accumulation of multiple HP reflections onto the mirrors, thus resulting in $T_{\perp,+}<T_{\perp,-}$ for transmitted light outside the cavity.
\textit{In other words, the blue regions constitute regions of efficient partial helicity-preservation induced by the cavity mirrors.}
In addition, we note that the obtained $\mathcal{DCT}_\perp$-signal generated by this HP cavity is orders of magnitudes higher that the one obtained for the LHIC medium computed in Sec. \ref{FPsilver}. 
We have shown in our previous ref. \cite{PhysRevA.107.L021501}, that this enhanced chiral discrimination property of the HP cavity might be useful to design a new generation of efficient chiral sensors. 
%

\subsection{Onset of Chiral Polaritons}
%
\begin{figure}[ht!]
\begin{center}
\includegraphics[width=\linewidth]{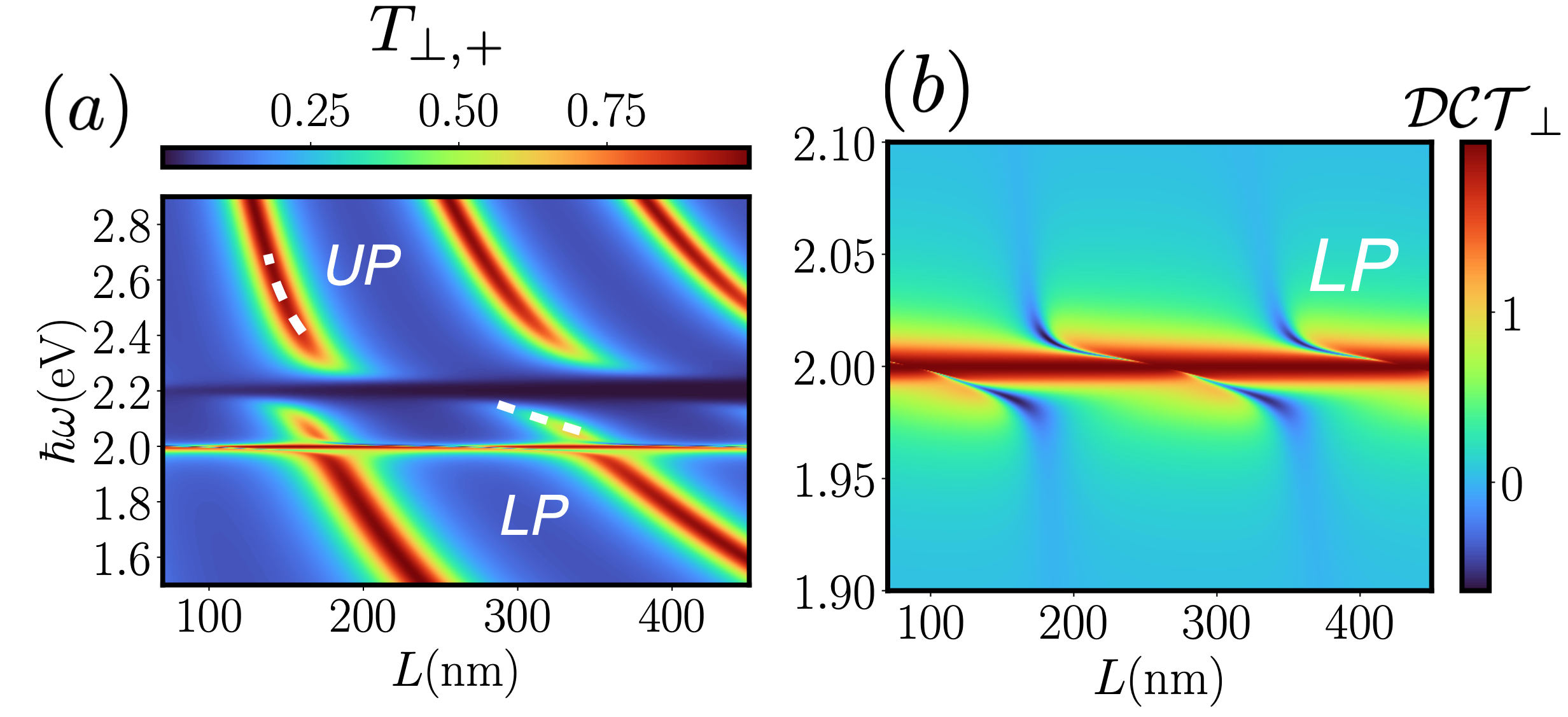}
\caption{($a$) Transmittance at normal incidence ($\theta=0$) $T_{\perp,+}$ of a FP cavity made by [HP mirror-LHIC-HP mirror] as a function of $\hbar\omega$ in eV and the thickness $L$ of a LHIC medium in nm. 
%
%
%
\textcolor{black}{The dashed white lines indicate the UP for the mode $p=1$ and the LP for the mode $p=2$ of the FP cavity.} 
%
%
%
($b$) Differential circular transmission at normal incidence $\mathcal{DCT}_{\perp}$ for the same system. 
The parameters for the dielectric photonic crystal mirrors are as follows: $\hbar\omega_{\mathrm{HP}}=\SI{2.0}{eV}$, $\hbar\gamma_{\mathrm{HP}}=\SI{0.01}{eV}$ and $\phi_{t}=\pi/2$. The parameters for the LHIC medium are : $\varepsilon_{\infty}=2.89$, $\hbar\omega_{p}\sqrt{f}=\SI{0.5}{eV}$, $\hbar\omega_{0}=\SI{2.2}{eV}$, $\hbar\gamma=\SI{0.05}{eV}$ and $\kappa=10^{-3}$.
}
\label{Plot5}
\end{center}
\end{figure}
%
We now fill the HP cavity with a Pasteur medium with an absorption frequency $\omega_{0}$ that is detuned with respect to the central frequency $\omega_{\mathrm{HP}}$ of the HP band.
The intrinsic contribution of the Pasteur medium to the cavity $\mathcal{DCT}_{\perp}$-signal (with respect to an achiral medium) was investigated in depth in our recent ref. \cite{PhysRevA.107.L021501}.
We also reported for the first time the formation of \textit{chiral polaritons} in such HP cavities.
In this final section of the paper, we expand the analysis of chiral polaritons that are formed upon increasing the medium oscillator strength.
%
%
\textcolor{black}{We show in Figure \ref{Plot5}-($a$) the computed transmittance $T_{\perp,+}$ in the vicinity of the material absorption frequency $\omega_0$, with the onset of the UP and LP branches corresponding to the $p=1$ and $p=2$ modes of the FP cavity (see dashed white lines).}
%
%
%
The zoom at frequencies close to $\omega_{\rm{HP}}$ in Figure \ref{Plot5}-($b$), reveals that, similarly to the bare optical mode of Fig.\ref{Plot4}-($b$), the LP branch inherits from the HP cavity an imbalance in its polarization content that is visible as a strong negative (blue) signal in $\mathcal{DCT}_{\perp}$.
We note that the strongest asymmetry is obtained in the regime of intermediate detuning $\delta\approx \sqrt{2}$, for which the cavity is optimally helicity-preserving. 
The formation of such a chiral polariton and the analysis of its related signature on the cavity optical properties is one of the main results of this paper.
%
%
\textcolor{black}{\textit{It is interesting to notice that the $\mathcal{DCT}_{\perp}$-signal associated to the chiral polaritons is dominated by the background optical activity induced by the dielectric photonic crystal mirrors.}
The contribution of the Pasteur medium is of lower contribution, and was obtained and investigated in Ref.\cite{PhysRevA.107.L021501} after subtracting the background contribution of the mirrors.}
%
%
%
%
\begin{figure}[ht!]
\begin{center}
\includegraphics[width=0.8\linewidth]{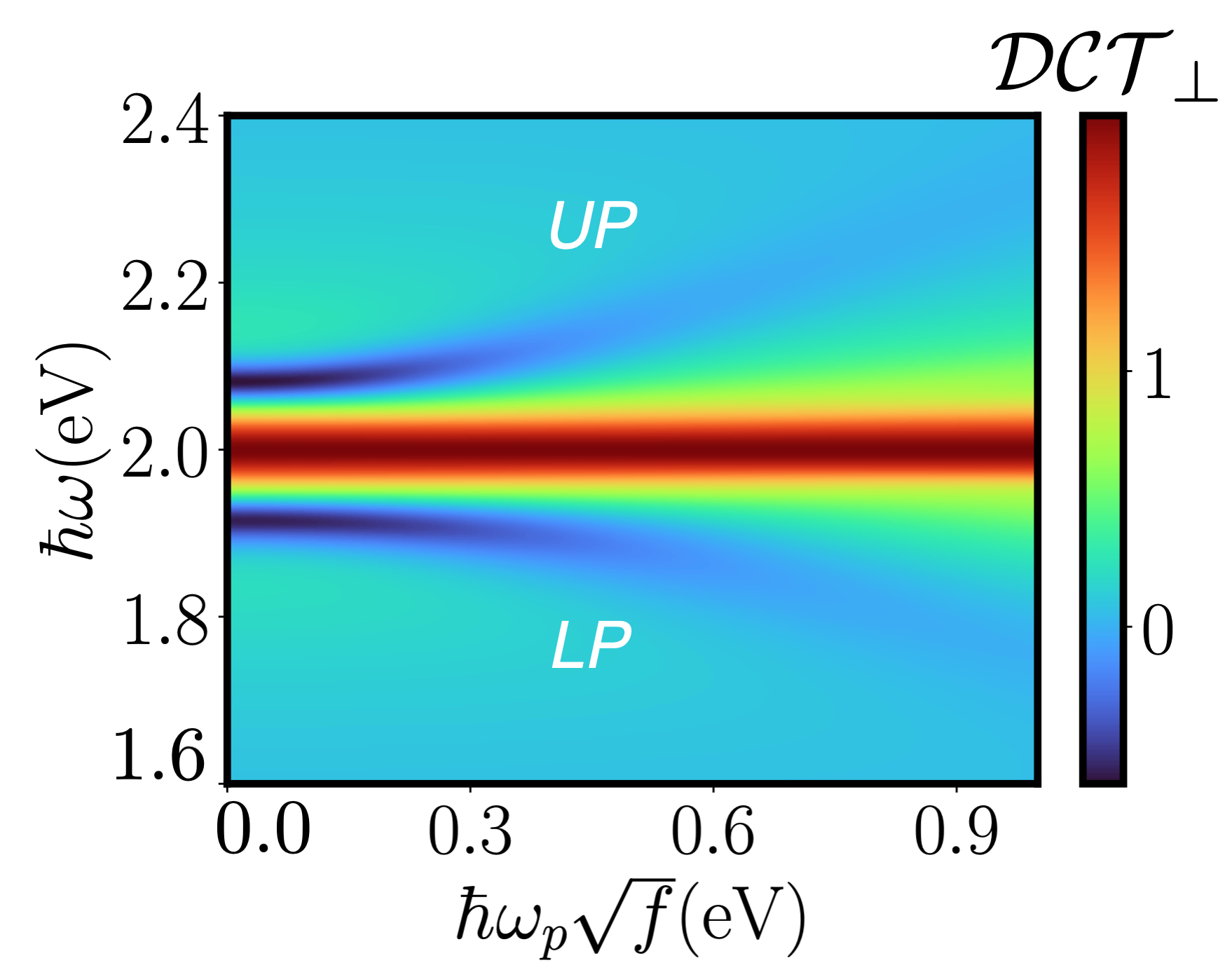}
\caption{Differential circular transmission at normal incidence $\mathcal{DCT}_{\perp}$ as a function of $\omega$ in eV and $\omega_{p}\sqrt{f}$ of a LHIC medium in eV. The parameters for the dielectric photonic crystal mirrors are those of Figure \ref{Plot5} except that $\hbar\gamma_{\mathrm{HP}}=\SI{0.05}{eV}$. The parameters for the LHIC medium are those of Figure \ref{Plot5} except that $L=\SI{180}{nm}$ and $\hbar\omega_{0}=\SI{2.0}{eV}$.
}
\label{Plot6}
\end{center}
\end{figure}
%
For completeness, we show in Figure \ref{Plot6}, the evolution of $\mathcal{DCT}_{\perp}$ as a function of frequency $\omega$ and coupling strength $\omega_{p}\sqrt{f}$, obtained in the case where the absorption band of the material properties is at resonance with the mirrors HP band ($\omega_{0}=\omega_{\mathrm{HP}}$).
The UP and LP branches (blue lines) are clearly resolved in this case and split symmetrically with respect to the HP band. 
Similarly to Fig. \ref{Plot5}, they inherit from the proximity of the HP band an imbalance in their polarization content, seen as a strong negative $\mathcal{DCT}_{\perp}$-signal. 
\textcolor{black}{For the range of parameters shown in Fig.\ref{Plot6}, the $\mathcal{DCT}_{\perp}$-signals associated to the LP and UP do not reverse sign upon changing the sign of the Pasteur coefficient $\kappa$ and thus the chirality of the molecules embedded inside cavity.
This is due to the fact that, chiroptical properties of the cavity at the HP-band frequency (which is resonant here with the absorption band of the Pasteur medium) are given by the mirrors property which select perfectly one helicity of the wave propagating inside the cavity, the other helicity being not transmitted (see Fig.\ref{Fig8}-(a)).}
%

%
\section{Conclusion and perspectives}
\label{Conc}
We have developed a theoretical approach based on classical electromagnetism, and implemented a numerical code based on transfer-matrices, to compute the optical properties of multilayered, linear and chiral materials embedded inside a Fabry-P\'erot cavity.
%
%
We have shown how Lorentz's reciprocity constrains the modelling and impacts the chiroptical scattering properties of such cavities. 
Our numerical method was shown to be equivalent (and we successfully compared it) to other existing theoretical approaches based on scattering-matrices or on the use of electromagnetic Green's functions. 
In particular, we have derived formally an explicit relation connecting the transmission-matrix of the cavity to the Green's function of the material inside the cavity, in presence of scattering by the mirrors.
This relation enabled to compute analytically the polarization-dependent transmittance and differential circular transmission ($\mathcal{DCT}$) for various Fabry-P\'erot cavities.
%
%
To add interpretative value to our approach, we observe its analogy with coherent electronic quantum transport, broadening the context of our results beyond the field of nanophotonics.
We then applied those methods to the case of standard Fabry-P\'{e}rot cavities with metallic silver mirrors, showing the inefficiency of such cavities to enhance significantly the $\mathcal{DCT}$-signal, either by changing the incidence angle of the incoming wave or by entering the polaritonic strong-coupling regime.
This effect was interpreted (at normal incidence) by the reversal in helicity of the electromagnetic waves upon reflection at each mirror interface.
The latter generates a mixing in left and right circular-polarization of the cavity electromagnetic field.
This means that any path containing a number of reflections provides the same relative imbalance as the direct transmission path (without reflection) across the cavity, and the $\mathcal{DCT}$-signal thus is not modified in the presence of mirrors.
We have derived a simple Beer-Lambert absorption law for the $\mathcal{DCT}$-signal that quantitatively fits with this phenomenon and shows that $\mathcal{DCT}$ is mainly independent of the nature of the metallic mirrors, thus is not modified significantly by the cavity.
To circumvent this bottleneck, following and complementing our previous ref. \cite{PhysRevA.107.L021501}, we thus proposed, modelled and investigated chiroptical properties of helicity-preserving Fabry-P\'erot cavities.
The latter are made of dielectric photonic crystal mirrors designed to preserve the helicity of reflected waves at each mirror interface and to enable the accumulation of a significant polarization imbalance of the electromagnetic field stored inside cavity.
In such cavities, we reported the appearance of \textit{chiral cavity-polaritons} upon entering the light-matter strong coupling regime.
We analyzed in depth the mechanism of formation of such chiral excitation, and related it to the 
specifically designed properties of multiple reflections of electromagnetic waves at the mirror interface.
%
%

%
Our approach is highly relevant and timely in the emerging field of \textit{chiral polaritonics}, 
for which it is mandatory to design new electromagnetic cavities for chiral-sensing purposes, or for alterating significantly the material stereochemical properties upon entering the polaritonic regime. 
A natural perspective and still open question raised by the present work, would be to investigate how vacuum quantum fluctuations of the chiral cavity-modes would couple to the material electronic properties.
Recent works started to propose several approaches along that direction 
\cite{riso2022strong,doi:10.1021/acs.jpclett.3c00286,salij2022chiral}.
We raise the point that any realistic microscopic model dealing with this issue, would have to incorporate as a limit case, and to encode information about the spatial structure of the classical electromagnetic field stored inside the cavity.
In particular, as we have shown, a fine tuning in energy of the cavity mirror properties has to be achieved to reach a significant imbalance in polarization-content of the chiral cavity-mode. 
This can be seen as a delicate adaptation of electromagnetic impedance between the external medium and the cavity material through the engineering of transmission and reflection matrices of the mirrors.
%
%
%
%
We hope our work to be useful for stimulating new theoretical and experimental developments along the research directions discussed in this paper.
We made the numerical code open source, available on
GitHub \cite{Code}.
%

%
\begin{acknowledgments}
L. Mauro and R. Avriller acknowledge financial support by Agence Nationale de la Recherche project CERCa, ANR-18-CE30-0006, EUR Light S\&T Graduate Program (PIA3 Program “Investment for the Future”, ANR-17-EURE-0027), IdEx of the University of Bordeaux / Grand Research Program GPR LIGHT, and Quantum Matter Bordeaux.
J. Fregoni and J. Feist acknowledge financial support by European Research Council through Grant ERC-2016-StG-714870 and by the Spanish Ministry for Science, Innovation, and Universities – Agencia Estatal de Investigaci\'on through Grants RTI2018-099737-B-I00, No PID2021125894NB-I00, and No CEX2018-000805-M (through the Mar\'ia de Maeztu Program for Units of Excellence in R\&D).
\end{acknowledgments}

%
\bibliography{biblio}

\begin{thebibliography}{86}%
\makeatletter
\providecommand \@ifxundefined [1]{%
 \@ifx{#1\undefined}
}%
\providecommand \@ifnum [1]{%
 \ifnum #1\expandafter \@firstoftwo
 \else \expandafter \@secondoftwo
 \fi
}%
\providecommand \@ifx [1]{%
 \ifx #1\expandafter \@firstoftwo
 \else \expandafter \@secondoftwo
 \fi
}%
\providecommand \natexlab [1]{#1}%
\providecommand \enquote  [1]{``#1''}%
\providecommand \bibnamefont  [1]{#1}%
\providecommand \bibfnamefont [1]{#1}%
\providecommand \citenamefont [1]{#1}%
\providecommand \href@noop [0]{\@secondoftwo}%
\providecommand \href [0]{\begingroup \@sanitize@url \@href}%
\providecommand \@href[1]{\@@startlink{#1}\@@href}%
\providecommand \@@href[1]{\endgroup#1\@@endlink}%
\providecommand \@sanitize@url [0]{\catcode `\\12\catcode `\$12\catcode
  `\&12\catcode `\#12\catcode `\^12\catcode `\_12\catcode `\%12\relax}%
\providecommand \@@startlink[1]{}%
\providecommand \@@endlink[0]{}%
\providecommand \url  [0]{\begingroup\@sanitize@url \@url }%
\providecommand \@url [1]{\endgroup\@href {#1}{\urlprefix }}%
\providecommand \urlprefix  [0]{URL }%
\providecommand \Eprint [0]{\href }%
\providecommand \doibase [0]{http://dx.doi.org/}%
\providecommand \selectlanguage [0]{\@gobble}%
\providecommand \bibinfo  [0]{\@secondoftwo}%
\providecommand \bibfield  [0]{\@secondoftwo}%
\providecommand \translation [1]{[#1]}%
\providecommand \BibitemOpen [0]{}%
\providecommand \bibitemStop [0]{}%
\providecommand \bibitemNoStop [0]{.\EOS\space}%
\providecommand \EOS [0]{\spacefactor3000\relax}%
\providecommand \BibitemShut  [1]{\csname bibitem#1\endcsname}%
\let\auto@bib@innerbib\@empty
\bibitem [{\citenamefont {Collet}\ \emph {et~al.}(2012)\citenamefont {Collet},
  \citenamefont {Crassous},\ and\ \citenamefont {Dutasta}}]{Collet2012}%
  \BibitemOpen
  \bibfield  {author} {\bibinfo {author} {\bibfnamefont {A.}~\bibnamefont
  {Collet}}, \bibinfo {author} {\bibfnamefont {J.}~\bibnamefont {Crassous}}, \
  and\ \bibinfo {author} {\bibfnamefont {J.}~\bibnamefont {Dutasta}},\ }\href
  {https://books.google.fr/books?id=M0umkoKaomcC} {\emph {\bibinfo {title}
  {Mol{\'e}cules chirales}}},\ Savoirs actuels\ (\bibinfo  {publisher} {EDP
  Sciences, Les Ulis (France)},\ \bibinfo {year} {2012})\BibitemShut {NoStop}%
\bibitem [{\citenamefont {Landau}\ \emph {et~al.}(1984)\citenamefont {Landau},
  \citenamefont {Bell}, \citenamefont {Kearsley}, \citenamefont {Pitaevskii},
  \citenamefont {Lifshitz},\ and\ \citenamefont
  {Sykes}}]{landau2013electrodynamics}%
  \BibitemOpen
  \bibfield  {author} {\bibinfo {author} {\bibfnamefont {L.~D.}\ \bibnamefont
  {Landau}}, \bibinfo {author} {\bibfnamefont {J.}~\bibnamefont {Bell}},
  \bibinfo {author} {\bibfnamefont {M.}~\bibnamefont {Kearsley}}, \bibinfo
  {author} {\bibfnamefont {L.}~\bibnamefont {Pitaevskii}}, \bibinfo {author}
  {\bibfnamefont {E.}~\bibnamefont {Lifshitz}}, \ and\ \bibinfo {author}
  {\bibfnamefont {J.}~\bibnamefont {Sykes}},\ }\href@noop {} {\emph {\bibinfo
  {title} {Electrodynamics of continuous media}}},\ Vol.~\bibinfo {volume} {8}\
  (\bibinfo  {publisher} {Elsevier},\ \bibinfo {year} {1984})\BibitemShut
  {NoStop}%
\bibitem [{\citenamefont {Craig}\ and\ \citenamefont
  {Thirunamachandran}(1984)}]{Craig1998}%
  \BibitemOpen
  \bibfield  {author} {\bibinfo {author} {\bibfnamefont {D.~P.}\ \bibnamefont
  {Craig}}\ and\ \bibinfo {author} {\bibfnamefont {T.}~\bibnamefont
  {Thirunamachandran}},\ }\href@noop {} {\emph {\bibinfo {title} {{Molecular}
  {Quantum} {Electrodynamics}: {An} {Introduction} to {Radiation}-{Molecule}
  {Interactions}}}}\ (\bibinfo  {publisher} {Academic Press, INC., London},\
  \bibinfo {year} {1984})\BibitemShut {NoStop}%
\bibitem [{\citenamefont {Barron}(2004)}]{Barron2004}%
  \BibitemOpen
  \bibfield  {author} {\bibinfo {author} {\bibfnamefont {L.~D.}\ \bibnamefont
  {Barron}},\ }\href {\doibase 10.1017/CBO9780511535468} {\emph {\bibinfo
  {title} {Molecular Light Scattering and Optical Activity}}},\ \bibinfo
  {edition} {2nd}\ ed.\ (\bibinfo  {publisher} {Cambridge University Press},\
  \bibinfo {year} {2004})\BibitemShut {NoStop}%
\bibitem [{\citenamefont {Tang}\ and\ \citenamefont {Cohen}(2010)}]{Tang2010}%
  \BibitemOpen
  \bibfield  {author} {\bibinfo {author} {\bibfnamefont {Y.}~\bibnamefont
  {Tang}}\ and\ \bibinfo {author} {\bibfnamefont {A.~E.}\ \bibnamefont
  {Cohen}},\ }\href {\doibase 10.1103/PhysRevLett.104.163901} {\bibfield
  {journal} {\bibinfo  {journal} {Phys. Rev. Lett.}\ }\textbf {\bibinfo
  {volume} {104}},\ \bibinfo {pages} {163901} (\bibinfo {year}
  {2010})}\BibitemShut {NoStop}%
\bibitem [{\citenamefont {V\'azquez-Lozano}\ and\ \citenamefont
  {Mart\'{\i}nez}(2018)}]{Lozano2018}%
  \BibitemOpen
  \bibfield  {author} {\bibinfo {author} {\bibfnamefont {J.~E.}\ \bibnamefont
  {V\'azquez-Lozano}}\ and\ \bibinfo {author} {\bibfnamefont {A.}~\bibnamefont
  {Mart\'{\i}nez}},\ }\href {\doibase 10.1103/PhysRevLett.121.043901}
  {\bibfield  {journal} {\bibinfo  {journal} {Phys. Rev. Lett.}\ }\textbf
  {\bibinfo {volume} {121}},\ \bibinfo {pages} {043901} (\bibinfo {year}
  {2018})}\BibitemShut {NoStop}%
\bibitem [{\citenamefont {Purdie}\ and\ \citenamefont
  {Brittain}(1994)}]{Purdie1989}%
  \BibitemOpen
  \bibfield  {author} {\bibinfo {author} {\bibfnamefont {N.}~\bibnamefont
  {Purdie}}\ and\ \bibinfo {author} {\bibfnamefont {H.}~\bibnamefont
  {Brittain}},\ }\href {https://books.google.fr/books?id=6BcYlAEACAAJ} {\emph
  {\bibinfo {title} {Analytical Applications of Circular Dichroism}}},\
  Developments in Food Science\ (\bibinfo  {publisher} {Elsevier},\ \bibinfo
  {year} {1994})\BibitemShut {NoStop}%
\bibitem [{\citenamefont {Snatzke}(1968)}]{snatzke_circular_1968}%
  \BibitemOpen
  \bibfield  {author} {\bibinfo {author} {\bibfnamefont {G.}~\bibnamefont
  {Snatzke}},\ }\href {\doibase 10.1002/anie.196800141} {\bibfield  {journal}
  {\bibinfo  {journal} {Angew. Chem., Int. Ed. Engl.}\ }\textbf {\bibinfo
  {volume} {7}},\ \bibinfo {pages} {14} (\bibinfo {year} {1968})}\BibitemShut
  {NoStop}%
\bibitem [{\citenamefont {Hecht}\ and\ \citenamefont
  {Barron}(1994)}]{Hecht1994}%
  \BibitemOpen
  \bibfield  {author} {\bibinfo {author} {\bibfnamefont {L.}~\bibnamefont
  {Hecht}}\ and\ \bibinfo {author} {\bibfnamefont {L.~D.}\ \bibnamefont
  {Barron}},\ }\href {\doibase https://doi.org/10.1016/0009-2614(94)87122-1}
  {\bibfield  {journal} {\bibinfo  {journal} {Chem. Phys. Lett.}\ }\textbf
  {\bibinfo {volume} {225}},\ \bibinfo {pages} {525} (\bibinfo {year}
  {1994})}\BibitemShut {NoStop}%
\bibitem [{\citenamefont {Mun}\ \emph {et~al.}(2020)\citenamefont {Mun},
  \citenamefont {Kim}, \citenamefont {Yang}, \citenamefont {Badloe},
  \citenamefont {Ni}, \citenamefont {Chen}, \citenamefont {Qiu},\ and\
  \citenamefont {Rho}}]{Mun2020}%
  \BibitemOpen
  \bibfield  {author} {\bibinfo {author} {\bibfnamefont {J.}~\bibnamefont
  {Mun}}, \bibinfo {author} {\bibfnamefont {M.}~\bibnamefont {Kim}}, \bibinfo
  {author} {\bibfnamefont {Y.}~\bibnamefont {Yang}}, \bibinfo {author}
  {\bibfnamefont {T.}~\bibnamefont {Badloe}}, \bibinfo {author} {\bibfnamefont
  {J.}~\bibnamefont {Ni}}, \bibinfo {author} {\bibfnamefont {Y.}~\bibnamefont
  {Chen}}, \bibinfo {author} {\bibfnamefont {C.-W.}\ \bibnamefont {Qiu}}, \
  and\ \bibinfo {author} {\bibfnamefont {J.}~\bibnamefont {Rho}},\ }\href
  {https://www.nature.com/articles/s41377-020-00367-8} {\bibfield  {journal}
  {\bibinfo  {journal} {Light Sci. App.}\ }\textbf {\bibinfo {volume} {9}}
  (\bibinfo {year} {2020})}\BibitemShut {NoStop}%
\bibitem [{\citenamefont {Condon}(1937)}]{Condon1937}%
  \BibitemOpen
  \bibfield  {author} {\bibinfo {author} {\bibfnamefont {E.~U.}\ \bibnamefont
  {Condon}},\ }\href {\doibase 10.1103/RevModPhys.9.432} {\bibfield  {journal}
  {\bibinfo  {journal} {Rev. Mod. Phys.}\ }\textbf {\bibinfo {volume} {9}},\
  \bibinfo {pages} {432} (\bibinfo {year} {1937})}\BibitemShut {NoStop}%
\bibitem [{\citenamefont {Post}(1962)}]{Post}%
  \BibitemOpen
  \bibfield  {author} {\bibinfo {author} {\bibfnamefont {E.~J.}\ \bibnamefont
  {Post}},\ }\href@noop {} {\emph {\bibinfo {title} {Formal Structure of
  Electromagnetics}}}\ (\bibinfo  {publisher} {North-Holland Publishing
  Company, Amsterdam},\ \bibinfo {year} {1962})\BibitemShut {NoStop}%
\bibitem [{\citenamefont {Jaggard}\ \emph {et~al.}(1979)\citenamefont
  {Jaggard}, \citenamefont {Mickelson},\ and\ \citenamefont
  {Papas}}]{Jaggard1979}%
  \BibitemOpen
  \bibfield  {author} {\bibinfo {author} {\bibfnamefont {D.~L.}\ \bibnamefont
  {Jaggard}}, \bibinfo {author} {\bibfnamefont {A.~R.}\ \bibnamefont
  {Mickelson}}, \ and\ \bibinfo {author} {\bibfnamefont {C.~J.}\ \bibnamefont
  {Papas}},\ }\href {\doibase 10.1007/BF00934418} {\bibfield  {journal}
  {\bibinfo  {journal} {Appl. Phys.}\ }\textbf {\bibinfo {volume} {18}},\
  \bibinfo {pages} {211} (\bibinfo {year} {1979})}\BibitemShut {NoStop}%
\bibitem [{\citenamefont {Lindell}\ \emph {et~al.}(1994)\citenamefont
  {Lindell}, \citenamefont {Sihvola}, \citenamefont {Viitanen},\ and\
  \citenamefont {Tretyakov}}]{Lindell1994}%
  \BibitemOpen
  \bibfield  {author} {\bibinfo {author} {\bibfnamefont {I.~V.}\ \bibnamefont
  {Lindell}}, \bibinfo {author} {\bibfnamefont {A.}~\bibnamefont {Sihvola}},
  \bibinfo {author} {\bibfnamefont {A.}~\bibnamefont {Viitanen}}, \ and\
  \bibinfo {author} {\bibfnamefont {S.}~\bibnamefont {Tretyakov}},\ }\href@noop
  {} {\emph {\bibinfo {title} {Electromagnetic waves in chiral and bi-isotropic
  media}}}\ (\bibinfo  {publisher} {Artech House Publishers},\ \bibinfo {year}
  {1994})\BibitemShut {NoStop}%
\bibitem [{\citenamefont {Bassiri}\ \emph {et~al.}(1990)\citenamefont
  {Bassiri}, \citenamefont {Papas},\ and\ \citenamefont
  {Engheta}}]{Bassiri1990}%
  \BibitemOpen
  \bibfield  {author} {\bibinfo {author} {\bibfnamefont {S.}~\bibnamefont
  {Bassiri}}, \bibinfo {author} {\bibfnamefont {C.~H.}\ \bibnamefont {Papas}},
  \ and\ \bibinfo {author} {\bibfnamefont {N.}~\bibnamefont {Engheta}},\ }\href
  {\doibase 10.1364/JOSAA.7.002154} {\bibfield  {journal} {\bibinfo  {journal}
  {J. Opt. Soc. Am. A}\ }\textbf {\bibinfo {volume} {7}},\ \bibinfo {pages}
  {2154} (\bibinfo {year} {1990})}\BibitemShut {NoStop}%
\bibitem [{\citenamefont {Silverman}\ and\ \citenamefont
  {Badoz}(1994)}]{Silverman1994}%
  \BibitemOpen
  \bibfield  {author} {\bibinfo {author} {\bibfnamefont {M.~P.}\ \bibnamefont
  {Silverman}}\ and\ \bibinfo {author} {\bibfnamefont {J.}~\bibnamefont
  {Badoz}},\ }\href {\doibase 10.1364/JOSAA.11.001894} {\bibfield  {journal}
  {\bibinfo  {journal} {J. Opt. Soc. Am. A}\ }\textbf {\bibinfo {volume}
  {11}},\ \bibinfo {pages} {1894} (\bibinfo {year} {1994})}\BibitemShut
  {NoStop}%
\bibitem [{\citenamefont {Silverman}(1986)}]{Silverman1986}%
  \BibitemOpen
  \bibfield  {author} {\bibinfo {author} {\bibfnamefont {M.~P.}\ \bibnamefont
  {Silverman}},\ }\href {\doibase 10.1364/JOSAA.3.000830} {\bibfield  {journal}
  {\bibinfo  {journal} {J. Opt. Soc. Am. A}\ }\textbf {\bibinfo {volume} {3}},\
  \bibinfo {pages} {830} (\bibinfo {year} {1986})}\BibitemShut {NoStop}%
\bibitem [{\citenamefont {Silverman}\ and\ \citenamefont
  {Badoz}(1990)}]{Silverman1990}%
  \BibitemOpen
  \bibfield  {author} {\bibinfo {author} {\bibfnamefont {M.~P.}\ \bibnamefont
  {Silverman}}\ and\ \bibinfo {author} {\bibfnamefont {J.}~\bibnamefont
  {Badoz}},\ }\href {\doibase 10.1364/JOSAA.7.001163} {\bibfield  {journal}
  {\bibinfo  {journal} {J. Opt. Soc. Am. A}\ }\textbf {\bibinfo {volume} {7}},\
  \bibinfo {pages} {1163} (\bibinfo {year} {1990})}\BibitemShut {NoStop}%
\bibitem [{\citenamefont {Jaggard}\ and\ \citenamefont
  {Sun}(1992)}]{Jaggard1992}%
  \BibitemOpen
  \bibfield  {author} {\bibinfo {author} {\bibfnamefont {D.~L.}\ \bibnamefont
  {Jaggard}}\ and\ \bibinfo {author} {\bibfnamefont {X.}~\bibnamefont {Sun}},\
  }\href {\doibase 10.1364/JOSAA.9.000804} {\bibfield  {journal} {\bibinfo
  {journal} {J. Opt. Soc. Am. A}\ }\textbf {\bibinfo {volume} {9}},\ \bibinfo
  {pages} {804} (\bibinfo {year} {1992})}\BibitemShut {NoStop}%
\bibitem [{\citenamefont {Andrews}(2018)}]{andrews2018quantum}%
  \BibitemOpen
  \bibfield  {author} {\bibinfo {author} {\bibfnamefont {D.~L.}\ \bibnamefont
  {Andrews}},\ }\href {\doibase 10.1088/2040-8986/aaaa56} {\bibfield  {journal}
  {\bibinfo  {journal} {J. Opt.}\ }\textbf {\bibinfo {volume} {20}},\ \bibinfo
  {pages} {033003} (\bibinfo {year} {2018})}\BibitemShut {NoStop}%
\bibitem [{\citenamefont {Rhee}\ \emph {et~al.}(2013)\citenamefont {Rhee},
  \citenamefont {Choi}, \citenamefont {Starling}, \citenamefont {Howell},\ and\
  \citenamefont {Cho}}]{Choi2013}%
  \BibitemOpen
  \bibfield  {author} {\bibinfo {author} {\bibfnamefont {H.}~\bibnamefont
  {Rhee}}, \bibinfo {author} {\bibfnamefont {J.~S.}\ \bibnamefont {Choi}},
  \bibinfo {author} {\bibfnamefont {D.~J.}\ \bibnamefont {Starling}}, \bibinfo
  {author} {\bibfnamefont {J.~C.}\ \bibnamefont {Howell}}, \ and\ \bibinfo
  {author} {\bibfnamefont {M.}~\bibnamefont {Cho}},\ }\href {\doibase
  10.1039/C3SC51255J} {\bibfield  {journal} {\bibinfo  {journal} {Chem. Sci.}\
  }\textbf {\bibinfo {volume} {4}},\ \bibinfo {pages} {4107} (\bibinfo {year}
  {2013})}\BibitemShut {NoStop}%
\bibitem [{\citenamefont {Yoo}\ and\ \citenamefont {Park}(2019)}]{Park2019}%
  \BibitemOpen
  \bibfield  {author} {\bibinfo {author} {\bibfnamefont {S.}~\bibnamefont
  {Yoo}}\ and\ \bibinfo {author} {\bibfnamefont {Q.-H.}\ \bibnamefont {Park}},\
  }\href {\doibase doi:10.1515/nanoph-2018-0167} {\bibfield  {journal}
  {\bibinfo  {journal} {Nanophotonics}\ }\textbf {\bibinfo {volume} {8}},\
  \bibinfo {pages} {249} (\bibinfo {year} {2019})}\BibitemShut {NoStop}%
\bibitem [{\citenamefont {Gao}\ \emph {et~al.}(2020)\citenamefont {Gao},
  \citenamefont {Wu}, \citenamefont {Lemaire}, \citenamefont {Carvalho},
  \citenamefont {Nlate}, \citenamefont {Buffeteau}, \citenamefont {Oda},
  \citenamefont {Battie}, \citenamefont {Pauly},\ and\ \citenamefont
  {Pouget}}]{Gao2020}%
  \BibitemOpen
  \bibfield  {author} {\bibinfo {author} {\bibfnamefont {J.}~\bibnamefont
  {Gao}}, \bibinfo {author} {\bibfnamefont {W.}~\bibnamefont {Wu}}, \bibinfo
  {author} {\bibfnamefont {V.}~\bibnamefont {Lemaire}}, \bibinfo {author}
  {\bibfnamefont {A.}~\bibnamefont {Carvalho}}, \bibinfo {author}
  {\bibfnamefont {S.}~\bibnamefont {Nlate}}, \bibinfo {author} {\bibfnamefont
  {T.}~\bibnamefont {Buffeteau}}, \bibinfo {author} {\bibfnamefont
  {R.}~\bibnamefont {Oda}}, \bibinfo {author} {\bibfnamefont {Y.}~\bibnamefont
  {Battie}}, \bibinfo {author} {\bibfnamefont {M.}~\bibnamefont {Pauly}}, \
  and\ \bibinfo {author} {\bibfnamefont {E.}~\bibnamefont {Pouget}},\ }\href
  {\doibase 10.1021/acsnano.9b08823} {\bibfield  {journal} {\bibinfo  {journal}
  {ACS Nano}\ }\textbf {\bibinfo {volume} {14}},\ \bibinfo {pages} {4111}
  (\bibinfo {year} {2020})}\BibitemShut {NoStop}%
\bibitem [{\citenamefont {Chen}\ \emph {et~al.}(2022)\citenamefont {Chen},
  \citenamefont {Chen}, \citenamefont {Kong}, \citenamefont {Wu}, \citenamefont
  {Chu},\ and\ \citenamefont {Qiu}}]{Chen2022}%
  \BibitemOpen
  \bibfield  {author} {\bibinfo {author} {\bibfnamefont {Y.}~\bibnamefont
  {Chen}}, \bibinfo {author} {\bibfnamefont {W.}~\bibnamefont {Chen}}, \bibinfo
  {author} {\bibfnamefont {X.}~\bibnamefont {Kong}}, \bibinfo {author}
  {\bibfnamefont {D.}~\bibnamefont {Wu}}, \bibinfo {author} {\bibfnamefont
  {J.}~\bibnamefont {Chu}}, \ and\ \bibinfo {author} {\bibfnamefont {C.-W.}\
  \bibnamefont {Qiu}},\ }\href {\doibase 10.1103/PhysRevLett.128.146102}
  {\bibfield  {journal} {\bibinfo  {journal} {Phys. Rev. Lett.}\ }\textbf
  {\bibinfo {volume} {128}},\ \bibinfo {pages} {146102} (\bibinfo {year}
  {2022})}\BibitemShut {NoStop}%
\bibitem [{\citenamefont {Kondratov}\ \emph {et~al.}(2016)\citenamefont
  {Kondratov}, \citenamefont {Gorkunov}, \citenamefont {Darinskii},
  \citenamefont {Gainutdinov}, \citenamefont {Rogov}, \citenamefont {Ezhov},\
  and\ \citenamefont {Artemov}}]{Kondraton2016}%
  \BibitemOpen
  \bibfield  {author} {\bibinfo {author} {\bibfnamefont {A.~V.}\ \bibnamefont
  {Kondratov}}, \bibinfo {author} {\bibfnamefont {M.~V.}\ \bibnamefont
  {Gorkunov}}, \bibinfo {author} {\bibfnamefont {A.~N.}\ \bibnamefont
  {Darinskii}}, \bibinfo {author} {\bibfnamefont {R.~V.}\ \bibnamefont
  {Gainutdinov}}, \bibinfo {author} {\bibfnamefont {O.~Y.}\ \bibnamefont
  {Rogov}}, \bibinfo {author} {\bibfnamefont {A.~A.}\ \bibnamefont {Ezhov}}, \
  and\ \bibinfo {author} {\bibfnamefont {V.~V.}\ \bibnamefont {Artemov}},\
  }\href {\doibase 10.1103/PhysRevB.93.195418} {\bibfield  {journal} {\bibinfo
  {journal} {Phys. Rev. B}\ }\textbf {\bibinfo {volume} {93}},\ \bibinfo
  {pages} {195418} (\bibinfo {year} {2016})}\BibitemShut {NoStop}%
\bibitem [{\citenamefont {Collins}\ \emph {et~al.}(2017)\citenamefont
  {Collins}, \citenamefont {Kuppe}, \citenamefont {Hooper}, \citenamefont
  {Sibilia}, \citenamefont {Centini},\ and\ \citenamefont
  {Valev}}]{Collins2017}%
  \BibitemOpen
  \bibfield  {author} {\bibinfo {author} {\bibfnamefont {J.~T.}\ \bibnamefont
  {Collins}}, \bibinfo {author} {\bibfnamefont {C.}~\bibnamefont {Kuppe}},
  \bibinfo {author} {\bibfnamefont {D.~C.}\ \bibnamefont {Hooper}}, \bibinfo
  {author} {\bibfnamefont {C.}~\bibnamefont {Sibilia}}, \bibinfo {author}
  {\bibfnamefont {M.}~\bibnamefont {Centini}}, \ and\ \bibinfo {author}
  {\bibfnamefont {V.~K.}\ \bibnamefont {Valev}},\ }\href {\doibase
  https://doi.org/10.1002/adom.201700182} {\bibfield  {journal} {\bibinfo
  {journal} {Adv. Opt. Mater.}\ }\textbf {\bibinfo {volume} {5}},\ \bibinfo
  {pages} {1700182} (\bibinfo {year} {2017})}\BibitemShut {NoStop}%
\bibitem [{\citenamefont {Wu}\ \emph {et~al.}(2018)\citenamefont {Wu},
  \citenamefont {Chen}, \citenamefont {Wang}, \citenamefont {Dong},\ and\
  \citenamefont {Zheng}}]{Wu2018}%
  \BibitemOpen
  \bibfield  {author} {\bibinfo {author} {\bibfnamefont {Z.}~\bibnamefont
  {Wu}}, \bibinfo {author} {\bibfnamefont {X.}~\bibnamefont {Chen}}, \bibinfo
  {author} {\bibfnamefont {M.}~\bibnamefont {Wang}}, \bibinfo {author}
  {\bibfnamefont {J.}~\bibnamefont {Dong}}, \ and\ \bibinfo {author}
  {\bibfnamefont {Y.}~\bibnamefont {Zheng}},\ }\href {\doibase
  10.1021/acsnano.8b02566} {\bibfield  {journal} {\bibinfo  {journal} {ACS
  Nano}\ }\textbf {\bibinfo {volume} {12}},\ \bibinfo {pages} {5030} (\bibinfo
  {year} {2018})}\BibitemShut {NoStop}%
\bibitem [{\citenamefont {Mohammadi}\ \emph {et~al.}(2018)\citenamefont
  {Mohammadi}, \citenamefont {Tsakmakidis}, \citenamefont {Askarpour},
  \citenamefont {Dehkhoda}, \citenamefont {Tavakoli},\ and\ \citenamefont
  {Altug}}]{Mohammadi2018}%
  \BibitemOpen
  \bibfield  {author} {\bibinfo {author} {\bibfnamefont {E.}~\bibnamefont
  {Mohammadi}}, \bibinfo {author} {\bibfnamefont {K.~L.}\ \bibnamefont
  {Tsakmakidis}}, \bibinfo {author} {\bibfnamefont {A.~N.}\ \bibnamefont
  {Askarpour}}, \bibinfo {author} {\bibfnamefont {P.}~\bibnamefont {Dehkhoda}},
  \bibinfo {author} {\bibfnamefont {A.}~\bibnamefont {Tavakoli}}, \ and\
  \bibinfo {author} {\bibfnamefont {H.}~\bibnamefont {Altug}},\ }\href
  {\doibase 10.1021/acsphotonics.8b00270} {\bibfield  {journal} {\bibinfo
  {journal} {ACS Photonics}\ }\textbf {\bibinfo {volume} {5}},\ \bibinfo
  {pages} {2669} (\bibinfo {year} {2018})}\BibitemShut {NoStop}%
\bibitem [{\citenamefont {Lieberman}\ \emph {et~al.}(2008)\citenamefont
  {Lieberman}, \citenamefont {Shemer}, \citenamefont {Fried}, \citenamefont
  {Kosower},\ and\ \citenamefont {Markovich}}]{Lieberman2008}%
  \BibitemOpen
  \bibfield  {author} {\bibinfo {author} {\bibfnamefont {I.}~\bibnamefont
  {Lieberman}}, \bibinfo {author} {\bibfnamefont {G.}~\bibnamefont {Shemer}},
  \bibinfo {author} {\bibfnamefont {T.}~\bibnamefont {Fried}}, \bibinfo
  {author} {\bibfnamefont {E.}~\bibnamefont {Kosower}}, \ and\ \bibinfo
  {author} {\bibfnamefont {G.}~\bibnamefont {Markovich}},\ }\href {\doibase
  https://doi.org/10.1002/anie.200800231} {\bibfield  {journal} {\bibinfo
  {journal} {Angew. Chem. Int. Ed.}\ }\textbf {\bibinfo {volume} {47}},\
  \bibinfo {pages} {4855} (\bibinfo {year} {2008})}\BibitemShut {NoStop}%
\bibitem [{\citenamefont {Zhang}\ and\ \citenamefont
  {Govorov}(2013)}]{Govorov2013}%
  \BibitemOpen
  \bibfield  {author} {\bibinfo {author} {\bibfnamefont {H.}~\bibnamefont
  {Zhang}}\ and\ \bibinfo {author} {\bibfnamefont {A.~O.}\ \bibnamefont
  {Govorov}},\ }\href {\doibase 10.1103/PhysRevB.87.075410} {\bibfield
  {journal} {\bibinfo  {journal} {Phys. Rev. B}\ }\textbf {\bibinfo {volume}
  {87}},\ \bibinfo {pages} {075410} (\bibinfo {year} {2013})}\BibitemShut
  {NoStop}%
\bibitem [{\citenamefont {Nesterov}\ \emph {et~al.}(2016)\citenamefont
  {Nesterov}, \citenamefont {Yin}, \citenamefont {Schäferling}, \citenamefont
  {Giessen},\ and\ \citenamefont {Weiss}}]{Weiss2016}%
  \BibitemOpen
  \bibfield  {author} {\bibinfo {author} {\bibfnamefont {M.~L.}\ \bibnamefont
  {Nesterov}}, \bibinfo {author} {\bibfnamefont {X.}~\bibnamefont {Yin}},
  \bibinfo {author} {\bibfnamefont {M.}~\bibnamefont {Schäferling}}, \bibinfo
  {author} {\bibfnamefont {H.}~\bibnamefont {Giessen}}, \ and\ \bibinfo
  {author} {\bibfnamefont {T.}~\bibnamefont {Weiss}},\ }\href {\doibase
  10.1021/acsphotonics.5b00637} {\bibfield  {journal} {\bibinfo  {journal} {ACS
  Photonics}\ }\textbf {\bibinfo {volume} {3}},\ \bibinfo {pages} {578}
  (\bibinfo {year} {2016})}\BibitemShut {NoStop}%
\bibitem [{\citenamefont {Wang}\ and\ \citenamefont
  {Xiao}(2021)}]{chiralthiophene}%
  \BibitemOpen
  \bibfield  {author} {\bibinfo {author} {\bibfnamefont {K.}~\bibnamefont
  {Wang}}\ and\ \bibinfo {author} {\bibfnamefont {Y.}~\bibnamefont {Xiao}},\
  }\href {\doibase https://doi.org/10.1002/chir.23333} {\bibfield  {journal}
  {\bibinfo  {journal} {Chirality}\ }\textbf {\bibinfo {volume} {33}},\
  \bibinfo {pages} {424} (\bibinfo {year} {2021})}\BibitemShut {NoStop}%
\bibitem [{\citenamefont {Vestler}\ \emph {et~al.}(2019)\citenamefont
  {Vestler}, \citenamefont {Ben-Moshe},\ and\ \citenamefont
  {Markovich}}]{Vestler2019}%
  \BibitemOpen
  \bibfield  {author} {\bibinfo {author} {\bibfnamefont {D.}~\bibnamefont
  {Vestler}}, \bibinfo {author} {\bibfnamefont {A.}~\bibnamefont {Ben-Moshe}},
  \ and\ \bibinfo {author} {\bibfnamefont {G.}~\bibnamefont {Markovich}},\
  }\href {\doibase 10.1021/acs.jpcc.8b10975} {\bibfield  {journal} {\bibinfo
  {journal} {J. Phys. Chem. C}\ }\textbf {\bibinfo {volume} {123}},\ \bibinfo
  {pages} {5017} (\bibinfo {year} {2019})}\BibitemShut {NoStop}%
\bibitem [{\citenamefont {Kneer}\ \emph {et~al.}(2018)\citenamefont {Kneer},
  \citenamefont {Roller}, \citenamefont {Besteiro}, \citenamefont {Schreiber},
  \citenamefont {Govorov},\ and\ \citenamefont {Liedl}}]{Kneer2018}%
  \BibitemOpen
  \bibfield  {author} {\bibinfo {author} {\bibfnamefont {L.~M.}\ \bibnamefont
  {Kneer}}, \bibinfo {author} {\bibfnamefont {E.-M.}\ \bibnamefont {Roller}},
  \bibinfo {author} {\bibfnamefont {L.~V.}\ \bibnamefont {Besteiro}}, \bibinfo
  {author} {\bibfnamefont {R.}~\bibnamefont {Schreiber}}, \bibinfo {author}
  {\bibfnamefont {A.~O.}\ \bibnamefont {Govorov}}, \ and\ \bibinfo {author}
  {\bibfnamefont {T.}~\bibnamefont {Liedl}},\ }\href {\doibase
  10.1021/acsnano.8b03146} {\bibfield  {journal} {\bibinfo  {journal} {ACS
  Nano}\ }\textbf {\bibinfo {volume} {12}},\ \bibinfo {pages} {9110} (\bibinfo
  {year} {2018})}\BibitemShut {NoStop}%
\bibitem [{\citenamefont {Graf}\ \emph {et~al.}(2019)\citenamefont {Graf},
  \citenamefont {Feis}, \citenamefont {Garcia-Santiago}, \citenamefont
  {Wegener}, \citenamefont {Rockstuhl},\ and\ \citenamefont
  {Fernandez-Corbaton}}]{Corbaton2019}%
  \BibitemOpen
  \bibfield  {author} {\bibinfo {author} {\bibfnamefont {F.}~\bibnamefont
  {Graf}}, \bibinfo {author} {\bibfnamefont {J.}~\bibnamefont {Feis}}, \bibinfo
  {author} {\bibfnamefont {X.}~\bibnamefont {Garcia-Santiago}}, \bibinfo
  {author} {\bibfnamefont {M.}~\bibnamefont {Wegener}}, \bibinfo {author}
  {\bibfnamefont {C.}~\bibnamefont {Rockstuhl}}, \ and\ \bibinfo {author}
  {\bibfnamefont {I.}~\bibnamefont {Fernandez-Corbaton}},\ }\href {\doibase
  10.1021/acsphotonics.8b01454} {\bibfield  {journal} {\bibinfo  {journal} {ACS
  Photonics}\ }\textbf {\bibinfo {volume} {6}},\ \bibinfo {pages} {482}
  (\bibinfo {year} {2019})}\BibitemShut {NoStop}%
\bibitem [{\citenamefont {Menzel}\ \emph {et~al.}(2010)\citenamefont {Menzel},
  \citenamefont {Rockstuhl},\ and\ \citenamefont
  {Lederer}}]{menzel_advanced_2010}%
  \BibitemOpen
  \bibfield  {author} {\bibinfo {author} {\bibfnamefont {C.}~\bibnamefont
  {Menzel}}, \bibinfo {author} {\bibfnamefont {C.}~\bibnamefont {Rockstuhl}}, \
  and\ \bibinfo {author} {\bibfnamefont {F.}~\bibnamefont {Lederer}},\ }\href
  {\doibase 10.1103/PhysRevA.82.053811} {\bibfield  {journal} {\bibinfo
  {journal} {Phys. Rev. A}\ }\textbf {\bibinfo {volume} {82}},\ \bibinfo
  {pages} {053811} (\bibinfo {year} {2010})}\BibitemShut {NoStop}%
\bibitem [{\citenamefont {Kuwata-Gonokami}\ \emph {et~al.}(2005)\citenamefont
  {Kuwata-Gonokami}, \citenamefont {Saito}, \citenamefont {Ino}, \citenamefont
  {Kauranen}, \citenamefont {Jefimovs}, \citenamefont {Vallius}, \citenamefont
  {Turunen},\ and\ \citenamefont {Svirko}}]{Kuwata2005}%
  \BibitemOpen
  \bibfield  {author} {\bibinfo {author} {\bibfnamefont {M.}~\bibnamefont
  {Kuwata-Gonokami}}, \bibinfo {author} {\bibfnamefont {N.}~\bibnamefont
  {Saito}}, \bibinfo {author} {\bibfnamefont {Y.}~\bibnamefont {Ino}}, \bibinfo
  {author} {\bibfnamefont {M.}~\bibnamefont {Kauranen}}, \bibinfo {author}
  {\bibfnamefont {K.}~\bibnamefont {Jefimovs}}, \bibinfo {author}
  {\bibfnamefont {T.}~\bibnamefont {Vallius}}, \bibinfo {author} {\bibfnamefont
  {J.}~\bibnamefont {Turunen}}, \ and\ \bibinfo {author} {\bibfnamefont
  {Y.}~\bibnamefont {Svirko}},\ }\href {\doibase 10.1103/PhysRevLett.95.227401}
  {\bibfield  {journal} {\bibinfo  {journal} {Phys. Rev. Lett.}\ }\textbf
  {\bibinfo {volume} {95}},\ \bibinfo {pages} {227401} (\bibinfo {year}
  {2005})}\BibitemShut {NoStop}%
\bibitem [{\citenamefont {Solomon}\ \emph {et~al.}(2019)\citenamefont
  {Solomon}, \citenamefont {Hu}, \citenamefont {Lawrence}, \citenamefont
  {García-Etxarri},\ and\ \citenamefont
  {Dionne}}]{solomon_enantiospecific_2019}%
  \BibitemOpen
  \bibfield  {author} {\bibinfo {author} {\bibfnamefont {M.~L.}\ \bibnamefont
  {Solomon}}, \bibinfo {author} {\bibfnamefont {J.}~\bibnamefont {Hu}},
  \bibinfo {author} {\bibfnamefont {M.}~\bibnamefont {Lawrence}}, \bibinfo
  {author} {\bibfnamefont {A.}~\bibnamefont {García-Etxarri}}, \ and\ \bibinfo
  {author} {\bibfnamefont {J.~A.}\ \bibnamefont {Dionne}},\ }\href {\doibase
  10.1021/acsphotonics.8b01365} {\bibfield  {journal} {\bibinfo  {journal} {ACS
  Photonics}\ }\textbf {\bibinfo {volume} {6}},\ \bibinfo {pages} {43}
  (\bibinfo {year} {2019})}\BibitemShut {NoStop}%
\bibitem [{\citenamefont {Li}\ \emph {et~al.}(2013)\citenamefont {Li},
  \citenamefont {Mutlu},\ and\ \citenamefont {Ozbay}}]{Li2013}%
  \BibitemOpen
  \bibfield  {author} {\bibinfo {author} {\bibfnamefont {Z.}~\bibnamefont
  {Li}}, \bibinfo {author} {\bibfnamefont {M.}~\bibnamefont {Mutlu}}, \ and\
  \bibinfo {author} {\bibfnamefont {E.}~\bibnamefont {Ozbay}},\ }\href
  {\doibase 10.1088/2040-8978/15/2/023001} {\bibfield  {journal} {\bibinfo
  {journal} {J. Opt.}\ }\textbf {\bibinfo {volume} {15}},\ \bibinfo {pages}
  {023001} (\bibinfo {year} {2013})}\BibitemShut {NoStop}%
\bibitem [{\citenamefont {Oh}\ and\ \citenamefont {Hess}(2015)}]{Oh2015}%
  \BibitemOpen
  \bibfield  {author} {\bibinfo {author} {\bibfnamefont {S.~S.}\ \bibnamefont
  {Oh}}\ and\ \bibinfo {author} {\bibfnamefont {O.}~\bibnamefont {Hess}},\
  }\href {\doibase 10.1186/s40580-015-0058-2} {\bibfield  {journal} {\bibinfo
  {journal} {Nano Converg.}\ }\textbf {\bibinfo {volume} {2}},\ \bibinfo
  {pages} {24} (\bibinfo {year} {2015})}\BibitemShut {NoStop}%
\bibitem [{\citenamefont {Papakostas}\ \emph {et~al.}(2003)\citenamefont
  {Papakostas}, \citenamefont {Potts}, \citenamefont {Bagnall}, \citenamefont
  {Prosvirnin}, \citenamefont {Coles},\ and\ \citenamefont
  {Zheludev}}]{Papakostas2003}%
  \BibitemOpen
  \bibfield  {author} {\bibinfo {author} {\bibfnamefont {A.}~\bibnamefont
  {Papakostas}}, \bibinfo {author} {\bibfnamefont {A.}~\bibnamefont {Potts}},
  \bibinfo {author} {\bibfnamefont {D.~M.}\ \bibnamefont {Bagnall}}, \bibinfo
  {author} {\bibfnamefont {S.~L.}\ \bibnamefont {Prosvirnin}}, \bibinfo
  {author} {\bibfnamefont {H.~J.}\ \bibnamefont {Coles}}, \ and\ \bibinfo
  {author} {\bibfnamefont {N.~I.}\ \bibnamefont {Zheludev}},\ }\href {\doibase
  10.1103/PhysRevLett.90.107404} {\bibfield  {journal} {\bibinfo  {journal}
  {Phys. Rev. Lett.}\ }\textbf {\bibinfo {volume} {90}},\ \bibinfo {pages}
  {107404} (\bibinfo {year} {2003})}\BibitemShut {NoStop}%
\bibitem [{\citenamefont {Zhang}\ \emph {et~al.}(2009)\citenamefont {Zhang},
  \citenamefont {Park}, \citenamefont {Li}, \citenamefont {Lu}, \citenamefont
  {Zhang},\ and\ \citenamefont {Zhang}}]{Zhang2009}%
  \BibitemOpen
  \bibfield  {author} {\bibinfo {author} {\bibfnamefont {S.}~\bibnamefont
  {Zhang}}, \bibinfo {author} {\bibfnamefont {Y.-S.}\ \bibnamefont {Park}},
  \bibinfo {author} {\bibfnamefont {J.}~\bibnamefont {Li}}, \bibinfo {author}
  {\bibfnamefont {X.}~\bibnamefont {Lu}}, \bibinfo {author} {\bibfnamefont
  {W.}~\bibnamefont {Zhang}}, \ and\ \bibinfo {author} {\bibfnamefont
  {X.}~\bibnamefont {Zhang}},\ }\href {\doibase 10.1103/PhysRevLett.102.023901}
  {\bibfield  {journal} {\bibinfo  {journal} {Phys. Rev. Lett.}\ }\textbf
  {\bibinfo {volume} {102}},\ \bibinfo {pages} {023901} (\bibinfo {year}
  {2009})}\BibitemShut {NoStop}%
\bibitem [{\citenamefont {Decker}\ \emph {et~al.}(2009)\citenamefont {Decker},
  \citenamefont {Ruther}, \citenamefont {Kriegler}, \citenamefont {Zhou},
  \citenamefont {Soukoulis}, \citenamefont {Linden},\ and\ \citenamefont
  {Wegener}}]{Decker2009}%
  \BibitemOpen
  \bibfield  {author} {\bibinfo {author} {\bibfnamefont {M.}~\bibnamefont
  {Decker}}, \bibinfo {author} {\bibfnamefont {M.}~\bibnamefont {Ruther}},
  \bibinfo {author} {\bibfnamefont {C.~E.}\ \bibnamefont {Kriegler}}, \bibinfo
  {author} {\bibfnamefont {J.}~\bibnamefont {Zhou}}, \bibinfo {author}
  {\bibfnamefont {C.~M.}\ \bibnamefont {Soukoulis}}, \bibinfo {author}
  {\bibfnamefont {S.}~\bibnamefont {Linden}}, \ and\ \bibinfo {author}
  {\bibfnamefont {M.}~\bibnamefont {Wegener}},\ }\href {\doibase
  10.1364/OL.34.002501} {\bibfield  {journal} {\bibinfo  {journal} {Opt.
  Lett.}\ }\textbf {\bibinfo {volume} {34}},\ \bibinfo {pages} {2501} (\bibinfo
  {year} {2009})}\BibitemShut {NoStop}%
\bibitem [{\citenamefont {Ni}\ \emph {et~al.}(2021)\citenamefont {Ni},
  \citenamefont {Liu}, \citenamefont {Hu}, \citenamefont {Hu}, \citenamefont
  {Lao}, \citenamefont {Li}, \citenamefont {Zhang}, \citenamefont {Wu},
  \citenamefont {Dong}, \citenamefont {Chu},\ and\ \citenamefont
  {Qiu}}]{Ni2021}%
  \BibitemOpen
  \bibfield  {author} {\bibinfo {author} {\bibfnamefont {J.}~\bibnamefont
  {Ni}}, \bibinfo {author} {\bibfnamefont {S.}~\bibnamefont {Liu}}, \bibinfo
  {author} {\bibfnamefont {G.}~\bibnamefont {Hu}}, \bibinfo {author}
  {\bibfnamefont {Y.}~\bibnamefont {Hu}}, \bibinfo {author} {\bibfnamefont
  {Z.}~\bibnamefont {Lao}}, \bibinfo {author} {\bibfnamefont {J.}~\bibnamefont
  {Li}}, \bibinfo {author} {\bibfnamefont {Q.}~\bibnamefont {Zhang}}, \bibinfo
  {author} {\bibfnamefont {D.}~\bibnamefont {Wu}}, \bibinfo {author}
  {\bibfnamefont {S.}~\bibnamefont {Dong}}, \bibinfo {author} {\bibfnamefont
  {J.}~\bibnamefont {Chu}}, \ and\ \bibinfo {author} {\bibfnamefont {C.-W.}\
  \bibnamefont {Qiu}},\ }\href {\doibase 10.1021/acsnano.0c08941} {\bibfield
  {journal} {\bibinfo  {journal} {ACS Nano}\ }\textbf {\bibinfo {volume}
  {15}},\ \bibinfo {pages} {2893} (\bibinfo {year} {2021})}\BibitemShut
  {NoStop}%
\bibitem [{\citenamefont {Carminati}\ \emph {et~al.}(1998)\citenamefont
  {Carminati}, \citenamefont {Nieto-Vesperinas},\ and\ \citenamefont
  {Greffet}}]{Carminati:98}%
  \BibitemOpen
  \bibfield  {author} {\bibinfo {author} {\bibfnamefont {R.}~\bibnamefont
  {Carminati}}, \bibinfo {author} {\bibfnamefont {M.}~\bibnamefont
  {Nieto-Vesperinas}}, \ and\ \bibinfo {author} {\bibfnamefont {J.-J.}\
  \bibnamefont {Greffet}},\ }\href {\doibase 10.1364/JOSAA.15.000706}
  {\bibfield  {journal} {\bibinfo  {journal} {J. Opt. Soc. Am. A}\ }\textbf
  {\bibinfo {volume} {15}},\ \bibinfo {pages} {706} (\bibinfo {year}
  {1998})}\BibitemShut {NoStop}%
\bibitem [{\citenamefont {Carminati}\ \emph {et~al.}(2000)\citenamefont
  {Carminati}, \citenamefont {S\'aenz}, \citenamefont {Greffet},\ and\
  \citenamefont {Nieto-Vesperinas}}]{carminati2000reciprocity}%
  \BibitemOpen
  \bibfield  {author} {\bibinfo {author} {\bibfnamefont {R.}~\bibnamefont
  {Carminati}}, \bibinfo {author} {\bibfnamefont {J.~J.}\ \bibnamefont
  {S\'aenz}}, \bibinfo {author} {\bibfnamefont {J.-J.}\ \bibnamefont
  {Greffet}}, \ and\ \bibinfo {author} {\bibfnamefont {M.}~\bibnamefont
  {Nieto-Vesperinas}},\ }\href
  {https://link.aps.org/doi/10.1103/PhysRevA.62.012712} {\bibfield  {journal}
  {\bibinfo  {journal} {Phys. Rev. A}\ }\textbf {\bibinfo {volume} {62}},\
  \bibinfo {pages} {012712} (\bibinfo {year} {2000})}\BibitemShut {NoStop}%
\bibitem [{\citenamefont {Sigwarth}\ and\ \citenamefont
  {Miniatura}(2022)}]{sigwarth_time_2022}%
  \BibitemOpen
  \bibfield  {author} {\bibinfo {author} {\bibfnamefont {O.}~\bibnamefont
  {Sigwarth}}\ and\ \bibinfo {author} {\bibfnamefont {C.}~\bibnamefont
  {Miniatura}},\ }\href {\doibase 10.1007/s43673-022-00053-4} {\bibfield
  {journal} {\bibinfo  {journal} {AAPPS bull.}\ }\textbf {\bibinfo {volume}
  {32}},\ \bibinfo {pages} {23} (\bibinfo {year} {2022})}\BibitemShut {NoStop}%
\bibitem [{\citenamefont {Drezet}\ and\ \citenamefont
  {Genet}(2015)}]{drezet_reciprocity_2017}%
  \BibitemOpen
  \bibfield  {author} {\bibinfo {author} {\bibfnamefont {A.}~\bibnamefont
  {Drezet}}\ and\ \bibinfo {author} {\bibfnamefont {C.}~\bibnamefont {Genet}},\
  }\href
  {https://www.jennystanford.com/9789814613187/singular-and-chiral-nanoplasmonics/}
  {\emph {\bibinfo {title} {Singular and Chiral Nano Plasmonics}}}\ (\bibinfo
  {publisher} {PanStanford Publishing},\ \bibinfo {year} {2015})\BibitemShut
  {NoStop}%
\bibitem [{\citenamefont {Born}\ and\ \citenamefont {Wolf}(1980)}]{Born1980}%
  \BibitemOpen
  \bibfield  {author} {\bibinfo {author} {\bibfnamefont {M.}~\bibnamefont
  {Born}}\ and\ \bibinfo {author} {\bibfnamefont {E.}~\bibnamefont {Wolf}},\
  }\href@noop {} {\emph {\bibinfo {title} {Principles of Optics}}}\ (\bibinfo
  {publisher} {Pergamon, New York},\ \bibinfo {year} {1980})\BibitemShut
  {NoStop}%
\bibitem [{\citenamefont {Ebbesen}(2016)}]{ebbesen2016hybrid}%
  \BibitemOpen
  \bibfield  {author} {\bibinfo {author} {\bibfnamefont {T.~W.}\ \bibnamefont
  {Ebbesen}},\ }\href {\doibase 10.1021/acs.accounts.6b00295} {\bibfield
  {journal} {\bibinfo  {journal} {Acc. Chem. Res.}\ }\textbf {\bibinfo {volume}
  {49}},\ \bibinfo {pages} {2403} (\bibinfo {year} {2016})}\BibitemShut
  {NoStop}%
\bibitem [{\citenamefont {Schwartz}\ \emph {et~al.}(2011)\citenamefont
  {Schwartz}, \citenamefont {Hutchison}, \citenamefont {Genet},\ and\
  \citenamefont {Ebbesen}}]{schwartz2011reversible}%
  \BibitemOpen
  \bibfield  {author} {\bibinfo {author} {\bibfnamefont {T.}~\bibnamefont
  {Schwartz}}, \bibinfo {author} {\bibfnamefont {J.~A.}\ \bibnamefont
  {Hutchison}}, \bibinfo {author} {\bibfnamefont {C.}~\bibnamefont {Genet}}, \
  and\ \bibinfo {author} {\bibfnamefont {T.~W.}\ \bibnamefont {Ebbesen}},\
  }\href {\doibase 10.1103/PhysRevLett.106.196405} {\bibfield  {journal}
  {\bibinfo  {journal} {Phys. Rev. Lett.}\ }\textbf {\bibinfo {volume} {106}},\
  \bibinfo {pages} {196405} (\bibinfo {year} {2011})}\BibitemShut {NoStop}%
\bibitem [{\citenamefont {Jackson}(1999)}]{jackson1999classical}%
  \BibitemOpen
  \bibfield  {author} {\bibinfo {author} {\bibfnamefont {J.~D.}\ \bibnamefont
  {Jackson}},\ }\href@noop {} {\emph {\bibinfo {title} {Classical
  electrodynamics}}}\ (\bibinfo  {publisher} {Wiley, California},\ \bibinfo
  {year} {1999})\BibitemShut {NoStop}%
\bibitem [{\citenamefont {Voronin}\ \emph {et~al.}(2022)\citenamefont
  {Voronin}, \citenamefont {Taradin}, \citenamefont {Gorkunov},\ and\
  \citenamefont {Baranov}}]{voronin2021single}%
  \BibitemOpen
  \bibfield  {author} {\bibinfo {author} {\bibfnamefont {K.}~\bibnamefont
  {Voronin}}, \bibinfo {author} {\bibfnamefont {A.~S.}\ \bibnamefont
  {Taradin}}, \bibinfo {author} {\bibfnamefont {M.~V.}\ \bibnamefont
  {Gorkunov}}, \ and\ \bibinfo {author} {\bibfnamefont {D.~G.}\ \bibnamefont
  {Baranov}},\ }\href {\doibase 10.1021/acsphotonics.2c00134} {\bibfield
  {journal} {\bibinfo  {journal} {ACS Photonics}\ }\textbf {\bibinfo {volume}
  {9}},\ \bibinfo {pages} {2652} (\bibinfo {year} {2022})}\BibitemShut
  {NoStop}%
\bibitem [{\citenamefont {Feis}\ \emph {et~al.}(2020)\citenamefont {Feis},
  \citenamefont {Beutel}, \citenamefont {K\"opfler}, \citenamefont
  {Garcia-Santiago}, \citenamefont {Rockstuhl}, \citenamefont {Wegener},\ and\
  \citenamefont {Fernandez-Corbaton}}]{Corbaton2020-2}%
  \BibitemOpen
  \bibfield  {author} {\bibinfo {author} {\bibfnamefont {J.}~\bibnamefont
  {Feis}}, \bibinfo {author} {\bibfnamefont {D.}~\bibnamefont {Beutel}},
  \bibinfo {author} {\bibfnamefont {J.}~\bibnamefont {K\"opfler}}, \bibinfo
  {author} {\bibfnamefont {X.}~\bibnamefont {Garcia-Santiago}}, \bibinfo
  {author} {\bibfnamefont {C.}~\bibnamefont {Rockstuhl}}, \bibinfo {author}
  {\bibfnamefont {M.}~\bibnamefont {Wegener}}, \ and\ \bibinfo {author}
  {\bibfnamefont {I.}~\bibnamefont {Fernandez-Corbaton}},\ }\href {\doibase
  10.1103/PhysRevLett.124.033201} {\bibfield  {journal} {\bibinfo  {journal}
  {Phys. Rev. Lett.}\ }\textbf {\bibinfo {volume} {124}},\ \bibinfo {pages}
  {033201} (\bibinfo {year} {2020})}\BibitemShut {NoStop}%
\bibitem [{\citenamefont {Sun}\ \emph {et~al.}(2022)\citenamefont {Sun},
  \citenamefont {Gu},\ and\ \citenamefont {Mukamel}}]{Sun2022}%
  \BibitemOpen
  \bibfield  {author} {\bibinfo {author} {\bibfnamefont {S.}~\bibnamefont
  {Sun}}, \bibinfo {author} {\bibfnamefont {B.}~\bibnamefont {Gu}}, \ and\
  \bibinfo {author} {\bibfnamefont {S.}~\bibnamefont {Mukamel}},\ }\href
  {\doibase 10.1039/D1SC04341B} {\bibfield  {journal} {\bibinfo  {journal}
  {Chem. Sci.}\ }\textbf {\bibinfo {volume} {13}},\ \bibinfo {pages} {1037}
  (\bibinfo {year} {2022})}\BibitemShut {NoStop}%
\bibitem [{\citenamefont {Scott}\ \emph {et~al.}(2020)\citenamefont {Scott},
  \citenamefont {Garcia-Santiago}, \citenamefont {Beutel}, \citenamefont
  {Rockstuhl}, \citenamefont {Wegener},\ and\ \citenamefont
  {Fernandez-Corbaton}}]{Corbaton2020}%
  \BibitemOpen
  \bibfield  {author} {\bibinfo {author} {\bibfnamefont {P.}~\bibnamefont
  {Scott}}, \bibinfo {author} {\bibfnamefont {X.}~\bibnamefont
  {Garcia-Santiago}}, \bibinfo {author} {\bibfnamefont {D.}~\bibnamefont
  {Beutel}}, \bibinfo {author} {\bibfnamefont {C.}~\bibnamefont {Rockstuhl}},
  \bibinfo {author} {\bibfnamefont {M.}~\bibnamefont {Wegener}}, \ and\
  \bibinfo {author} {\bibfnamefont {I.}~\bibnamefont {Fernandez-Corbaton}},\
  }\href {\doibase 10.1063/5.0025006} {\bibfield  {journal} {\bibinfo
  {journal} {Appl. Phys. Rev.}\ }\textbf {\bibinfo {volume} {7}},\ \bibinfo
  {pages} {041413} (\bibinfo {year} {2020})}\BibitemShut {NoStop}%
\bibitem [{\citenamefont {Yoo}\ and\ \citenamefont {Park}(2015)}]{Yoo2015}%
  \BibitemOpen
  \bibfield  {author} {\bibinfo {author} {\bibfnamefont {S.~J.}\ \bibnamefont
  {Yoo}}\ and\ \bibinfo {author} {\bibfnamefont {Q.~H.}\ \bibnamefont {Park}},\
  }\href {\doibase 10.1103/PhysRevLett.114.203003} {\bibfield  {journal}
  {\bibinfo  {journal} {Phys. Rev. Lett.}\ }\textbf {\bibinfo {volume} {114}},\
  \bibinfo {pages} {203003} (\bibinfo {year} {2015})}\BibitemShut {NoStop}%
\bibitem [{\citenamefont {Semnani}\ \emph {et~al.}(2020)\citenamefont
  {Semnani}, \citenamefont {Flannery}, \citenamefont {Al~Maruf},\ and\
  \citenamefont {Bajcsy}}]{Maruf2020}%
  \BibitemOpen
  \bibfield  {author} {\bibinfo {author} {\bibfnamefont {B.}~\bibnamefont
  {Semnani}}, \bibinfo {author} {\bibfnamefont {J.}~\bibnamefont {Flannery}},
  \bibinfo {author} {\bibfnamefont {R.}~\bibnamefont {Al~Maruf}}, \ and\
  \bibinfo {author} {\bibfnamefont {M.}~\bibnamefont {Bajcsy}},\ }\href
  {https://www.nature.com/articles/s41377-020-0256-5} {\bibfield  {journal}
  {\bibinfo  {journal} {Light Sci. App.}\ }\textbf {\bibinfo {volume} {9}}
  (\bibinfo {year} {2020})}\BibitemShut {NoStop}%
\bibitem [{\citenamefont {Bao}\ \emph {et~al.}(2020)\citenamefont {Bao},
  \citenamefont {Liu}, \citenamefont {Tian}, \citenamefont {Wang},
  \citenamefont {Cui}, \citenamefont {Jiang}, \citenamefont {Zhang},\ and\
  \citenamefont {Cao}}]{Cao2020}%
  \BibitemOpen
  \bibfield  {author} {\bibinfo {author} {\bibfnamefont {J.}~\bibnamefont
  {Bao}}, \bibinfo {author} {\bibfnamefont {N.}~\bibnamefont {Liu}}, \bibinfo
  {author} {\bibfnamefont {H.}~\bibnamefont {Tian}}, \bibinfo {author}
  {\bibfnamefont {Q.}~\bibnamefont {Wang}}, \bibinfo {author} {\bibfnamefont
  {T.}~\bibnamefont {Cui}}, \bibinfo {author} {\bibfnamefont {W.}~\bibnamefont
  {Jiang}}, \bibinfo {author} {\bibfnamefont {S.}~\bibnamefont {Zhang}}, \ and\
  \bibinfo {author} {\bibfnamefont {T.}~\bibnamefont {Cao}},\ }\href {\doibase
  10.34133/2020/7873581} {\bibfield  {journal} {\bibinfo  {journal} {Research}\
  }\textbf {\bibinfo {volume} {2020}},\ \bibinfo {pages} {7873581} (\bibinfo
  {year} {2020})}\BibitemShut {NoStop}%
\bibitem [{\citenamefont {Gautier}\ \emph {et~al.}(2022)\citenamefont
  {Gautier}, \citenamefont {Li}, \citenamefont {Ebbesen},\ and\ \citenamefont
  {Genet}}]{Gautier2021}%
  \BibitemOpen
  \bibfield  {author} {\bibinfo {author} {\bibfnamefont {J.}~\bibnamefont
  {Gautier}}, \bibinfo {author} {\bibfnamefont {M.}~\bibnamefont {Li}},
  \bibinfo {author} {\bibfnamefont {T.~W.}\ \bibnamefont {Ebbesen}}, \ and\
  \bibinfo {author} {\bibfnamefont {C.}~\bibnamefont {Genet}},\ }\href
  {\doibase 10.1021/acsphotonics.1c00780} {\bibfield  {journal} {\bibinfo
  {journal} {ACS Photonics}\ }\textbf {\bibinfo {volume} {9}},\ \bibinfo
  {pages} {778} (\bibinfo {year} {2022})}\BibitemShut {NoStop}%
\bibitem [{\citenamefont {Yuan}\ \emph {et~al.}(2021)\citenamefont {Yuan},
  \citenamefont {Zhou}, \citenamefont {Qiao}, \citenamefont {Eng~Aik},
  \citenamefont {Tu}, \citenamefont {Wu},\ and\ \citenamefont
  {Chen}}]{Yuan2021}%
  \BibitemOpen
  \bibfield  {author} {\bibinfo {author} {\bibfnamefont {Z.}~\bibnamefont
  {Yuan}}, \bibinfo {author} {\bibfnamefont {Y.}~\bibnamefont {Zhou}}, \bibinfo
  {author} {\bibfnamefont {Z.}~\bibnamefont {Qiao}}, \bibinfo {author}
  {\bibfnamefont {C.}~\bibnamefont {Eng~Aik}}, \bibinfo {author} {\bibfnamefont
  {W.-C.}\ \bibnamefont {Tu}}, \bibinfo {author} {\bibfnamefont
  {X.}~\bibnamefont {Wu}}, \ and\ \bibinfo {author} {\bibfnamefont {Y.-C.}\
  \bibnamefont {Chen}},\ }\href {\doibase 10.1021/acsnano.1c01805} {\bibfield
  {journal} {\bibinfo  {journal} {ACS Nano}\ }\textbf {\bibinfo {volume}
  {15}},\ \bibinfo {pages} {8965} (\bibinfo {year} {2021})}\BibitemShut
  {NoStop}%
\bibitem [{\citenamefont {Mauro}\ \emph {et~al.}(2023)\citenamefont {Mauro},
  \citenamefont {Fregoni}, \citenamefont {Feist},\ and\ \citenamefont
  {Avriller}}]{PhysRevA.107.L021501}%
  \BibitemOpen
  \bibfield  {author} {\bibinfo {author} {\bibfnamefont {L.}~\bibnamefont
  {Mauro}}, \bibinfo {author} {\bibfnamefont {J.}~\bibnamefont {Fregoni}},
  \bibinfo {author} {\bibfnamefont {J.}~\bibnamefont {Feist}}, \ and\ \bibinfo
  {author} {\bibfnamefont {R.}~\bibnamefont {Avriller}},\ }\href {\doibase
  10.1103/PhysRevA.107.L021501} {\bibfield  {journal} {\bibinfo  {journal}
  {Phys. Rev. A}\ }\textbf {\bibinfo {volume} {107}},\ \bibinfo {pages}
  {L021501} (\bibinfo {year} {2023})}\BibitemShut {NoStop}%
\bibitem [{\citenamefont {Onsager}(1931{\natexlab{a}})}]{Onsager1931}%
  \BibitemOpen
  \bibfield  {author} {\bibinfo {author} {\bibfnamefont {L.}~\bibnamefont
  {Onsager}},\ }\href {\doibase 10.1103/PhysRev.37.405} {\bibfield  {journal}
  {\bibinfo  {journal} {Phys. Rev.}\ }\textbf {\bibinfo {volume} {37}},\
  \bibinfo {pages} {405} (\bibinfo {year} {1931}{\natexlab{a}})}\BibitemShut
  {NoStop}%
\bibitem [{\citenamefont {Onsager}(1931{\natexlab{b}})}]{Onsager2}%
  \BibitemOpen
  \bibfield  {author} {\bibinfo {author} {\bibfnamefont {L.}~\bibnamefont
  {Onsager}},\ }\href {\doibase 10.1103/PhysRev.38.2265} {\bibfield  {journal}
  {\bibinfo  {journal} {Phys. Rev.}\ }\textbf {\bibinfo {volume} {38}},\
  \bibinfo {pages} {2265} (\bibinfo {year} {1931}{\natexlab{b}})}\BibitemShut
  {NoStop}%
\bibitem [{\citenamefont {Casimir}(1945)}]{RevModPhys.17.343}%
  \BibitemOpen
  \bibfield  {author} {\bibinfo {author} {\bibfnamefont {H.~B.~G.}\
  \bibnamefont {Casimir}},\ }\href {\doibase 10.1103/RevModPhys.17.343}
  {\bibfield  {journal} {\bibinfo  {journal} {Rev. Mod. Phys.}\ }\textbf
  {\bibinfo {volume} {17}},\ \bibinfo {pages} {343} (\bibinfo {year}
  {1945})}\BibitemShut {NoStop}%
\bibitem [{\citenamefont {Newton}(2013)}]{newton2013scattering}%
  \BibitemOpen
  \bibfield  {author} {\bibinfo {author} {\bibfnamefont {R.~G.}\ \bibnamefont
  {Newton}},\ }\href@noop {} {\emph {\bibinfo {title} {Scattering theory of
  waves and particles}}}\ (\bibinfo  {publisher} {Springer Science \& Business
  Media, New York},\ \bibinfo {year} {2013})\BibitemShut {NoStop}%
\bibitem [{\citenamefont {Schäfer}\ \emph {et~al.}(2020)\citenamefont
  {Schäfer}, \citenamefont {Ruggenthaler}, \citenamefont {Rokaj},\ and\
  \citenamefont {Rubio}}]{Schafer2020}%
  \BibitemOpen
  \bibfield  {author} {\bibinfo {author} {\bibfnamefont {C.}~\bibnamefont
  {Schäfer}}, \bibinfo {author} {\bibfnamefont {M.}~\bibnamefont
  {Ruggenthaler}}, \bibinfo {author} {\bibfnamefont {V.}~\bibnamefont {Rokaj}},
  \ and\ \bibinfo {author} {\bibfnamefont {A.}~\bibnamefont {Rubio}},\ }\href
  {\doibase 10.1021/acsphotonics.9b01649} {\bibfield  {journal} {\bibinfo
  {journal} {ACS Photonics}\ }\textbf {\bibinfo {volume} {7}},\ \bibinfo
  {pages} {975} (\bibinfo {year} {2020})}\BibitemShut {NoStop}%
\bibitem [{\citenamefont {Hopfield}(1958)}]{PhysRev.112.1555}%
  \BibitemOpen
  \bibfield  {author} {\bibinfo {author} {\bibfnamefont {J.~J.}\ \bibnamefont
  {Hopfield}},\ }\href {\doibase 10.1103/PhysRev.112.1555} {\bibfield
  {journal} {\bibinfo  {journal} {Phys. Rev.}\ }\textbf {\bibinfo {volume}
  {112}},\ \bibinfo {pages} {1555} (\bibinfo {year} {1958})}\BibitemShut
  {NoStop}%
\bibitem [{\citenamefont {Landauer}(1957)}]{Landauer1957}%
  \BibitemOpen
  \bibfield  {author} {\bibinfo {author} {\bibfnamefont {R.}~\bibnamefont
  {Landauer}},\ }\href {\doibase 10.1147/rd.13.0223} {\bibfield  {journal}
  {\bibinfo  {journal} {IBM J. Res. Dev.}\ }\textbf {\bibinfo {volume} {1}},\
  \bibinfo {pages} {223} (\bibinfo {year} {1957})}\BibitemShut {NoStop}%
\bibitem [{\citenamefont {B\"uttiker}(1986)}]{Buttiker1986}%
  \BibitemOpen
  \bibfield  {author} {\bibinfo {author} {\bibfnamefont {M.}~\bibnamefont
  {B\"uttiker}},\ }\href {\doibase 10.1103/PhysRevLett.57.1761} {\bibfield
  {journal} {\bibinfo  {journal} {Phys. Rev. Lett.}\ }\textbf {\bibinfo
  {volume} {57}},\ \bibinfo {pages} {1761} (\bibinfo {year}
  {1986})}\BibitemShut {NoStop}%
\bibitem [{\citenamefont {Fisher}\ and\ \citenamefont
  {Lee}(1981)}]{fisher1981relation}%
  \BibitemOpen
  \bibfield  {author} {\bibinfo {author} {\bibfnamefont {D.~S.}\ \bibnamefont
  {Fisher}}\ and\ \bibinfo {author} {\bibfnamefont {P.~A.}\ \bibnamefont
  {Lee}},\ }\href {\doibase 10.1103/PhysRevB.23.6851} {\bibfield  {journal}
  {\bibinfo  {journal} {Phys. Rev. B}\ }\textbf {\bibinfo {volume} {23}},\
  \bibinfo {pages} {6851} (\bibinfo {year} {1981})}\BibitemShut {NoStop}%
\bibitem [{\citenamefont {Pedrotti}\ \emph {et~al.}(2017)\citenamefont
  {Pedrotti}, \citenamefont {Pedrotti},\ and\ \citenamefont
  {Pedrotti}}]{IntroOptics}%
  \BibitemOpen
  \bibfield  {author} {\bibinfo {author} {\bibfnamefont {F.~L.}\ \bibnamefont
  {Pedrotti}}, \bibinfo {author} {\bibfnamefont {L.~M.}\ \bibnamefont
  {Pedrotti}}, \ and\ \bibinfo {author} {\bibfnamefont {L.~S.}\ \bibnamefont
  {Pedrotti}},\ }\href@noop {} {\emph {\bibinfo {title} {Introduction to
  Optics}}}\ (\bibinfo  {publisher} {Addison-Wesley},\ \bibinfo {year}
  {2017})\BibitemShut {NoStop}%
\bibitem [{\citenamefont {Baranov}\ \emph {et~al.}(2020)\citenamefont
  {Baranov}, \citenamefont {Munkhbat}, \citenamefont {Länk}, \citenamefont
  {Verre}, \citenamefont {Käll},\ and\ \citenamefont {Shegai}}]{Baranov2020}%
  \BibitemOpen
  \bibfield  {author} {\bibinfo {author} {\bibfnamefont {D.~G.}\ \bibnamefont
  {Baranov}}, \bibinfo {author} {\bibfnamefont {B.}~\bibnamefont {Munkhbat}},
  \bibinfo {author} {\bibfnamefont {N.~O.}\ \bibnamefont {Länk}}, \bibinfo
  {author} {\bibfnamefont {R.}~\bibnamefont {Verre}}, \bibinfo {author}
  {\bibfnamefont {M.}~\bibnamefont {Käll}}, \ and\ \bibinfo {author}
  {\bibfnamefont {T.}~\bibnamefont {Shegai}},\ }\href {\doibase
  doi:10.1515/nanoph-2019-0372} {\bibfield  {journal} {\bibinfo  {journal}
  {Nanophotonics}\ }\textbf {\bibinfo {volume} {9}},\ \bibinfo {pages} {283}
  (\bibinfo {year} {2020})}\BibitemShut {NoStop}%
\bibitem [{\citenamefont {Munkhbat}\ \emph {et~al.}(2021)\citenamefont
  {Munkhbat}, \citenamefont {Canales}, \citenamefont {Küçüköz},
  \citenamefont {Baranov},\ and\ \citenamefont {Shegai}}]{Munkhbat2021}%
  \BibitemOpen
  \bibfield  {author} {\bibinfo {author} {\bibfnamefont {B.}~\bibnamefont
  {Munkhbat}}, \bibinfo {author} {\bibfnamefont {A.}~\bibnamefont {Canales}},
  \bibinfo {author} {\bibfnamefont {B.}~\bibnamefont {Küçüköz}}, \bibinfo
  {author} {\bibfnamefont {D.~G.}\ \bibnamefont {Baranov}}, \ and\ \bibinfo
  {author} {\bibfnamefont {T.~O.}\ \bibnamefont {Shegai}},\ }\href
  {https://www.nature.com/articles/s41586-021-03826-3} {\bibfield  {journal}
  {\bibinfo  {journal} {Nature}\ }\textbf {\bibinfo {volume} {597}},\ \bibinfo
  {pages} {214} (\bibinfo {year} {2021})}\BibitemShut {NoStop}%
\bibitem [{\citenamefont {Zhu}\ \emph {et~al.}(1990)\citenamefont {Zhu},
  \citenamefont {Gauthier}, \citenamefont {Morin}, \citenamefont {Wu},
  \citenamefont {Carmichael},\ and\ \citenamefont {Mossberg}}]{Zhu1990}%
  \BibitemOpen
  \bibfield  {author} {\bibinfo {author} {\bibfnamefont {Y.}~\bibnamefont
  {Zhu}}, \bibinfo {author} {\bibfnamefont {D.~J.}\ \bibnamefont {Gauthier}},
  \bibinfo {author} {\bibfnamefont {S.~E.}\ \bibnamefont {Morin}}, \bibinfo
  {author} {\bibfnamefont {Q.}~\bibnamefont {Wu}}, \bibinfo {author}
  {\bibfnamefont {H.~J.}\ \bibnamefont {Carmichael}}, \ and\ \bibinfo {author}
  {\bibfnamefont {T.~W.}\ \bibnamefont {Mossberg}},\ }\href {\doibase
  10.1103/PhysRevLett.64.2499} {\bibfield  {journal} {\bibinfo  {journal}
  {Phys. Rev. Lett.}\ }\textbf {\bibinfo {volume} {64}},\ \bibinfo {pages}
  {2499} (\bibinfo {year} {1990})}\BibitemShut {NoStop}%
\bibitem [{\citenamefont {Weisbuch}\ \emph {et~al.}(1992)\citenamefont
  {Weisbuch}, \citenamefont {Nishioka}, \citenamefont {Ishikawa},\ and\
  \citenamefont {Arakawa}}]{PhysRevLett.69.3314}%
  \BibitemOpen
  \bibfield  {author} {\bibinfo {author} {\bibfnamefont {C.}~\bibnamefont
  {Weisbuch}}, \bibinfo {author} {\bibfnamefont {M.}~\bibnamefont {Nishioka}},
  \bibinfo {author} {\bibfnamefont {A.}~\bibnamefont {Ishikawa}}, \ and\
  \bibinfo {author} {\bibfnamefont {Y.}~\bibnamefont {Arakawa}},\ }\href
  {\doibase 10.1103/PhysRevLett.69.3314} {\bibfield  {journal} {\bibinfo
  {journal} {Phys. Rev. Lett.}\ }\textbf {\bibinfo {volume} {69}},\ \bibinfo
  {pages} {3314} (\bibinfo {year} {1992})}\BibitemShut {NoStop}%
\bibitem [{\citenamefont {Gorkunov}\ \emph {et~al.}(2020)\citenamefont
  {Gorkunov}, \citenamefont {Antonov},\ and\ \citenamefont
  {Kivshar}}]{Gorkunov2020}%
  \BibitemOpen
  \bibfield  {author} {\bibinfo {author} {\bibfnamefont {M.~V.}\ \bibnamefont
  {Gorkunov}}, \bibinfo {author} {\bibfnamefont {A.~A.}\ \bibnamefont
  {Antonov}}, \ and\ \bibinfo {author} {\bibfnamefont {Y.~S.}\ \bibnamefont
  {Kivshar}},\ }\href {\doibase 10.1103/PhysRevLett.125.093903} {\bibfield
  {journal} {\bibinfo  {journal} {Phys. Rev. Lett.}\ }\textbf {\bibinfo
  {volume} {125}},\ \bibinfo {pages} {093903} (\bibinfo {year}
  {2020})}\BibitemShut {NoStop}%
\bibitem [{\citenamefont {Fan}\ \emph {et~al.}(2003)\citenamefont {Fan},
  \citenamefont {Suh},\ and\ \citenamefont {Joannopoulos}}]{Fan2003}%
  \BibitemOpen
  \bibfield  {author} {\bibinfo {author} {\bibfnamefont {S.}~\bibnamefont
  {Fan}}, \bibinfo {author} {\bibfnamefont {W.}~\bibnamefont {Suh}}, \ and\
  \bibinfo {author} {\bibfnamefont {J.~D.}\ \bibnamefont {Joannopoulos}},\
  }\href {\doibase 10.1364/JOSAA.20.000569} {\bibfield  {journal} {\bibinfo
  {journal} {J. Opt. Soc. Am. A}\ }\textbf {\bibinfo {volume} {20}},\ \bibinfo
  {pages} {569} (\bibinfo {year} {2003})}\BibitemShut {NoStop}%
\bibitem [{\citenamefont {Fink}(1993)}]{Fink1993}%
  \BibitemOpen
  \bibfield  {author} {\bibinfo {author} {\bibfnamefont {M.}~\bibnamefont
  {Fink}},\ }\href {\doibase 10.1088/0022-3727/26/9/001} {\bibfield  {journal}
  {\bibinfo  {journal} {J. Phys. D}\ }\textbf {\bibinfo {volume} {26}},\
  \bibinfo {pages} {1333} (\bibinfo {year} {1993})}\BibitemShut {NoStop}%
\bibitem [{\citenamefont {de~Rosny}\ \emph {et~al.}(2010)\citenamefont
  {de~Rosny}, \citenamefont {Lerosey},\ and\ \citenamefont {Fink}}]{Rosny2010}%
  \BibitemOpen
  \bibfield  {author} {\bibinfo {author} {\bibfnamefont {J.}~\bibnamefont
  {de~Rosny}}, \bibinfo {author} {\bibfnamefont {G.}~\bibnamefont {Lerosey}}, \
  and\ \bibinfo {author} {\bibfnamefont {M.}~\bibnamefont {Fink}},\ }\href
  {\doibase 10.1109/TAP.2010.2052567} {\bibfield  {journal} {\bibinfo
  {journal} {IEEE Trans. Antennas Propag.}\ }\textbf {\bibinfo {volume} {58}},\
  \bibinfo {pages} {3139} (\bibinfo {year} {2010})}\BibitemShut {NoStop}%
\bibitem [{\citenamefont {Fernandez-Corbaton}\ \emph
  {et~al.}(2016)\citenamefont {Fernandez-Corbaton}, \citenamefont {Fruhnert},\
  and\ \citenamefont {Rockstuhl}}]{Corbaton2016}%
  \BibitemOpen
  \bibfield  {author} {\bibinfo {author} {\bibfnamefont {I.}~\bibnamefont
  {Fernandez-Corbaton}}, \bibinfo {author} {\bibfnamefont {M.}~\bibnamefont
  {Fruhnert}}, \ and\ \bibinfo {author} {\bibfnamefont {C.}~\bibnamefont
  {Rockstuhl}},\ }\href {\doibase 10.1103/PhysRevX.6.031013} {\bibfield
  {journal} {\bibinfo  {journal} {Phys. Rev. X}\ }\textbf {\bibinfo {volume}
  {6}},\ \bibinfo {pages} {031013} (\bibinfo {year} {2016})}\BibitemShut
  {NoStop}%
\bibitem [{\citenamefont {Riso}\ \emph {et~al.}(2022)\citenamefont {Riso},
  \citenamefont {Grazioli}, \citenamefont {Ronca}, \citenamefont {Giovannini},\
  and\ \citenamefont {Koch}}]{riso2022strong}%
  \BibitemOpen
  \bibfield  {author} {\bibinfo {author} {\bibfnamefont {R.~R.}\ \bibnamefont
  {Riso}}, \bibinfo {author} {\bibfnamefont {L.}~\bibnamefont {Grazioli}},
  \bibinfo {author} {\bibfnamefont {E.}~\bibnamefont {Ronca}}, \bibinfo
  {author} {\bibfnamefont {T.}~\bibnamefont {Giovannini}}, \ and\ \bibinfo
  {author} {\bibfnamefont {H.}~\bibnamefont {Koch}},\ }\href {\doibase
  10.48550/ARXIV.2209.01987} {\enquote {\bibinfo {title} {Strong coupling in
  chiral cavities: nonperturbative framework for enantiomer discrimination},}\
  } (\bibinfo {year} {2022})\BibitemShut {NoStop}%
\bibitem [{\citenamefont {Schäfer}\ and\ \citenamefont
  {Baranov}(2023)}]{doi:10.1021/acs.jpclett.3c00286}%
  \BibitemOpen
  \bibfield  {author} {\bibinfo {author} {\bibfnamefont {C.}~\bibnamefont
  {Schäfer}}\ and\ \bibinfo {author} {\bibfnamefont {D.~G.}\ \bibnamefont
  {Baranov}},\ }\href {\doibase 10.1021/acs.jpclett.3c00286} {\bibfield
  {journal} {\bibinfo  {journal} {The Journal of Physical Chemistry Letters}\
  }\textbf {\bibinfo {volume} {14}},\ \bibinfo {pages} {3777} (\bibinfo {year}
  {2023})}\BibitemShut {NoStop}%
\bibitem [{\citenamefont {Salij}\ \emph {et~al.}(2024)\citenamefont {Salij},
  \citenamefont {Goldsmith},\ and\ \citenamefont
  {Tempelaar}}]{salij2022chiral}%
  \BibitemOpen
  \bibfield  {author} {\bibinfo {author} {\bibfnamefont {A.~H.}\ \bibnamefont
  {Salij}}, \bibinfo {author} {\bibfnamefont {R.~H.}\ \bibnamefont
  {Goldsmith}}, \ and\ \bibinfo {author} {\bibfnamefont {R.}~\bibnamefont
  {Tempelaar}},\ }\href {\doibase 10.1038/s41467-023-44523-1} {\bibfield
  {journal} {\bibinfo  {journal} {Nature communications}\ }\textbf {\bibinfo
  {volume} {15}},\ \bibinfo {pages} {340} (\bibinfo {year} {2024})}\BibitemShut
  {NoStop}%
\bibitem [{Cod()}]{Code}%
  \BibitemOpen
  \href {\doibase https://github.com/jfeist/chiral-transfermatrix} {\
  https://github.com/jfeist/chiral-transfermatrix}\BibitemShut {NoStop}%
\bibitem [{\citenamefont {Atkins}\ and\ \citenamefont
  {Friedman}(2011)}]{Atkins}%
  \BibitemOpen
  \bibfield  {author} {\bibinfo {author} {\bibfnamefont {P.}~\bibnamefont
  {Atkins}}\ and\ \bibinfo {author} {\bibfnamefont {R.}~\bibnamefont
  {Friedman}},\ }\href@noop {} {\emph {\bibinfo {title} {Molecular Quantum
  Mechanics}}}\ (\bibinfo  {publisher} {Oxford University Press},\ \bibinfo
  {year} {2011})\BibitemShut {NoStop}%
\end{thebibliography}%
%

\appendix
%
\section{Reciprocity beyond dipole-approximation}
\label{BeCustoState}
Using the expansion of eq. \eqref{DA1} (main text) at linear order in the spatial derivative of the electric field, we obtain in the left-hand side of eq. \eqref{LR12} 
\begin{eqnarray}
\int_{\mathcal{V}_{1}}d^3\vec{r}_{1}\;\vec{j}_{1}(\vec{r}_{1}) \cdot\vec{E}_{2}(\vec{r}_1)
&\approx& -i\omega\vec{\mu}_{1}\cdot\vec{E}_{2}(\vec{0})
\label{firstexpenscust} \\
&+&\frac{\partial E_{2,i}}{\partial x_{1,j}}(\vec{0})\int_{\mathcal{V}_{1}}d^3\vec{r}_{1}\;x_{1,j}j_{1,i}(\vec{r}_{1}),
\nonumber
\end{eqnarray}
where $\vec{j}_{l}(\vec{r})=-i\omega\vec{\mu}\delta^{(3)}(\vec{r}-\vec{r}_{l})$ is the current-dipole source, and where Einstein convention has been adopted for summation on equal components of the involved tensors. 
The last term in eq. \eqref{firstexpenscust} gives rise to both contributions of the magnetic dipole and the electric quadrupole moments.
We remark that any second-rank tensor $T_{ij}$ can be written as the sum of its symmetric (s) part and antisymmetric (a) part as follows:
\begin{equation}
T_{ij}=T_{s,ij}+T_{a,ij}=\frac{T_{ij}+T_{ji}}{2}+\frac{T_{ij}-T_{ji}}{2},
\end{equation}
such that the product of the tensor $T_{1,ij}=\frac{\partial E_{2,i}}{\partial x_{1,j}}(\vec{0})$ by the tensor $T_{2,ij}=\int_{\mathcal{V}_{1}}d^3\vec{r}_{1}\;x_{1,j}j_{1,i}(\vec{r}_{1})$ in eq. \eqref{firstexpenscust} becomes 
\begin{equation}
\begin{split}
T_{1,ij}T_{2,ij}=T_{1s,ij}T_{2s,ij}+T_{1a,ij}T_{2a,ij}
\end{split} \, .
\label{symmantisymm}
\end{equation} 
The symmetric and antisymmetric parts of eq. \eqref{symmantisymm} are proportional to the contribution of the electric quadrupole moment $Q_{ij}$ and magnetic dipole $\vec{m}$, respectively given by  \cite{jackson1999classical}
\begin{eqnarray}
\vec{m} &=& \frac12\int_{\mathcal{V}}d^3\vec{r}\;\left[\vec{r}\wedge\vec{j}\left(\vec{r}\right)\right]
\label{qmagdip} \, , \\
Q_{ij} &=& \frac{i}{\omega}\int_{\mathcal{V}}d^3\vec{r}\left[3\left(x_{i}j_{j}+x_{j}j_{i}\right)-2\delta_{ij}\vec{r}\cdot\vec{j}\right] \, .
\label{qmagdipQ}
\end{eqnarray}
Introducing eq. \eqref{qmagdip} and eq. \eqref{qmagdipQ} into eq. \eqref{symmantisymm}, one can rewrite eq. \eqref{firstexpenscust} in the compact form, including also the contribution of the electric quadrupole moment
\begin{equation}
\begin{split}
&\vec{\mu}_1\cdot \vec{E}_{2}(\vec{r}_{1})-\vec{m}_1\cdot \vec{B}_{2}(\vec{r}_{1})+\frac{\stackrel{\leftrightarrow}{Q}_{1}}{12}\frac{\partial \vec{E}_{2}}{\partial \vec{x}_{1}}(\vec{r}_{1})=\\
&\vec{\mu}_2\cdot \vec{E}_{1}(\vec{r}_{2})-\vec{m}_2\cdot \vec{B}_{1}(\vec{r}_{2})+\frac{\stackrel{\leftrightarrow}{Q}_{2}}{12} \frac{\partial \vec{E}_{1}}{\partial \vec{x}_{2}}(\vec{r}_{2})
\label{StateQuadrupole}
\end{split}
\end{equation}
%
%
%
This relation generalizes eq. \eqref{DA3} in the main text.
%

%

\section{Derivation of the Condon constitutive relations}
\label{appInducMicr}

The molecules inside the cavity are supposed to be coupled to the interaction Hamiltonian $\hat{\mathcal{V}}(t)$ of eq. \eqref{qedham}, with the classical time-dependent vector-potential expressed in Coulomb gauge $\vec{A}(\vec{r},t)=(1/2)\vec{A}_{0}(\vec{r})\left[\exp\left(i\epsilon t/\hbar\right)+\exp\left(-i \epsilon t/\hbar\right)\right]$, and $\epsilon=\hbar\omega$.
Using first-order perturbation theory in the interaction Hamiltonian, the molecular state that was far in the past in the eigenstate $\phi_{a}\left(\vec{r}\right)\equiv \langle{\vec{r}}\ket{a}$ of $\hat{\mathcal{H}}_0$, becomes at time $t$
\begin{equation}
\begin{split}
\psi_a\left(\vec{r},t\right)&=\phi_{a}\left(\vec{r}\right) e^{-i\frac{\epsilon_{a}}{\hbar}t} + \sum_{b}c_{b}(t)\phi_{b}\left(\vec{r}\right)e^{-i\frac{\epsilon_{b}}{\hbar}t} ,
\label{pertw}
\end{split}
\end{equation} 
with
\begin{equation}
\begin{split}
c_{b}\left(t\right)&=\frac{1}{2\hbar}\left[i\Omega_{ba}\bra{b}\hat{\vec{\mu}}\ket{a}\cdot\vec{A}+\bra{b}\hat{\vec{m}}\ket{a}\cdot\left(\nabla\wedge\vec{A}\right)\right]\times\\
&\times\left[\frac{e^{i\left(\Omega_{ba}+\omega\right)t}}{\Omega_{ba}+\omega}+\frac{e^{i\left(\Omega_{ba}-\omega\right)t}}{\Omega_{ba}-\omega}\right].
\label{cb}
\end{split}
\end{equation}
We introduced in eq. \eqref{cb}, the Bohr-frequency $\Omega_{ba}\equiv\left(\epsilon_{b}-\epsilon_{a}\right)/\hbar$ for the $a-b$ transition, the electric-dipole transition $\bra{b}\hat{\vec{\mu}}\ket{a}\equiv\bra{b}q\hat{\vec{r}}\ket{a}$ and the magnetic-dipole transition $\bra{b}\hat{\vec{m}}\ket{a}\equiv\bra{b}\frac{q}{2m}\left(\hat{\vec{r}}\wedge\hat{\vec{p}}\right)\ket{a}$.
Following Condon's original paper \cite{Condon1937}, we define the induced electric dipole $\vec{\mu}_{a}(t)$ and magnetic dipole $\vec{m}_{a}(t)$ as
\begin{equation}
\begin{split}
\vec{\mu}_{a}(t)&=2\mathrm{Re}\{\sum_{b}c_{b}\left(t\right)\bra{a}\hat{\vec{\mu}}\ket{b}e^{-i\Omega_{ba}t}\},\\
\vec{m}_{a}(t)&=2\mathrm{Re}\{\sum_{b}c_{b}\left(t\right)\bra{a}\hat{\vec{m}}\ket{b}e^{-i\Omega_{ba}t}\}.
\end{split}
\label{idip}
\end{equation}
We introduce eq. \eqref{cb} into the latter expression, and get for the induced electric dipole moment
\begin{equation}
\begin{split}
\vec{\mu}_{a}(t)&=\frac{2}{\hbar}\mathrm{Re}\{\sum_{b}\bra{a}\hat{\mu}_{i}\ket{b}\bra{b}\hat{\mu}_{j}\ket{a}\times\\
&\times\left[\frac{i\Omega^{2}_{ba}}{\left(\Omega_{ba}^{2}-\omega^{2}\right)\omega^{2}}\frac{\partial E_{j}}{\partial t}+\frac{\Omega_{ba}}{\Omega_{ba}^{2}-\omega^{2}}E_{j}\right]+\\
&+\sum_{b}\bra{a}\hat{\mu}_{i}\ket{b}\bra{b}\hat{m}_{j}\ket{a}\times\\
&\times\left[\frac{\Omega_{ba}}{\Omega_{ba}^{2}-\omega^{2}}\mu_{0}H_{j}+\frac{i}{\Omega_{ba}^{2}-\omega^{2}}\frac{\partial \left(\mu_{0}H_{j}\right)}{\partial t}\right]\},
\end{split}
\label{mi}
\end{equation}
with the electric field $\vec{E}=-\frac{\partial\vec{A}}{\partial t}$ and magnetic induction field $\vec{B}=\vec{\nabla}\wedge\vec{A}$ expressed in Coulomb gauge. 
An analogous calculation can be performed for the induced magnetic dipole moment, leading to
\begin{equation}
\begin{split}
\vec{m}_{a}(t)&=\frac{2}{\hbar}\mathrm{Re}\{\sum_{b}\bra{a}\hat{m}_{i}\ket{b}\bra{b}\hat{\mu}_{j}\ket{a}\times\\
&\times\left[\frac{i\Omega^{2}_{ba}}{\left(\Omega_{ba}^{2}-\omega^{2}\right)\omega^{2}}\frac{\partial E_{j}}{\partial t}+\frac{\Omega_{ba}}{\Omega_{ba}^{2}-\omega^{2}}E_{j}\right]+\\
&+\sum_{b}\bra{a}\hat{m}_{i}\ket{b}\bra{b}\hat{m}_{j}\ket{a}\times\\
&\times\left[\frac{\Omega_{ba}}{\Omega_{ba}^{2}-\omega^{2}}\mu_{0}H_{j}+\frac{i}{\Omega_{ba}^{2}-\omega^{2}}\frac{\partial\left(\mu_{0}H_{j}\right)}{\partial t}\right]\}.
\end{split}
\label{pi}
\end{equation}
We now perform an average of the tensorial quantities appearing in eq. \eqref{mi} and eq. \eqref{pi}, over all equiprobable orientations of the molecular system with respect to the $\vec{A}$-field direction.
We obtain, after taking the real parts
\begin{equation}
\begin{split}
\langle\vec{\mu}_{a}(t)\rangle &=\frac{2}{3\hbar}\sum_{b} \frac{\Omega_{ba}\lVert\bra{a}\hat{\vec{\mu}}\ket{b}\rVert^{2}}{\Omega_{ba}^{2}-\omega^{2}}\vec{E}(\vec{0},t) - \\
&- \frac{2}{3\hbar}\sum_{b}\frac{\mathrm{Im}\left(\bra{a}\hat{\vec{\mu}}\ket{b}\cdot\bra{b}\hat{\vec{m}}\ket{a}\right)}{\Omega_{ba}^{2}-\omega^{2}}\mu_{0}\frac{\partial\vec{H}}{\partial t}(\vec{0},t),
\end{split}
\end{equation}
and similarly,
\begin{equation}
\begin{split}
\langle\vec{m}_{a}(t)\rangle &=\frac{2}{3\hbar}
 \sum_{b}\frac{\Omega^{2}_{ba}\mathrm{Im}\left(\bra{a}\hat{\vec{\mu}}\ket{b}\cdot\bra{b}\hat{\vec{m}}\ket{a}\right)}{\left(\Omega_{ba}^{2}-\omega^{2}\right)\omega^{2}}\frac{\partial\vec{E}}{\partial t}(\vec{0},t)+\\
&+\frac{2}{3\hbar}\sum_{b}\frac{\Omega_{ba}\lVert\bra{a}\hat{\vec{m}}\ket{b}\rVert^{2}}{\Omega_{ba}^{2}-\omega^{2}}\mu_{0}\vec{H}(\vec{0},t).
\label{avid}
\end{split}
\end{equation}
The latter two expressions are the ones originally found by Condon \cite{Condon1937}.
After Fourier transforming them, we obtain back eq. \eqref{constAver} and eq. \eqref{polariz} used in the main text.
%

\section{Continuity relations at a single interface}
\label{TmatrixBoundary}
Here, we consider a half-infinite LHIC medium characterized by $\kappa$, $\varepsilon$ and $\mu$, which creates an interface with a different half-infinite LHIC medium with $\kappa^{\prime}$, $\varepsilon^{\prime}$ and $\mu^{\prime}$ (see Figure \ref{Fig5} in the main text). 
The boundary conditions for the electromagnetic fields $(\vec{D},\vec{E},\vec{B},\vec{H})$ at this interface (chosen at $z=0$) are given in absence of external free sources by \cite{jackson1999classical}
\begin{equation}
\begin{split}
\vec{n}\cdot\left(\vec{D}-\vec{D}^{\prime}\right)=0,\hspace{0,2cm}\vec{n}\cdot\left(\vec{B}-\vec{B}^{\prime}\right)=0,\\
\vec{n}\wedge\left(\vec{E}-\vec{E}^{\prime}\right)=0,\hspace{0,2cm}\vec{n}\wedge\left(\vec{H}-\vec{H}^{\prime}\right)=0,\\
\end{split}
\label{bc}
\end{equation}
with $\vec{n}$, the unit-vector normal to the interface.
The boundary conditions written in eq. \eqref{bc} imply continuity of the phase of the propagating electromagnetic fields whatever is the chosen $(x,y)$ position at the planar interface, thus leading to
\begin{equation}
\begin{split}
\left(\vec{k}_{\pm}\cdot\vec{r}\right)\big|_{z=0}=\left(-\vec{k}_{\pm}\cdot\vec{r}\right)\big|_{z=0}=\left(\vec{k}^{\prime}_{\pm}\cdot\vec{r}\right)\big|_{z=0}\,,\\
\left(-\vec{k}^{\prime}_{\pm}\cdot\vec{r}\right)\big|_{z=0}=\left(\vec{k}^{\prime}_{\pm}\cdot\vec{r}\right)\big|_{z=0}=\left(-\vec{k}_{\pm}\cdot\vec{r}\right)\big|_{z=0}\,,\\
\end{split}
\end{equation}
with the first (second) line due to incident waves arriving from the left (right).
These relations imply that the wave-vectors must lie in a plane and fulfill the following chiral Descartes-Snell's laws \cite{Jaggard1992}
\begin{equation}
\begin{split}
k_{+}\sin\theta_{+}=k_{-}\sin\theta_{-}=k^{\prime}_{+}\sin\theta^{\prime}_{+}=k^{\prime}_{-}\sin\theta^{\prime}_{-}.
\end{split}
\label{csnell}
\end{equation}
We consider the electric eigenfields $\vec{E}^{\pm}$ that are connected to the other eigenfields through Condon's constitutive relations (see eq. \eqref{ccr} in the main text) as
\begin{equation}
\begin{split}
\vec{D}^{\pm}&=i\frac{\kappa}{c\,\omega\mu}\vec{k}_{\pm}\wedge\vec{E}^{\pm}+\left(\varepsilon-\frac{\kappa^{2}}{c^{2}\mu}\right)\vec{E}^{\pm},\\
\vec{B}^{\pm}&=\frac{1}{\omega}\vec{k}_{\pm}\wedge\vec{E}^{\pm},\\
\vec{H}^{\pm}&=\frac{1}{\omega\mu}\vec{k}_{\pm}\wedge\vec{E}^{\pm}+i\frac{\kappa}{c\mu}\vec{E}^{\pm}.
\end{split}
\label{fieldRelations}
\end{equation}
They fulfill the boundary conditions (see eq. \eqref{bc}), which take the form
\begin{equation}
\begin{split}
&\sum_{\alpha=\pm} \alpha \left( E^{\alpha}_{\rightarrow}
+ E^{\alpha}_{\leftarrow} \right) = 
\sum_{\alpha=\pm} \alpha \left( E'^{\alpha}_{\rightarrow}
+ E'^{\alpha}_{\leftarrow} \right) ,\\
&\sum_{\alpha=\pm} \left( E^{\alpha}_{\rightarrow}
+ E^{\alpha}_{\leftarrow} \right) =  \frac{\eta}{\eta^{\prime}} 
\sum_{\alpha=\pm} \left( E'^{\alpha}_{\rightarrow}
+ E'^{\alpha}_{\leftarrow} \right) ,\\
&\sum_{\alpha=\pm} \left( E^{\alpha}_{\rightarrow} - E^{\alpha}_{\leftarrow} \right)  \cos\theta_{\alpha} = 
\sum_{\alpha=\pm} \left( E'^{\alpha}_{\rightarrow} - E'^{\alpha}_{\leftarrow} \right)  \cos\theta'_{\alpha} ,\\
&\sum_{\alpha=\pm} \alpha \left( E^{\alpha}_{\rightarrow} - E^{\alpha}_{\leftarrow} \right)  \cos\theta_{\alpha} = \frac{\eta}{\eta^{\prime}} 
\sum_{\alpha=\pm} \alpha \left( E'^{\alpha}_{\rightarrow} - E'^{\alpha}_{\leftarrow} \right)  \cos\theta'_{\alpha} .
\end{split}
\label{eq}
\end{equation}
From eq. \eqref{eq}, it is straightforward to derive the expression of the transfer-matrix and related sub-matrices given in eq. \eqref{eq:mt} (main text).
Moreover, eq. \eqref{eq} is consistent with the results obtained in ref. \cite{Jaggard1992}, but with a different choice of constitutive relations. 

\section{Transmittance and Green function}
\label{GFappro}
We detail the computation of the forward transmittance $\mathcal{T}_\rightarrow$, based on the summation of all multiple scattering processes through which, an input electromagnetic-wave $\vec{\mathcal{E}}_{\rm{in}}$ propagates across the FP cavity (see Figure \ref{Fig6} in the main text).  
The first transmission path, representing the direct transmission of the input-wave without any reflection at the mirror interfaces, is given by
\begin{equation}
\vec{\mathcal{E}}_{1}=\mathcal{T}^{\rightarrow}_{12}\stackrel{\leftrightarrow}{M}_{\phi}\mathcal{T}^{\rightarrow}_{21}\vec{\mathcal{E}}_{0},
\end{equation}
where the notations of Sec. \ref{GenGreen} is adopted.
The second transmitted path takes into account one round-trip inside cavity with two reflection processes at the mirror interfaces
\begin{equation}
\vec{\mathcal{E}}_{2}=\mathcal{T}^{\rightarrow}_{12}\left( \stackrel{\leftrightarrow}{M}_{\phi}\mathcal{R}^{\rightarrow}_{22}\stackrel{\leftrightarrow}{M}_{\phi}\mathcal{R}^{\leftarrow}_{22} \right)\stackrel{\leftrightarrow}{M}_{\phi}\mathcal{T}^{\rightarrow}_{21}\vec{\mathcal{E}}_{0},
\end{equation}
The higher-order transmission paths, involves round-trips with a higher number of multiple reflection at the mirror interfaces.  
The total transmitted wave $\vec{\mathcal{E}}_{\mathrm{out}}$ across the cavity is obtained by 
resuming the geometric series $\vec{\mathcal{E}}_{\mathrm{out}}=\vec{\mathcal{E}}_{1}+\vec{\mathcal{E}}_{2}+\cdots$, taking into account all orders in the number of internal reflections
\begin{equation}
\begin{split}
\vec{\mathcal{E}}_{\mathrm{out}}&=\mathcal{T}^{\rightarrow}_{12}\left(\mathds{Id}-\stackrel{\leftrightarrow}{M}_{\phi}\mathcal{R}^{\rightarrow}_{22}\stackrel{\leftrightarrow}{M}_{\phi}\mathcal{R}^{\leftarrow}_{22}\right)^{-1}\stackrel{\leftrightarrow}{M}_{\phi}\mathcal{T}^{\rightarrow}_{21}\vec{\mathcal{E}}_{in},\\
&\equiv \mathcal{T}^{\rightarrow}_{12}\left(\stackrel{\leftrightarrow}{M}_{\phi}^{-1}-\mathcal{R}^{\rightarrow}_{22}\stackrel{\leftrightarrow}{M}_{\phi}\mathcal{R}^{\leftarrow}_{22}\right)^{-1}\mathcal{T}^{\rightarrow}_{21}\vec{\mathcal{E}}_{in}.\\
\end{split}
\end{equation}
This expression recovers the outcome of eqs. \eqref{calcJ1} and \eqref{calcJ2} in the main text.
%
 
\section{Expression of the macroscopic Pasteur coefficient}
\label{ModKappa}
The microscopic dielectric polarization and Pasteur coupling were obtained in eq. \eqref{polariz} (main text).
They can be rewritten by including a damping term $\Gamma_{ba}$, that takes into account in a phenomenological way, the presence of losses (or linewidth of the $a-b$ transition) \cite{Condon1937}
\begin{equation}
\begin{split}
\alpha_{a}(\omega)&=\frac{2}{3\hbar}\sum_{b}\frac{\Omega_{ba}}{\Omega_{ba}^{2}-\omega^{2}-i\Gamma_{ba}\omega}\lVert\bra{a}\vec{\mu}\ket{b}\rVert^{2},\\
g_{a}(\omega)&=\frac{2c}{3\hbar}\sum_{b}\frac{\mathrm{Im}\{\bra{a}\vec{\mu}\ket{b}\cdot \bra{b}\vec{m}\ket{a}\}}{\Omega_{ba}^{2}-\omega^{2}-i\Gamma_{ba}\omega}.
\label{PolApp}
\end{split}
\end{equation}
Introducing the dimensionless oscillator-strength $f_{ba}$, defined by
\begin{equation}
f_{ba}=\frac{2m}{3q^{2}\hbar}\Omega_{ba}\lVert\bra{a}\vec{\mu}\ket{b}\rVert^{2}\hspace{0,2cm}\mbox{with}\;\sum_{b}f_{ba}=1,
\label{oscills}
\end{equation}
we obtain for the microscopic dielectric polarization
\begin{equation}
\alpha_{a}(\omega)=\frac{q^{2}}{\varepsilon_{0}m}\sum_{b}\frac{f_{ba}}{\Omega_{ba}^{2}-\omega^{2}-i\Gamma_{ba}\omega}.
\label{AlphaMic}
\end{equation}
This formula provides an interpretation of the polarizability of molecules: if a molecule has an intense peak in its absorption spectrum, then it is highly polarizable \cite{Atkins}.
In the case where only the $a-b$ molecular-transition matters for describing the absorption spectrum in the range of frequency of interest, the macroscopic dielectric susceptibility $\chi_e(\omega)$ as defined in Sec. \ref{CondConst} (main text), follows directly from eq. \eqref{AlphaMic} 
\begin{equation}
\chi_e\left(\omega\right)=\frac{\omega^2_{p} f}{\left(\omega^{2}_{0}-\omega^{2}\right)-i\gamma\omega},
\end{equation}
where the plasma-frequency is $\omega_{p}\equiv \sqrt{Nq^{2}/\varepsilon_{0} m}$, the 
Bohr transition-frequency is $\omega_{0}\equiv \Omega_{ba}$, and the effective damping-rate is $\gamma \equiv \Gamma_{ba}$.
This relation recovers the expression given in eq. \eqref{ConstRel3} (main text) derived from a classical Drude-Lorentz model of the dielectric susceptibility.
Similar arguments are used to derive an explicit expression of the macroscopic Pasteur coefficient
$\kappa\left(\omega\right)=\omega g(\omega)$, with $g(\omega)$ defined in Sec. \ref{CondConst} (main text).
This involves the term $R_{ba}\equiv \mathrm{Im}\{\bra{a}\vec{\mu}\ket{b}\cdot\bra{b}\vec{m}\ket{a}\}$ in eq. \eqref{PolApp}, known as the rotational-strength of the given absorption line (or $a-b$ transition). 
The latter has the following properties
\begin{equation}
\begin{split}
R_{ba}=-R_{ab},\\
\sum_{b}R_{ba}=0.
\end{split}
\end{equation}
We further introduce $\kappa_{ba}$, the Pasteur coefficient associated to the $a-b$ transition, that is defined by the relation
\begin{equation}
R_{ba}=\kappa_{ba}\lVert\bra{a}\vec{\mu}\ket{b}\rVert^{2}.
\label{kab}
\end{equation}
Inserting eq. \eqref{kab}, and eq. \eqref{oscills}, into the expression of $g_a(\omega)$ provided by eq. \eqref{PolApp}, we finally obtain for the macroscopic Pasteur coefficient $\kappa(\omega)$ given by 
\begin{equation}
\kappa\left(\omega\right) = \kappa\frac{\omega^{2}_{p}\,\omega f_{0}}{\omega_{0}\left[\left(\omega^{2}_{0}-\omega^{2}\right)-i\gamma\omega\right]},
\label{kappapdisp1}
\end{equation}
where $\kappa \equiv \kappa_{ba}$.
This relation recovers the outcome of the classical model provided in eq. \eqref{kappapdisp} (main text).
%

%
\section{Time-reversal symmetry and scattering matrix}
\label{Treversability}
We consider the implication of time-reversal (TR) symmetry on the scattering-matrix of a TR-symmetric optical cavity. 
Under the transformation $t \rightarrow -t$, the charge and current-density in Maxwell equations transform as even and odd functions respectively
\begin{eqnarray}
\mbox{TR}\left\lbrack\rho\right\rbrack\left( \vec{r},t \right)
&=&
\rho\left( \vec{r},-t \right),
\label{TRS1}\\
\mbox{TR}\left\lbrack\vec{j}\right\rbrack\left( \vec{r},t \right)
&=&
-\vec{j}\left( \vec{r},-t \right).
\label{TRS2}
\end{eqnarray}
As a result of Maxwell equations, the electric field $\vec{E}$ and magnetic induction field $\vec{B}$ transform under TR as 
\begin{eqnarray}
\mbox{TR}\left\lbrack\vec{E}\right\rbrack\left( \vec{r},t \right)
&=&
\vec{E}\left( \vec{r},-t \right),
\label{TRS3}\\
\mbox{TR}\left\lbrack\vec{B}\right\rbrack\left( \vec{r},t \right)
&=&
-\vec{B}\left( \vec{r},-t \right).
\label{TRS4}
\end{eqnarray}
These laws of transformation can be expressed as a function of frequency $\omega$ rather than time $t$. 
Using the fact that the electromagnetic fields are real, we obtain after Fourier transformation \cite{Rosny2010}
\begin{eqnarray}
\mbox{TR}\left\lbrack\vec{E}\right\rbrack\left( \vec{r},\omega \right)
&=&
\vec{E}^{\star}\left( \vec{r},\omega \right),
\label{TRS5}\\
\mbox{TR}\left\lbrack\vec{B}\right\rbrack\left( \vec{r},\omega \right)
&=&
-\vec{B}^{\star}\left( \vec{r},\omega \right).
\label{TRS6} 
\end{eqnarray}
Using the above equations and the compact notation defined in eq. \eqref{genconst} (main text), we obtain
\begin{eqnarray}
\mbox{TR}\left\lbrack\vec{\mathcal{E}}\right\rbrack &=& \begin{bmatrix}
\mbox{TR}\left\lbrack\vec{E}\right\rbrack \\
\mbox{TR}\left\lbrack\vec{H}\right\rbrack 
\end{bmatrix}
=
\begin{bmatrix}
\vec{E}^{\star} \\
-\vec{H}^{\star}
\end{bmatrix}
\equiv 
\hat{\sigma}_z \vec{\mathcal{E}}^{\star}.
\label{TRScat1}
\end{eqnarray}
We now look for the transformed scattering-matrix $\mbox{TR}\left\lbrack\stackrel{\leftrightarrow}{\mathcal{S}}\right\rbrack$, defined as 
$\mbox{TR}\left\lbrack\vec{\mathcal{E}}_{\rm{out}}\right\rbrack=\mbox{TR}\left\lbrack\stackrel{\leftrightarrow}{\mathcal{S}}\right\rbrack \mbox{TR}\left\lbrack\vec{\mathcal{E}}_{\rm{in}}\right\rbrack$.
Using eq. \eqref{TRScat1} and the fact that output-fields become input-fields under TR and viceversa, we obtain the TR of eq. \eqref{SC3} (main text) as
\begin{eqnarray}
\begin{bmatrix}
\vec{\mathcal{E}}^{\star}_{1,\rm{in}}\\
\vec{\mathcal{E}}^{\star}_{2,\rm{in}}\\
\end{bmatrix}
&=&
\stackrel{\leftrightarrow}{\sigma}_z
\mbox{TR}\left\lbrack\stackrel{\leftrightarrow}{\mathcal{S}}\right\rbrack
\stackrel{\leftrightarrow}{\sigma}_z
\begin{bmatrix}
\vec{\mathcal{E}}^{\star}_{1,\rm{out}}\\
\vec{\mathcal{E}}^{\star}_{2,\rm{out}}\\
\end{bmatrix},
\label{TRScat3}
\end{eqnarray}
with $\hat{\sigma}_z$ defined in Sec. \ref{Greenscatt}.
We now invert and complex conjugate eq. \eqref{SC3}, yielding $\vec{\mathcal{E}}^{\star}_{\rm{in}}=\left(\stackrel{\leftrightarrow}{\mathcal{S}^{\star}}\right)^{-1}\vec{\mathcal{E}}^{\star}_{\rm{out}}$.
The latter expression coincides with eq. \eqref{TRScat3} for any input-field, which implies that the scattering-matrix is transformed under TR as
\begin{eqnarray}
\mbox{TR}\left\lbrack\stackrel{\leftrightarrow}{\mathcal{S}}\right\rbrack
&=&
\stackrel{\leftrightarrow}{\sigma}_z
\left(\stackrel{\leftrightarrow}{\mathcal{S}^{\star}}\right)^{-1}
\stackrel{\leftrightarrow}{\sigma}_z.
\label{TRScat5}
\end{eqnarray}
Expressing the fact that an optical system is TR-symmetric if and only if $\mbox{TR}\left\lbrack\stackrel{\leftrightarrow}{\mathcal{S}}\right\rbrack=\stackrel{\leftrightarrow}{\mathcal{S}}$, provides back the general constraint on the scattering-matrix written in eq. \eqref{TRscattering} (main text).

\end{document}